\documentclass[aps,prd,tightenlines,showpacs,,floatfix,eqsecnum,%
amssymb,byrevtex,nofootinbib]{revtex4}  %endfloats

\usepackage{epsfig,bm,amsmath,amssymb}

\begin{document}

\preprint{hep-ph/0405010}

%%%%%%%%%%%%%%%%%%%%% Title %%%%%%%%%%%%%%%%%%%%%%

\title{Pentaquark baryons in SU(3) quark model}

%%%%%%%%%%%%%%%%%%%% Authors %%%%%%%%%%%%%%%%%%%%%
%%%%%%%%%%%%%%%%%%%% Addresses %%%%%%%%%%%%%%%%%%%%%

\author{Yongseok Oh}%
\email{yoh@physast.uga.edu}

\affiliation{Department of Physics and Astronomy,
University of Georgia, Athens, GA 30602, U.S.A.}

\author{Hungchong Kim}%
\email{hung@phya.yonsei.ac.kr}

\affiliation{Institute of Physics and Applied Physics,
Yonsei University, Seoul 120-749, Korea}

\date{\today}

%%%%%%%%%%%%%%%%%%%% Abstract %%%%%%%%%%%%%%%%%%%%%

\begin{abstract}

We study the SU(3) group structure of pentaquark baryons which are made
of four quarks and one antiquark.
The pentaquark baryons form $\bm{1}$, $\bm{8}$, $\bm{10}$,
$\overline{\bm{10}}$, $\bm{27}$, and $\bm{35}$ multiplets in SU(3) quark model.
First, the flavor wave functions of all the pentaquark baryons are
constructed in SU(3) quark model and then the flavor SU(3) symmetry
relations for the interactions of the pentaquarks with three-quark
baryons and pentaquark baryons are obtained.
Constructing the general interactions in SU(3) could be important
for understanding the pentaquark baryon properties from reaction mechanisms.
We also discuss possible pentaquarks in $\bm{27}$-plet and $\bm{35}$-plet
and their decay channels that can be used to identify them in 
future experiments.
The mass sum rules for the pentaquark baryons 
are also presented.

\end{abstract}


\maketitle

\section{Introduction}

The discovery of $\Theta^+(1540)$ by the LEPS Collaboration at SPring-8
\cite{LEPS03} and the observations of the subsequent experiments
\cite{DIANA03,CLAS03-b,SAPHIR03,CLAS03-c,ADK03,CLAS03-d,AMW03,HERMES-04,%
SVD04,AER04,Salur04,PHENIX-04}
have initiated great interests in exotic baryons in hadron physics.
Because of its positive strangeness, the minimal quark content 
must be $udud\bar{s}$ and hence 
$\Theta^+$ is an exotic pentaquark state.
The observation of $\Xi^{--}(1862)$ by NA49 Collaboration \cite{NA49-03}
may suggest that $\Xi(1862)$ forms pentaquark antidecuplet with
$\Theta^+(1540)$, which must be confirmed by other experiments
\cite{FW04}.%
\footnote{The recent report from WA89 Collaboration \cite{WA89-04,Pocho04}
shows no evidence for $\Xi(1860)$ in $\Sigma^-$-nucleus collisions.}
Recently, a pentaquark state containing anti-charm quark was observed by
H1 Collaboration \cite{H1-04}.
Because of the observation of pentaquark states in various reaction channels,
the existence of pentaquark baryons now becomes widely accepted.%
\footnote{However, higher statistics experiments are required to firmly
establish the observed resonances. 
See Ref.~\cite{Pocho04} for a compilation of positive and negative reports
on the existence of $\Theta^+(1540)$ in various experiments.
Another interpretation for the $\Theta^+(1540)$ peak was suggested by
Ref.~\cite{DKST03}.}
Thus it is natural to search for the other pentaquark baryons which have
been predicted by hadron models.

Pentaquark baryons may be pure exotic or crypto-exotic.
The pure exotic states can easily be identified by their unique quantum
numbers, but the crypto-exotic states are hard to be identified as their
quantum numbers can also be generated by three-quark states.
Therefore, it is crucial to have careful analyses for their decay channels.
Historically, there have been many efforts to find pentaquark states with
the development of quark models, which, however, failed to observe
$\Theta(1540)$.
The efforts to search for pentaquark baryons until 1980's were summarized in
Refs.~\cite{Golo71,PDG86}. (See also Ref.~\cite{GM99}.)%
\footnote{
In the literature we could find several resonances that were claimed to be
crypto-exotic states.
For example, $X(1340)$, $X(1450)$, and $X(1640)$ were reported by
Ref.~\cite{HKKK79} and $X(3520)$ by Ref.~\cite{KMCM91}.
$X(1390)$, $X(1480)$, and $X(1620)$ that have isospin $I \ge 5/2$ were
observed by Ref.~\cite{ABGG90}, and Ref.~\cite{BCGMP79} reported
$\Sigma(3170)$.
Most of them were found to have narrow widths, but their existence was not
confirmed and questioned by later experiments \cite{AGDD91,ACDD85}.
SPHINX Collaboration has reported the existence of $X(2000)$, $X(2050)$,
and $X(2400)$ that are expected to have the quark content of $uuds\bar{s}$
\cite{Lands99}, whose existence should be carefully re-examined by other
experiments.}
Early theoretical works on exotic baryons can be found, e.g., in
Refs.~\cite{early}.
Rigorous theoretical studies were then performed for heavy quark sector,
i.e., pentaquark baryons with one anti-charmed quark or anti-bottom
quark.
In the pioneering work of Lipkin \cite{Lip87} and Grenoble group
\cite{GSR87}, the anti-charmed pentaquark with one strange quark was
shown to have the same binding energy as the $H$ dibaryon in the heavy
quark mass limit and in the SU(3) limit.
Then it has been studied in more sophisticated quark models
\cite{FGRS89-ZR94,Stan98-GRSP97} and in Skyrme model
\cite{RS93,OPM94b-OPM94c-OP95}.
Following the first experimental search for heavy pentaquarks
\cite{E791-98-E791-99}, the observation of $\Theta^+(1540)$ and
$\Theta_c(3099)$ has brought new interests in this subject
\cite{Cheung03,CCH04-HL04,BKM04,SWW04,WM04c,KLO04}.

In the light quark sector, pentaquark states were anticipated in the
Skyrme model \cite{Chem85-Man84-Pras87,Weig98}.
The first detailed study on antidecuplet was made by Diakonov {\it
et al.\/} \cite{DPP97,PSTCG00}, which predicted a very narrow $\Theta^+$
with a mass around 1530 MeV by identifying $N(1710)$ as the nucleon
analogue of the antidecuplet.
After the discovery of $\Theta^+(1540)$ there have been lots of
theoretical models and ideas to explain the structure of pentaquark
baryons and to search for the other pentaquark states.
The subsequent theoretical studies include the soliton models
\cite{WK03,Pras03,IKOR03,JM03,BFK03}, QCD sum rules
\cite{Zhu03-MNNRL03-SDO03-Eidem04}, large $N_c$ QCD
\cite{Cohen03-CL04,Man04}, and
lattice calculation \cite{CFKK03,Sasaki03,CH04a}, etc.
As the quark models have provided a cornerstone for hadron physics, it
is legitimate to start with the quark models and study the structure
of pentaquark baryons.
In Ref.~\cite{KL03a}, Karliner and Lipkin suggested a triquark-diquark model,
where, for example, $\Theta^+$ is a system of $(ud)$-$(ud\bar{s})$.
In Ref.~\cite{JW03}, Jaffe and Wilczek advocated a
diquark-diquark-antiquark model so that $\Theta^+$ is $(ud)$-$(ud)$-$\bar{s}$.
 In this model, they also considered the mixing of the pentaquark
antidecuplet with the pentaquark octet, which makes it different from the
SU(3) soliton models where the octet describes the normal (three-quark)
baryon octet.
Assuming that the nucleon and $\Sigma$ analogues
are in the ideal mixing
of the octet and antidecuplet, the nucleon analogue is then identified as
the Roper resonance $N(1440)$.
In Ref.~\cite{OKL03b}, it was pointed out that the Roper resonance $N(1710)$
should be excluded as a pure antidecuplet state.
This is because, within SU(3) symmetry, antidecuplet does not couple to
decuplet and meson octet, whereas $N(1710)$ has a large branching ratio into
$\pi\Delta$ channel.
Therefore, mixing with other multiplets is required if one wants to identify
$N(1710)$ as a pentaquark crypto-exotic state.
However, recent study for the ideal mixing between antidecuplet and octet
states shows that the ideally mixed state still has vanishing coupling with
the $\pi\Delta$ channel \cite{CD04,LKO04}, which excludes $N(1440)$ as a
pentaquark state.
This shows the importance of reaction/decay studies in identifying
especially crypto-exotic pentaquark states.
More discussions on the quark model predictions based on the diquark
picture can be found, e.g., in Refs.~\cite{OKL03b,DP03b}.
Predictions on the antidecuplet spectrum in various quark models can be
found, e.g., in Refs.~\cite{BGS03,SR03,CCKN03a,CCKN04b,GK03b,KLV04}

In quark model, pentaquark baryons form six multiplets, $\bm{1}$,
$\bm{8}$, $\bm{10}$, $\overline{\bm{10}}$, $\bm{27}$, and $\bm{35}$.
The other type resonances are thus expected together with antidecuplet, 
particularly the isovector $\Theta$ belonging to $\bm{27}$-plet and 
isotensor $\Theta$ as a member of $\bm{35}$-plet. 
The interest in this direction has been growing
\cite{BGS03,Pras04,WM04a-WM04b,EKP04,DS04} and
it is important to know the interactions and decay channels to
search for the other pentaquark baryons.

Furthermore, understanding the $\Theta^+$ properties such as spin-parity
requires careful analyses of production mechanisms including $\gamma N
\to \bar{K} \Theta$ \cite{LK03a-LK03b,NHK03,OKL03a,ZA03,NL04,YCJ03},
$\gamma N \to \bar{K}^* \Theta$ \cite{OKL03d},
$\gamma N \to K \pi \Theta$ \cite{LKK03}
$\gamma N \to K^+ K^- \Theta$ \cite{NT03},
$NN \to Y\Theta$ \cite{OKL03c,THH03,HBEK03}, and
$K N \to K \pi N$ \cite{HHO03}.
Most model predictions for those production processes, however, do not
consider the intermediate pentaquark baryons in its production mechanisms
as the unknown inputs like the electromagnetic and strong couplings 
of pentaquark baryons are required.
Therefore, knowing the interaction Lagrangian of pentaquark baryons are
necessary.

The physical pentaquark states would be mixtures of various multiplets
as in the chiral soliton model \cite{EKP04}.
Such a representation mixing is induced by SU(3) symmetry breaking
and it can be studied in quark potential models.
Therefore, it is desirable to obtain the full set of pentaquark wave
functions in quark model for further investigation.
In this paper, we construct the flavor wave functions of pentaquark
baryons in SU(3) quark model.
There are several works in this direction and the flavor wave functions
of antidecuplet has been obtained in
Refs.~\cite{CCKN03a,CCKN04b,CCKN03c,CD04}.
(See also Ref.~\cite{Man04} for the relation between the wave functions of
pentaquark baryons in quark model and Skyrme model in the large $N_c$
limit.)
The SU(3) symmetric interactions for antidecuplet have been studied in
Refs.~\cite{OKL03b,LKO04,CD04}, which motivated the development of a
chiral Lagrangian for antidecuplet \cite{KLLP03,LZHD04}.
In this work, we extend the SU(3) quark model to pentaquark states and
obtain the flavor wave functions of {\it all\/} pentaquark states including
singlet, octet, decuplet, antidecuplet, $\bm{27}$-plet, and
$\bm{35}$-plet.
Then we obtain the SU(3) symmetric Lagrangian of
pentaquark--three-quark and pentaquark-pentaquark interactions with
meson octet.

This paper is organized as follows.
In the next Section, we start with the quark and antiquark operators and
form a diquark state.
Then by taking direct product of two diquarks and one antiquark, we form
the pentaquark states in tensor notation.
The physical baryon states are, therefore, represented by SU(3) tensors
as shown in Sect.~III.
This allows us to obtain the flavor wave functions of pentaquark
baryons by identifying each SU(3) tensor with physical pentaquark states.
Although the procedure in this paper follows the diquark-diquark-antiquark
picture, the obtained wave function is general as we do not impose any
dynamics to the quarks.
In Sect.~IV, we develop SU(3) symmetric Lagrangian for pentaquark
baryons.
We consider the interactions of pentaquarks with normal three-quark
baryons (octet and decuplet) and meson octet.
Then it is extended to construct the interactions of pentaquark baryons
with (other) pentaquark multiplets and meson octet.
The SU(3) symmetric case is considered in this work, but the symmetry
breaking can be included in a standard way \cite{PS04}.
In Sect. V, the mass relations among pentaquark baryons are obtained.
We found Gell-Mann--Okubo mass relation for $\bm{27}$-plet and equal
spacing rules not only for $\bm{10}$ and $\overline{\bm{10}}$ but also
for $\bm{27}$ and $\bm{35}$.
Section VI contains a summary.

\section{Wave functions of pentaquark baryons}

We start with the representations for quark and antiquark.
We denote a quark by $q_i$ and an antiquark by $q^i$ with $i=1,2,3$, so
that $q_1$, $q_2$, and $q_3$ are $u$, $d$, and $s$ quark, respectively.
The inner products of the quark and antiquark operators are normalized as
\begin{equation}
(q_i, q_j) = \delta_{ij}, \qquad (q^i, q^j) = \delta^{ij}, \qquad
(q_i, q^j) = 0.
\end{equation}

\subsection{Diquark}

\begin{figure}[t]
\centering
\epsfig{file=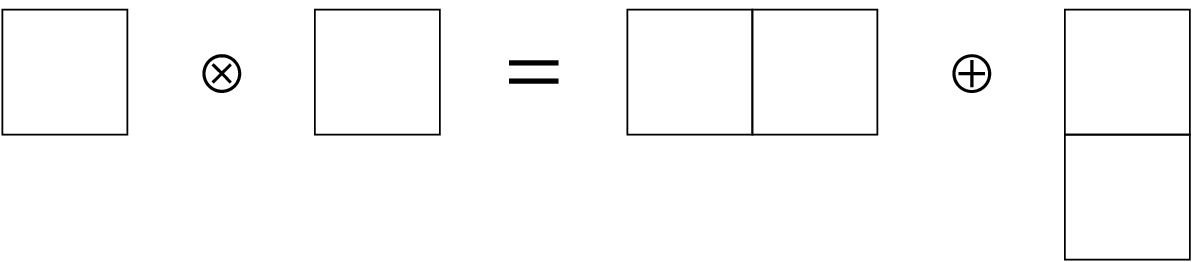, width=0.5\hsize}
\caption{Product of two quarks, $\bm{3} \otimes \bm{3} = \bm{6} \oplus
\overline{\bm{3}}$.}
\label{fig:diq}
\end{figure}

We first construct a diquark by a direct product of two quarks.
In flavor SU(3) group, the quarks $q_i$ are in the fundamental
representation $\bm{3}$, which is described by a box in Young tableau.
The direct product of two quarks then gives $\bm{3} \otimes \bm{3} =
\bm{6} \oplus \overline{\bm{3}}$ as shown in Fig.~\ref{fig:diq}.
In $(p,q)$ notation, $\bm{6}$ is $(2,0)$ type and $\overline{\bm{3}}$ is
$(0,1)$ type.
Generally, $(p,q)$ type can be represented by a tensor $T_{a_1, \dots,
a_p}^{b_1, \dots, b_q}$, which is completely symmetric in upper indices
and in lower indices, namely, 
\begin{equation}
T_{a_1, a_2, \dots, a_p}^{b_1, b_2, \dots, b_q} = 
T_{a_2, a_1, \dots, a_p}^{b_1, b_2, \dots, b_q} = 
T_{a_1, a_2, \dots, a_p}^{b_2, b_1, \dots, b_q} = 
T_{a_2, a_1, \dots, a_p}^{b_2, b_1, \dots, b_q},
\end{equation}
etc. It is also traceless on every pair of indices so that
\begin{equation}
T_{a_1, a_2, \dots, a_p}^{a_1, b_2, \dots, b_q} = 0.
\end{equation}
Therefore, $\bm{6}$ is represented by a tensor $T_{ij}$ and
$\overline{\bm{3}}$ is by $T^i$ like in the case of antiquarks.

\begin{table}[t]
\centering
\begin{tabular}{c|c|c} \hline\hline
irreducible tensor & type & notation \\ \hline
$S_{jk}$ & $(2,0)$ & $\bm{6}$ \\ 
$T^i$ & $(0,1)$ & $\overline{\bm{3}}$ \\ \hline\hline
\end{tabular}
\caption{Irreducible tensors of a diquark.}
\label{tab:diq}
\end{table}

Explicitly, by symmetrizing and anti-symmetrizing, the product of
two quarks is written as
\begin{equation}
q_j q_k = \frac{1}{\sqrt2} S_{jk} + \frac{1}{2\sqrt2} \epsilon_{ijk}
T^i,
\end{equation}
where
\begin{equation}
\left( \begin{array}{c} S_{jk} \\ A_{jk} \end{array} \right) =
\frac{1}{\sqrt2} \left( q_j q_k \pm q_k q_j \right),
\end{equation}
and
\begin{equation}
T^i = \epsilon^{ijk} A_{jk},
\end{equation}
so that $S_{jk}$ and $T^i$ represent $\bm{6}$ and $\overline{\bm{3}}$,
respectively.
The inner products of $S_{jk}$ and $T^i$ are then obtained as
\begin{equation}
(S_{jk}, S_{lm}) = \delta_{jl} \delta_{km} + \delta_{jm} \delta_{kl},
\qquad
(T^i, T^j) = 4 \delta^{ij}.
\label{diq}
\end{equation}
Since $S_{jk}$ and $T^i$ are irreducible representations, the inner
product $(S_{jk}, T^i)$ vanishes.

\subsection{Two diquarks}

\begin{figure}[t]
\centering
\epsfig{file=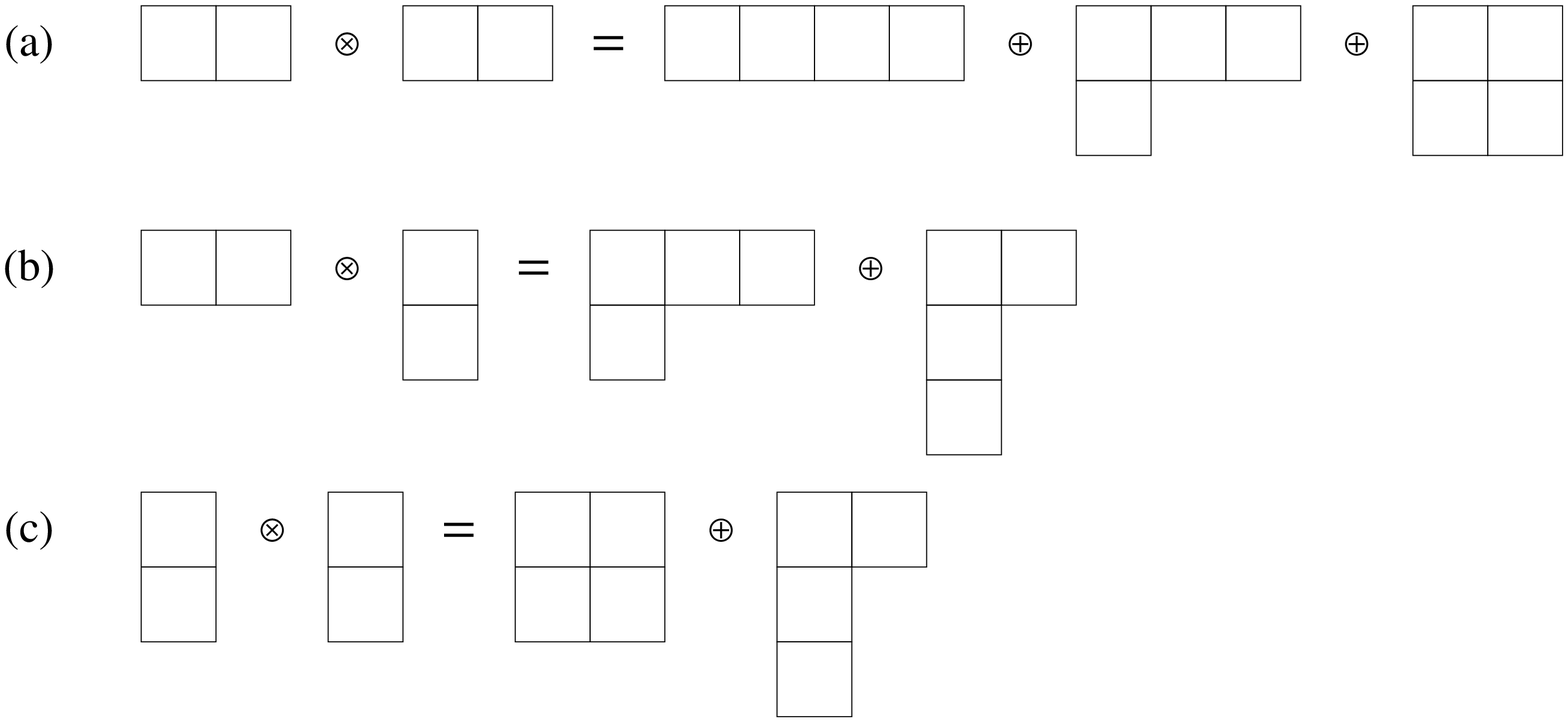, width=0.9\hsize}
\caption{Products of two diquarks. (a) $\bm{6} \otimes \bm{6} =
\bm{15}_1 \oplus \bm{15}_2 \oplus \overline{\bm{6}}$, (b) $\bm{6} \otimes
\overline{\bm{3}} = \bm{15}_2 \oplus \bm{3}$, and (c) $\overline{\bm{3}}
\otimes \overline{\bm{3}} = \overline{\bm{6}} \oplus \bm{3}$.}
\label{fig:2diq}
\end{figure}

Since one diquark is either in $\bm{6}$ or $\overline{\bm{3}}$, the direct
product of two diquarks contains
$\bm{6} \otimes \bm{6}$, $\bm{6} \otimes \overline{\bm{3}}$,
$\overline{\bm{3}} \otimes \bm{6}$, and
$\overline{\bm{3}} \otimes \overline{\bm{3}}$:
\begin{eqnarray}
(q_j q_k) (q_l q_m) &=& \frac12 S_{jk} S_{lm} + \frac14 \left(
\epsilon_{ijk} T^i S_{lm} + \epsilon_{ilm} S_{jk} T^i \right) + \frac18
\epsilon_{ijk} \epsilon_{nlm} T^i T^n.
\end{eqnarray}
The direct products of each multiplets are shown in Fig.~\ref{fig:2diq}.

We start with noting that the product
$\overline{\bm{3}} \otimes \overline{\bm{3}}$ is decomposed into two parts
as
\begin{equation}
T^i T^j = 2\sqrt{2} S^{ij} + 2 \epsilon^{ijk} T_k,
\end{equation}
where
\begin{eqnarray}
S^{ij} &=& \frac{1}{4\sqrt2} \left( T^i T^j + T^j T^i \right),
\nonumber \\
T_k &=& \frac{1}{8} \epsilon_{ijk} \left( T^i T^j - T^j T^i \right).
\label{eq:3x3f}
\end{eqnarray}
As shown in Fig.~\ref{fig:2diq}, this means that $\overline{\bm{3}}
\otimes \overline{\bm{3}} = \overline{\bm{6}} \oplus \bm{3}$.

In the case of $\bm{6} \otimes \overline{\bm{3}}$, we have
\begin{equation}
S_{jk} T^i = T^i_{jk} + \frac{1}{\sqrt2} \left( \delta^i_j \delta^m_k +
\delta^i_k \delta^m_j \right) Q_m,
\end{equation}
where
\begin{eqnarray}
Q_m &=& \frac{1}{\sqrt8} S_{ml} T^l, \nonumber \\
T^i_{jk} &=& S_{jk} T^i - \frac{1}{\sqrt2}  \left( \delta^i_j \delta^m_k +
\delta^i_k \delta^m_j \right) Q_m,
\label{eq:6x3f}
\end{eqnarray}
which shows that $\bm{6} \otimes \overline{\bm{3}} = \bm{15} \oplus
\bm{3}$. 
One can easily make sure that $T^i_{jk}$ represents $\bm{15}$.
Namely, the 6 states from the two (symmetric) lower indices and 3 states
from the upper index give 18 possible states, which are reduced to 15
when combined with three traceless conditions.

Similarly, $\overline{\bm{3}} \otimes \bm{6}$ is broken down to
\begin{equation}
T^i S_{jk} = \widetilde{T}^i_{jk} +
\frac{1}{\sqrt2} \left( \delta^i_j \delta^m_k
+ \delta^i_k \delta^m_j \right) \widetilde{Q}_m,
\end{equation}
where
\begin{eqnarray}
\widetilde{Q}_m &=& \frac{1}{\sqrt8} T^l S_{ml} , \nonumber \\
\widetilde{T}^i_{jk} &=& T^i S_{jk} - \frac{1}{\sqrt2} \left(
\delta^i_j \delta^m_k + \delta^i_k \delta^m_j \right) \widetilde{Q}_m.
\label{eq:6x3f2}
\end{eqnarray}

The product $\bm{6} \otimes \bm{6}$ is obtained as
\begin{equation}
S_{jk} S_{lm} = \frac{1}{\sqrt6} T_{jklm} + \frac{1}{2\sqrt2} \left(
\epsilon_{akl} \delta^b_m \delta^c_j + \epsilon_{ajm} \delta^b_l
\delta^c_k \right) S^a_{bc}
+ \frac{1}{\sqrt6} \left( \epsilon_{ajl} \epsilon_{bkm} + \epsilon_{akl}
\epsilon_{bjm} \right) T^{ab},
\label{eq:6x6}
\end{equation}
where
\begin{eqnarray}
T^{ij} &=& \frac{1}{\sqrt6} \epsilon^{iab} \epsilon^{jcd} S_{ac} S_{bd},
\nonumber \\
S^i_{jk} &=& \frac{1}{\sqrt2} \epsilon^{ilm} \left( S_{jl} S_{km} +
S_{kl} S_{jm} \right),
\nonumber \\
T_{jklm} &=& \frac{1}{\sqrt6} \left( S_{jk} S_{lm} + S_{lk} S_{jm} +
S_{jm} S_{kl} + S_{lj} S_{km} + S_{km} S_{jl} + S_{lm} S_{jk} \right),
\label{eq:6x6f}
\end{eqnarray}
as we have expected from $\bm{6} \otimes \bm{6} = \bm{15}_1 \oplus \bm{15}_2
\oplus \overline{\bm{6}}$.
Here, we have an identity,
\begin{equation}
\epsilon_{alk} T^a_{jm} + \epsilon_{akj} T^a_{lm} + \epsilon_{ajl}
T^a_{km} = 0,
\end{equation}
which is valid for $T^i_{jk}$ in Eq.~(\ref{eq:6x3f}), $\widetilde{T}^i_{jk}$
in Eq.~(\ref{eq:6x3f2}), and $S^i_{jk}$ in Eq.~(\ref{eq:6x6f}).

By collecting the above results, we have
\begin{eqnarray}
(q_j q_k) (q_l q_m) &=& \frac{1}{2\sqrt6} T_{jklm}
+ \frac{1}{4\sqrt2} \left( \epsilon_{akl} \delta^b_m \delta^c_j +
\epsilon_{ajm} \delta^b_l \delta^c_k \right) S^a_{bc}
\nonumber \\ && \mbox{}
+ \frac{1}{\sqrt6} \left( \epsilon_{ajl} \epsilon_{bkm} +
\epsilon_{akl}\epsilon_{bjm} \right) T^{ab}
+ \frac14 \epsilon_{ijk} \left\{ T^i_{lm} + \frac{1}{\sqrt2} \left(
\delta^i_l \delta^a_m + \delta^i_m \delta^a_l \right) Q_a \right\}
\nonumber \\ && \mbox{}
+ \frac14 \epsilon_{ijk} \left\{ \widetilde{T}^i_{lm} + \frac{1}{\sqrt2} \left(
\delta^i_l \delta^a_m + \delta^i_m \delta^a_l \right) \widetilde{Q}_a \right\}
+ \frac{1}{4\sqrt2} \epsilon_{ijk} \epsilon_{nlm} \left( 2 S^{in} +
\epsilon^{ina} T_a \right).
\end{eqnarray}
The obtained irreducible representations for four-quarks are summarized
in Table~\ref{tab:2diq}.

The inner products of the multiplets are obtained as follows.
First, for $T_a$, $Q_a$, and $\widetilde{Q}_a$ of $(1,0)$ type, we have
\begin{equation}
(T_a, T_b) = 2 \delta_{ab}.
\end{equation}
For $T^{ij}$ and $S^{ij}$ of $(0,2)$ type, we have
\begin{equation}
(T^{ij}, T^{lm}) = \delta^{il} \delta^{jm} + \delta^{im} \delta^{jl}.
\end{equation}
For $T^{i}_{jk}$, $\widetilde{T}^i_{jk}$, and $S^{i}_{jk}$ of $(2,1)$ type,
which satisfy the traceless condition, $T^i_{ik}=0$, we have
\begin{equation}
(T^{i}_{jk}, T^{l}_{mn}) = 4 \delta^{il} \left(\delta_{jm}\delta_{kn}
+ \delta_{jn} \delta_{km} \right) - \delta^i_j \left( \delta^l_m
\delta_{kn} + \delta^l_n \delta_{km} \right) - \delta^i_k \left(
\delta^l_m \delta_{jn} + \delta^l_n \delta_{jm} \right).
\end{equation}
For $T_{ijkl}$ of $(4,0)$ type, we have
\begin{eqnarray}
(T_{jklm}, T_{abcd}) &=&
(S_{jk}, S_{ab}) (S_{lm}, S_{cd}) + (S_{jk}, S_{bc}) (S_{lm}, S_{ad})
+(S_{jk}, S_{ad}) (S_{lm}, S_{bc})
\nonumber \\ && \mbox{}
+ (S_{jk}, S_{ac}) (S_{lm}, S_{bd})
+(S_{jk}, S_{bd}) (S_{lm}, S_{ac}) + (S_{jk}, S_{cd}) (S_{lm}, S_{ab}),
\nonumber \\ && \mbox{}
\end{eqnarray}
where
\begin{equation}
(S_{jk}, S_{lm}) = \delta_{jl} \delta_{km} + \delta_{jm} \delta_{kl},
\end{equation}
as in Eq.~(\ref{diq}).

\begin{table}[t]
\centering
\begin{tabular}{c|c|c|c} \hline\hline
irreducible tensor & type & notation & source \\ \hline
$T_i$ & $(1,0)$ & $\bm{3}$ & $\overline{\bm{3}} \otimes
\overline{\bm{3}}$ \\
$Q_i$ & $(1,0)$ & $\bm{3}$ & $\bm{6} \otimes
\overline{\bm{3}}$ \\
$\widetilde{Q}_i$ & $(1,0)$ & $\bm{3}$ & $\overline{\bm{3}} \otimes \bm{6}$ \\
$S^{ij}$ & $(0,2)$ & $\overline{\bm{6}}$ & $\overline{\bm{3}} \otimes
\overline{\bm{3}}$ \\ 
$T^{ij}$ & $(0,2)$ & $\overline{\bm{6}}$ & $\bm{6} \otimes \bm{6}$ \\
$T^i_{jk}$ & $(2,1)$ & $\bm{15}_2$ & $\bm{6} \otimes
\overline{\bm{3}}$ \\
$\widetilde{T}^i_{jk}$ & $(2,1)$ & $\bm{15}_2$ & $\overline{\bm{3}} \otimes
\bm{6}$ \\
$S^{i}_{jk}$ & $(2,1)$ & $\bm{15}_2$ & $\bm{6} \otimes \bm{6}$ \\
$T_{ijkl}$ & $(4,0)$ & $\bm{15}_1$ & $\bm{6} \otimes \bm{6}$ \\
\hline\hline
\end{tabular}
\caption{Irreducible tensors of the product of two diquarks.}
\label{tab:2diq}
\end{table}

\subsection{Pentaquarks}

\begin{figure}[t]
\centering
\epsfig{file=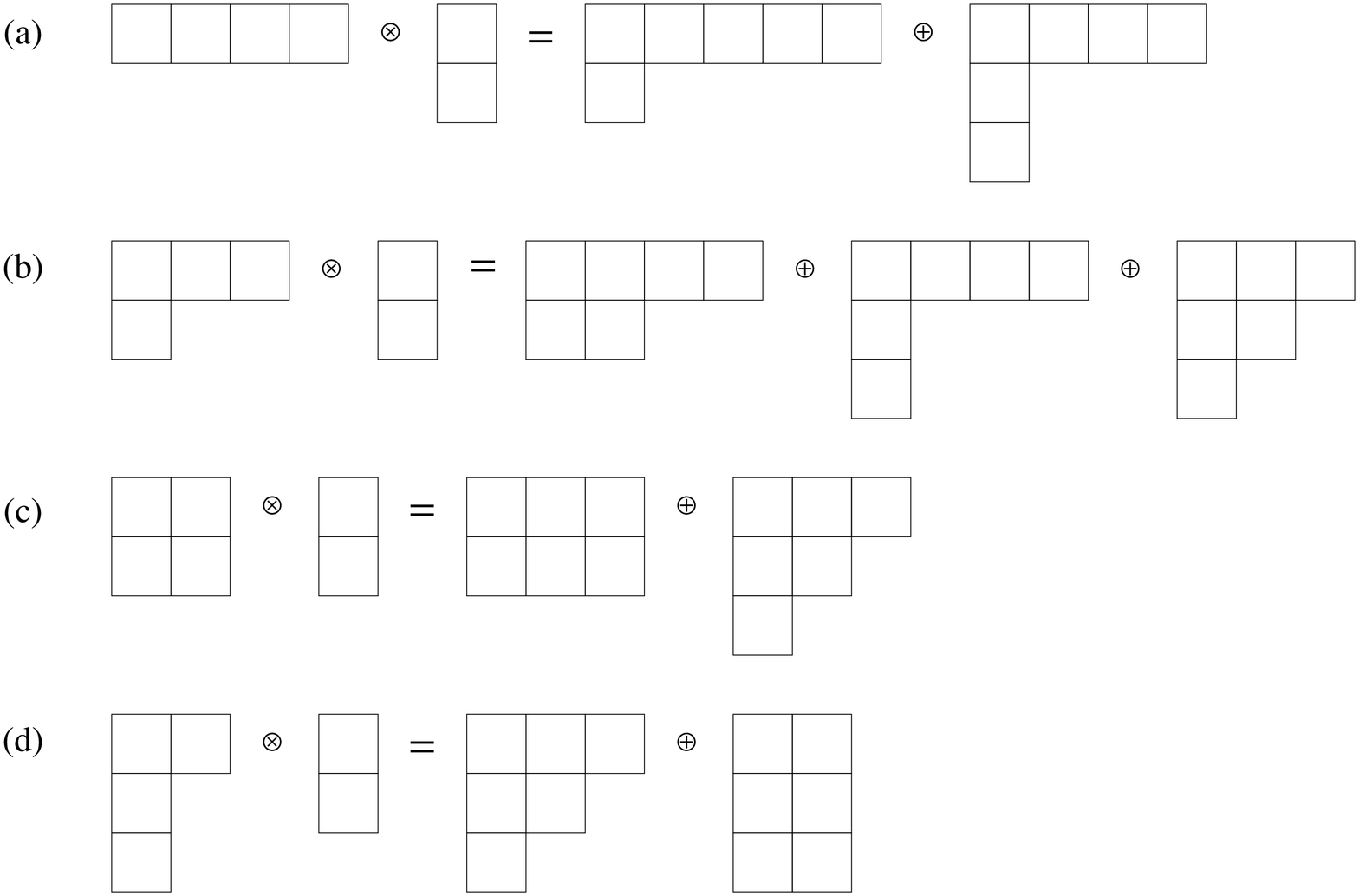, width=0.7\hsize}
\caption{Products of two diquarks and one antiquark.
(a) $\bm{15}_1 \otimes \overline{\bm{3}} = \bm{35} \oplus \bm{10}$,
(b) $\bm{15}_2 \otimes \overline{\bm{3}} = \bm{27} \oplus \bm{10} \oplus
\bm{8}$,
(c) $\overline{\bm{6}} \otimes \overline{\bm{3}} = \overline{\bm{10}}
\oplus \bm{8}$, and
(d) $\bm{3} \otimes \overline{\bm{3}} = \bm{8} \oplus \bm{1}$.}
\label{fig:5q}
\end{figure}

As the two diquarks can form $\bm{3}$, $\overline{\bm{6}}$, $\bm{15}_1$,
and $\bm{15}_2$ and the antiquarks form $\overline{\bm{3}}$, there are six
multiplets for pentaquarks, $\bm{1}$, $\bm{8}$, $\bm{10}$,
$\overline{\bm{10}}$, $\bm{27}$, and $\bm{35}$ as depicted in
Fig.~\ref{fig:5q}.
We now discuss the possible pentaquark multiplets in detail.

\subsubsection{$\bm{3} \otimes \overline{\bm{3}}$}

The two diquarks can form $\bm{3}$ when they are in $\overline{\bm{3}} \otimes
\overline{\bm{3}}$ or $\bm{6} \otimes \overline{\bm{3}}$.
Since
\begin{equation}
\bm{3} \otimes \overline{\bm{3}} = \bm{8} \oplus \bm{1},
\end{equation}
the pentaquarks are either in singlet or in octet representation in this
case.
Let $T_i$ stand for $T_i$, $Q_i$, or $\widetilde{Q}_i$ of $(1,0)$ type for
the four-quarks, then we
have
\begin{equation}
T_l^{}\, \bar{q}^q = \sqrt{2} \left( P^q_l + \frac{1}{\sqrt3} \delta^q_l S
\right),
\end{equation}
where
\begin{eqnarray}
S &=& \frac{1}{\sqrt{6}} T_m \bar{q}^m, \nonumber \\
P^q_l &=& \frac{1}{\sqrt2} \left( T_l \bar{q}^q - \sqrt{\frac{2}{3}}
\delta^q_l S \right).
\end{eqnarray}
Thus, $S$ and $P^q_l$ represent the pentaquark singlet and pentaquark octet,
respectively.

\subsubsection{$\overline{\bm{6}} \otimes \overline{\bm{3}}$}

When the two diquarks are in $\overline{\bm{3}} \otimes \overline{\bm{3}}$ or
in $\bm{6} \otimes \bm{6}$, they can form $\overline{\bm{6}}$.
In this case, we have antidecuplet and octet as
\begin{equation}
\overline{\bm{6}} \otimes \overline{\bm{3}} = \overline{\bm{10}} \oplus
\bm{8}.
\end{equation}
This is the only case where we can form the antidecuplet.
In other words, antidecuplet baryons can be formed only when both
diquarks are in $\overline{\bm{3}}$ or the both are in $\bm{6}$.

For $S^{ij}$ or $T^{ij}$ of $(0,2)$ type for the four-quark system, which we
write as $S^{ij}$ in general, we have
\begin{equation}
S^{ij} \bar{q}^q = \frac{1}{\sqrt3} \left\{ T^{ijq} + \left(
\epsilon^{ljq} P^i_l + \epsilon^{liq} P^j_l \right) \right\},
\end{equation}
where
\begin{eqnarray}
T^{ijk} &=& \frac{1}{\sqrt3} \left( S^{ij} \bar{q}^k + S^{ik} \bar{q}^j
+ S^{jk} \bar{q}^i \right),
\nonumber \\
P^i_l &=& \frac{1}{\sqrt3} \epsilon_{lab} S^{ia} \bar{q}^b,
\end{eqnarray}
which represent the pentaquark antidecuplet and pentaquark octet,
respectively.

\subsubsection{$\bm{15}_{1} \otimes \overline{\bm{3}}$}

If both of the two diquarks are in $\bm{6}$, then one can form
four-quark in $\bm{15}$-plet of $(4,0)$ type, which is represented by
$T_{ijkl}$.
Since
\begin{equation}
{\bm{15}}_1 \otimes \overline{\bm{3}} = {\bm{35}} \oplus \bm{10},
\end{equation}
we can form pentaquark $\bm{35}$ and $\bm{10}$.
Note also that this is the only combination that gives rise to pentaquark
$\bm{35}$-plet.

Explicitly, we have
\begin{eqnarray}
T_{ijkl} \bar{q}^q &=& T^q_{ijkl} + \frac{1}{\sqrt6} \left(
\delta^q_i\delta^b_j\delta^c_k\delta^d_l +
\delta^q_j\delta^b_i\delta^c_k\delta^d_l +
\delta^q_k\delta^b_i\delta^c_j\delta^d_l +
\delta^q_l\delta^b_i\delta^c_j\delta^d_k \right) D_{bcd},
\end{eqnarray}
where
\begin{eqnarray}
D_{ijk} &=& \frac{1}{\sqrt6} T_{ijkl} \bar{q}^l,
\nonumber \\
T^a_{ijkl} &=& T_{ijkl} \bar{q}^a - \frac{1}{\sqrt6} \left( \delta^a_i
D_{jkl} + \delta^a_j D_{ikl} + \delta^a_k D_{ijl} + \delta^a_l D_{ijk}
\right).
\end{eqnarray}

\subsubsection{$\bm{15}_{2} \otimes \overline{\bm{3}}$}

The $\bm{15}_2$ multiplet of two diquark system can be formed when one
of the diquarks is in $\bm{6}$ and the other diquark in $\overline{\bm{3}}$.
Since
\begin{equation}
{\bm{15}}_2 \otimes \overline{\bm{3}} = {\bm{27}} \oplus \bm{10} \oplus
\bm{8},
\end{equation}
the $\bm{27}$-plet can be formed only in this case.
Explicitly, if we write $T^i_{kl}$ for $T^i_{kl}$, $\widetilde{T}^i_{kl}$,
and $S^i_{kl}$, we obtain
\begin{eqnarray}
T^i_{kl} \bar{q}^q = \sqrt{2}\, T^{iq}_{kl} + \sqrt{\frac23}
\epsilon^{iqm} D_{mkl}
+ \frac{4}{\sqrt{15}} \left( \delta^q_l \delta^b_k + \delta^b_l
\delta^q_k \right) P^i_b
- \frac{1}{\sqrt{15}} \left( \delta^i_l \delta^b_k + \delta^b_l
\delta^i_k \right) P^q_b,
\end{eqnarray}
where
\begin{eqnarray}
P^i_j &=& \frac{1}{\sqrt{15}} T^i_{jk} \bar{q}^k,
\nonumber \\
D_{jkl} &=& \frac{1}{\sqrt{24}} \left( \epsilon_{jab} T^a_{kl} \bar{q}^b
+ \epsilon_{kab} T^a_{jl} \bar{q}^b + \epsilon_{lab} T^a_{jk} \bar{q}^b
\right),
\nonumber \\
T^{ij}_{kl} &=& \frac{1}{2\sqrt{2}} \left( T^i_{kl} \bar{q}^j + T^j_{kl}
\bar{q}^i \right) - \frac{1}{10\sqrt{2}} \left( \delta^i_l P^j_k +
\delta^j_l P^i_k + \delta^i_k P^j_l + \delta^j_k P^i_l \right).
\end{eqnarray}

\subsubsection{Inner products}

The results for the pentaquark multiplets obtained in this Section
are summarized in Table~\ref{tab:penta}.
Using the expressions given above, it is now straightforward to check the
traceless conditions, $P^i_i = T^{ij}_{ik} = T^i_{ijkl} = 0$.
Furthermore, the inner products are obtained as
\begin{eqnarray}
(S, S) &=& 1,
\\
(P^i_j, P^k_l) &=& \delta^{ik} \delta_{jl} - \frac13 \delta^i_j
\delta^k_l,
\\
(D_{ijk}, D_{lmn}) &=& ( S_{ij}, S_{lm}) \,\delta_{kn}
+ (S_{ij}, S_{ln})\, \delta_{km} + (S_{ij}, S_{mn}) \,\delta_{kl},
\\
(T^{ijk}, T^{lmn}) &=& (S^{ij}, S^{lm})\, \delta^{kn}
+ (S^{ij},S^{ln}) \, \delta^{km} +
(S^{ij},S^{mn})\, \delta^{kl},
\\
(T^{ij}_{kl}, T^{pq}_{rs}) &=& (S^{ij},S^{pq})\, (S_{kl},S_{rs})
- \frac15 \left( \delta^{jp} \delta^q_r + \delta^{jq}\delta^p_r \right)
  \left( \delta^i_k\delta_{ls} + \delta^i_l \delta_{ks} \right)
\nonumber \\ && \mbox{}
- \frac15 \left( \delta^{ip} \delta^q_r + \delta^{iq}\delta^p_r \right)
  \left( \delta^j_k\delta_{ls} + \delta^j_l \delta_{ks} \right)
- \frac15 \left( \delta^{jp} \delta^q_s + \delta^{jq}\delta^p_s \right)
  \left( \delta^i_k\delta_{lr} + \delta^i_l \delta_{kr} \right)
\nonumber \\ && \mbox{}
- \frac15 \left( \delta^{ip} \delta^q_s + \delta^{iq}\delta^p_s \right)
  \left( \delta^j_k\delta_{lr} + \delta^j_l \delta_{kr} \right)
+ \frac{1}{10} \left( \delta^i_k \delta^j_l + \delta^i_l\delta^j_k \right)
  \left( \delta^p_r\delta^q_s + \delta^p_s \delta^q_r \right),
\nonumber \\  \label{in-27} \\
(T^a_{ijkl}, T^b_{pqrs}) &=& \delta^{ab} \left\{ (S_{ij},S_{pq})\,
(S_{kl},S_{rs})
+ (S_{ij},S_{qr})\, (S_{kl},S_{ps}) + (S_{ij},S_{ps})\, (S_{kl},S_{qr})
\right.
\nonumber \\ && \mbox{} \qquad \left. + (S_{ij},S_{pr})\,
(S_{kl},S_{qs})
+ (S_{ij},S_{qs})\, (S_{kl},S_{pr}) + (S_{ij},S_{rs})\, (S_{kl},S_{pq})
\right\}
\nonumber \\ && \mbox{}
- \frac16 \delta^a_i \left\{ \delta^b_p (D_{jkl}, D_{qrs})
+ \delta^b_q (D_{jkl},D_{prs}) + \delta^b_r (D_{jkl}, D_{pqs})
+ \delta^b_s (D_{jkl}, D_{pqr}) \right\}
\nonumber \\ && \mbox{}
- \frac16 \delta^a_j \left\{ \delta^b_p (D_{ikl}, D_{qrs})
+ \delta^b_q (D_{ikl},D_{prs}) + \delta^b_r (D_{ikl}, D_{pqs})
+ \delta^b_s (D_{ikl}, D_{pqr}) \right\}
\nonumber \\ && \mbox{}
- \frac16 \delta^a_k \left\{ \delta^b_p (D_{ijl}, D_{qrs})
+ \delta^b_q (D_{ijl},D_{prs}) + \delta^b_r (D_{ijl}, D_{pqs})
+ \delta^b_s (D_{ijl}, D_{pqr}) \right\}
\nonumber \\ && \mbox{}
- \frac16 \delta^a_l \left\{ \delta^b_p (D_{ijk}, D_{qrs})
+ \delta^b_q (D_{ijk},D_{prs}) + \delta^b_r (D_{ijk}, D_{pqs})
+ \delta^b_s (D_{ijk}, D_{pqr}) \right\},
\nonumber \\
\end{eqnarray}
where
\begin{eqnarray}
(S^{ij},S^{kl}) &=& \delta^{ik} \delta^{jl} + \delta^{il}\delta^{jk}, 
\nonumber \\
(S_{ij},S_{kl}) &=& \delta_{ik} \delta_{jl} + \delta_{il}\delta_{jk}.
\end{eqnarray}

\begin{table}[t]
\centering
\begin{tabular}{c|c|c|c} \hline\hline
irreducible tensor & type & notation & source (two diquark state) \\ \hline
$S$ & $(0,0)$ & $\bm{1}$ & $\bm{3}$ \\
$P^i_j$ & $(1,1)$ & $\bm{8}$ & $\bm{3}$,
$\overline{\bm{6}}$, $\bm{15}_2$ \\
$D_{ijk}$ & $(3,0)$ & $\bm{10}$ & $\bm{15}_1$, $\bm{15}_2$ \\
$T^{ijk}$ & $(0,3)$ & $\overline{\bm{10}}$ & $\overline{\bm{6}}$ \\
$T^{ij}_{kl}$ & $(2,2)$ & $\bm{27}$ & $\bm{15}_2$ \\
$T^{a}_{ijkl}$ & $(4,1)$ & $\bm{35}$ & $\bm{15}_1$ \\
\hline\hline
\end{tabular}
\caption{Representations of the pentaquark multiplets.}
\label{tab:penta}
\end{table}

With these informations, we are ready to match the states with the
physical baryons.

\section{Flavor wave functions of Pentaquark baryons}

Now we identify the tensor representations for pentaquark baryons obtained
so far with the physical baryon states.
We first present our nomenclature for pentaquark particles.
Based on its hypercharge, we name the baryon as
\begin{eqnarray}
&& Y = 2 \qquad \Theta, \nonumber \\
&& Y = 1 \qquad N, \Delta, \nonumber \\
&& Y = 0 \qquad \Sigma, \Lambda, \nonumber \\
&& Y = -1 \qquad \Xi, \nonumber \\
&& Y = -2 \qquad \Omega, \nonumber \\
&& Y = -3 \qquad X,
\end{eqnarray}
where we denote the $Y=-3$ particle as $X$ following Ref.~\cite{HL64}.
We denote the isospin of the particle and the multiplet which it belongs to
as subscripts, for example,
\begin{equation}
\Sigma_{{35},2}
\end{equation}
represents a particle of $\bm{35}$-plet with hypercharge $0$ and isospin $2$.
In the case that the isospin or the multiplet is clear, we drop such
subscripts as in $\Theta^+$.
The superscript is reserved for the charge.

In tensor representation, the number of lower indices of
$T_{a_1, \dots, a_p}^{b_1, \dots, b_q}$ is $p$ and that of upper indices
is $q$.
Now suppose that among its lower indices the numbers of 1's, 2's, and 3's
are $p_1$, $p_2$, and $p_3$, respectively, and that among upper indices
it has $q_1$ 1's, $q_2$ 2's, and $q_3$ 3's.
Then we have $p_1+p_2+p_3 = p$ and $q_1+q_2+q_3=q$.
The irreducible tensor is an eigenstate of hypercharge $Y$ and the third
component of isospin $I_3$ with the eigenvalues \cite{Low}
\begin{eqnarray}
Y &=& p_1 - q_1 + p_2 - q_2 - \frac23 (p-q), \nonumber \\
I_3 &=& \frac12 (p_1-q_1) - \frac12 (p_2-q_2).
\end{eqnarray}
The charge of the particle is obtained from the Gell-Mann--Nishijima
formula, $Q = I_3 + Y/2$.
By this way, we can match the SU(3) tensors to the physical baryon
states.
We summarize the pentaquark particles in Table~\ref{tab:name}.

\begin{table}[t]
\centering
\begin{tabular}{c|cc|l} \hline\hline
multiplet & hypercharge & isospin & particle \\ \hline
$\bm{1}$ & $0$ & $0$ & $\Lambda_1^0$ \\ \hline
$\bm{8}$  & $1$ & $1/2$ & $N_{8}^{+},N_8^{0}$ \\
          & $0$ & $1$ & $\Sigma_8^{+},\Sigma_8^0,\Sigma_8^-$ \\
          & $0$ & $0$ & $\Lambda_8^{0}$ \\
          & $-1$ & $1/2$ & $\Xi_8^{0}, \Xi^-_{8}$ \\ \hline
$\bm{10}$ & $1$ & $3/2$ & $\Delta_{10}^{++},\Delta_{10}^+,\Delta_{10}^0,
                           \Delta_{10}^{-}$ \\
          & $0$ & $1$ & $\Sigma_{10}^{+},\Sigma_{10}^0, \Sigma_{10}^{-}$ \\
          & $-1$ & $1/2$ & $\Xi^{0}_{10},\Xi_{10}^-$ \\
          & $-2$ & $0$ & $\Omega_{10}^-$ \\ \hline
$\overline{\bm{10}}$ & $2$ & $0$ & $\Theta^{+}$ \\
                     & $1$ & $1/2$ & $N_{\overline{10}}^{+},
                                      N_{\overline{10}}^0$ \\
                     & $0$ & $1$ & $\Sigma_{\overline{10}}^{+},
\Sigma_{\overline{10}}^0,\Sigma_{\overline{10}}^-$ \\
                     & $-1$ & $3/2$ & $\Xi_{\overline{10},3/2}^{+}$,
     $\Xi_{\overline{10},3/2}^0$, $\Xi_{\overline{10},3/2}^-$,
     $\Xi_{\overline{10},3/2}^{--}$ \\ \hline
$\bm{27}$ & $2$ & $1$ & $\Theta_1^{++}, \Theta_1^+, \Theta_1^0$ \\
          & $1$ & $3/2$ & $\Delta_{27}^{++}$,  $\Delta_{27}^{+}$,
             $\Delta_{27}^{0}$,  $\Delta_{27}^{-}$ \\
          & $1$ & $1/2$ & $N_{27}^+$, $N_{27}^0$ \\
          & $0$ & $2$ & $\Sigma_{27,2}^{++}$,  $\Sigma_{27,2}^{+}$,
$\Sigma_{27,2}^0$, $\Sigma_{27,2}^-$, $\Sigma_{27,2}^{--}$\\
  & $0$ & $1$ & $\Sigma_{27}^+$, $\Sigma_{27}^0$, $\Sigma_{27}^-$ \\
          & $0$ & $0$ & $\Lambda_{27}^{0}$ \\
          & $-1$ & $3/2$ & $\Xi_{27,3/2}^+$, $\Xi_{27,3/2}^{0}$,
          $\Xi_{27,3/2}^-$, $\Xi_{27,3/2}^{--}$ \\
          & $-1$ & $1/2$ & $\Xi_{27}^{0}$, $\Xi_{27}^-$ \\
          & $-2$ & $1$ & $\Omega_{27,1}^0$, $\Omega_{27,1}^-$,
          $\Omega_{27,1}^{--}$ \\
\hline
$\bm{35}$ & $2$ & $2$ & $\Theta_2^{+++}$, $\Theta_2^{++}$, $\Theta_2^+$,
$\Theta_2^0$, $\Theta_2^-$ \\
  & $1$ & $5/2$ & $\Delta_{5/2}^{+++}$,
$\Delta_{5/2}^{++}$, $\Delta_{5/2}^{+}$, $\Delta_{5/2}^{0}$, $\Delta_{5/2}^-$, $\Delta_{5/2}^{--}$ \\
  & $1$ & $3/2$ & $\Delta_{35}^{++}$, $\Delta_{35}^{+}$,
$\Delta_{35}^{0}$, $\Delta_{35}^{-}$ \\
          & $0$ & $2$ & $\Sigma_{35,2}^{++}$,  $\Sigma_{35,2}^{+}$,
$\Sigma_{35,2}^0$, $\Sigma_{35,2}^-$, $\Sigma_{35,2}^{--}$\\
  & $0$ & $1$ & $\Sigma_{35}^+$, $\Sigma_{35}^0$, $\Sigma_{35}^-$ \\
          & $-1$ & $3/2$ & $\Xi_{35,3/2}^+$, $\Xi_{35,3/2}^{0}$,
          $\Xi_{35,3/2}^-$, $\Xi_{35,3/2}^{--}$ \\
          & $-1$ & $1/2$ & $\Xi^{0}_{35},\Xi_{35}^-$ \\
          & $-2$ & $1$ & $\Omega_{35,1}^0$, $\Omega_{35,1}^-$,
         $\Omega_{35,1}^{--}$ \\
          & $-2$ & $0$ & $\Omega_{35}^-$ \\
          & $-3$ & $1/2$ & $X^-$, $X^{--}$ \\
\hline\hline
\end{tabular}
\caption{Pentaquark baryons}
\label{tab:name}
\end{table}

\subsection{Singlet}

The pentaquark singlet is given by
\begin{equation}
S = \frac{1}{\sqrt6} T_m \bar{q}^m,
\end{equation}
which is normalized as $(S,S)=1$, where $T_m$ stands for $T_m$, $Q_m$,
and $\widetilde{Q}_m$ defined in Eqs.~(\ref{eq:3x3f}), (\ref{eq:6x3f}), and
(\ref{eq:6x3f2}), respectively.
This is identified as
\begin{equation}
S = -\Lambda_1^0,
\label{sing}
\end{equation}
where the phase is chosen to be consistent with the tables of de Swart
\cite{deS63}, i.e., the conventional phase.

\subsection{Octet}

\begin{figure}[t]
\centering
\epsfig{file=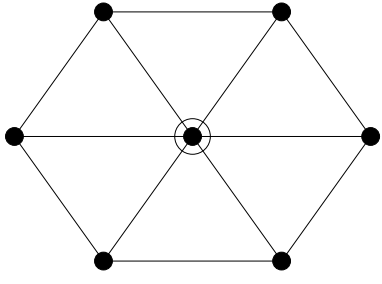, width=5cm}
\caption{Weight diagram for pentaquark octet.}
\label{fig:8}
\end{figure}

The weight diagram for pentaquark octet is shown in Fig.~\ref{fig:8}.
The octet tensor $P^i_j$ then represent the particles as follows,
\begin{eqnarray}
&&
P^3_1 = N_{8}^+ , \qquad
P^3_2 = N_{8}^0, \qquad
P^2_1 = \Sigma_{8}^+ , \qquad
\nonumber \\ &&
P^1_2 = \Sigma_{8}^-, \qquad
P^1_1 = 1/\sqrt2 \Sigma_{8}^0 + 1/\sqrt6 \Lambda_8^0 , \qquad
P^2_2 = -1/\sqrt2 \Sigma_{8}^0 + 1/\sqrt6 \Lambda_8^0,
\nonumber \\ &&
P^3_3 = -\sqrt{2/3} \Lambda_{8}^0 , \qquad
P^2_3 = \Xi_{8}^0, \qquad
P^1_3 = -\Xi_{8}^-.
\end{eqnarray}

The mixed states $P^1_1$, $P^2_2$, and $P^3_3$ are decomposed of the
physical states as follows \cite{Low,Close,FR}.
This is well-known, but here we give the procedure as a pedagogic
example.
First, we note that the inner product of the octet members is
\begin{equation}
(P^i_j, P^k_l) = \delta^{ik} \delta_{jl} - \frac13 \delta^i_j
\delta^k_l.
\end{equation}
With this normalization, we can see that all the off-diagonal components are
already normalized to one. For diagonal components, we have
\begin{eqnarray}
&& (P^1_1, P^1_1) = (P^2_2, P^2_2) = (P^3_3, P^3_3) = \frac23, 
\nonumber \\
&& (P^1_1, P^2_2) = (P^2_2, P^3_3) = (P^3_3, P^1_1) = -\frac13.
\end{eqnarray}
Here, one can find that $P^3_3$ does not contain $\Sigma^0_8$ part. This can
be seen from the fact that by isospin lowering or raising, $P^3_3$
cannot be obtained from $\Sigma^+_8$ or $\Sigma^-_8$, which are $P^2_1$ and
$P^1_2$.
Then we have
\begin{equation}
P^3_3 = -\sqrt{\frac23} \Lambda^0_8,
\end{equation}
with the conventional sign choice.
In order to know $P^1_1$ and $P^2_2$, we write
\begin{equation}
P^1_1 = a_1 \Sigma^0_8 + b_1 \Lambda^0_8, \qquad
P^2_2 = a_2 \Sigma^0_8 + b_2 \Lambda^0_8.
\end{equation}
Then the traceless condition reads
\begin{equation}
a_1 + a_2 = 0, \qquad b_1 + b_2 - \sqrt{\frac23} = 0.
\end{equation}
With the above condition and taking the inner products, one can obtain 
that
\begin{equation}
a_1 = -a_2 = 1/\sqrt2, \qquad b_1 = b_2 = 1/\sqrt6.
\end{equation}
Note that the phase convention for $a_{1,2}$ is chosen to be 
consistent with the tables of de Swart \cite{deS63}.

\subsection{Decuplet}

\begin{figure}[t]
\centering
\epsfig{file=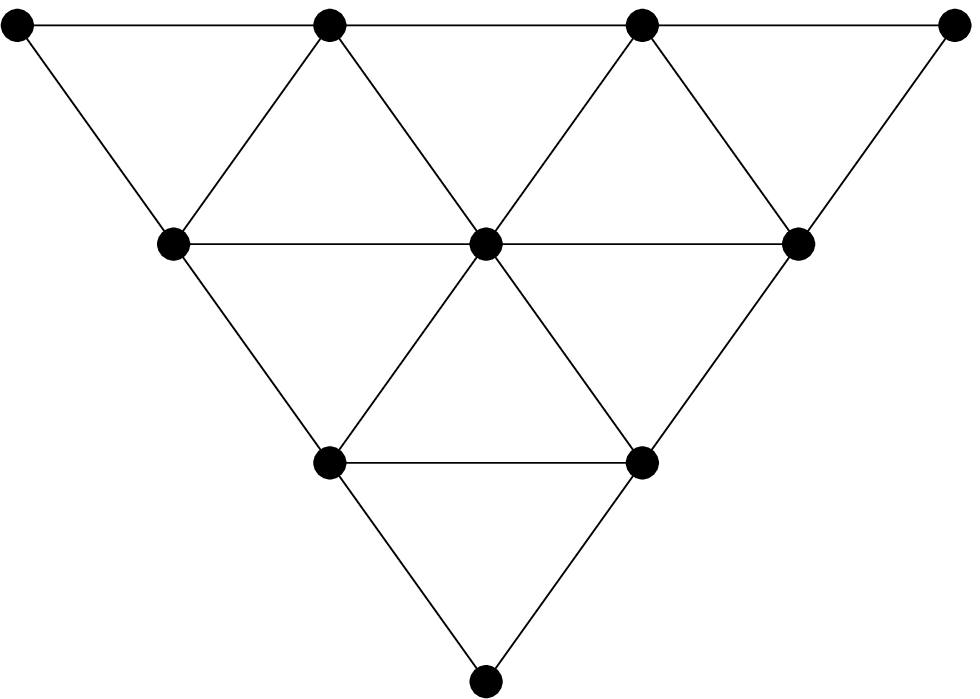, width=5cm}
\caption{Weight diagram for pentaquark decuplet.}
\label{fig:10}
\end{figure}

In the case of decuplet, we have
\begin{eqnarray}
&&
D_{111} = \sqrt6 \Delta_{10}^{++}, \qquad D_{112} = \sqrt2 \Delta_{10}^+, \qquad
D_{122} = \sqrt2 \Delta_{10}^0, \nonumber \\
&&
D_{222} = \sqrt6 \Delta_{10}^-, \qquad D_{113} = \sqrt2 \Sigma_{10}^+, \qquad
D_{123} = -\Sigma_{10}^0, \nonumber \\
&&
D_{223} = -\sqrt2 \Sigma_{10}^-, \qquad D_{133} = \sqrt2 \Xi_{10}^0, \qquad
D_{233} = \sqrt2 \Xi_{10}^-, \nonumber \\
&&
D_{333} = -\sqrt6 \Omega_{10}^-.
\end{eqnarray}
Its weight diagram is shown in Fig.~\ref{fig:10}.

\subsection{Antidecuplet}

\begin{figure}[t]
\centering
\epsfig{file=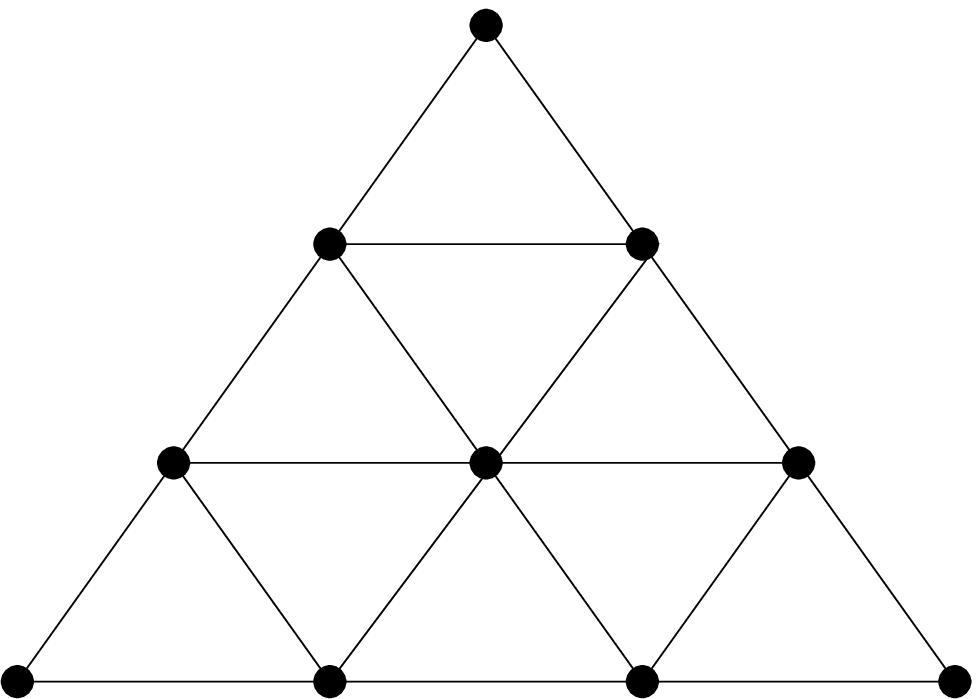, width=5cm}
\caption{Weight diagram for pentaquark antidecuplet.}
\label{fig:10bar}
\end{figure}

As can be seen from its weight diagram (Fig.~\ref{fig:10bar}),
antidecuplet wave functions can also be read from those for decuplet
baryons.
We have
\begin{eqnarray}
&&
T^{111} = \sqrt6 \Xi_{\overline{10},3/2}^{--}, \qquad
T^{112} = -\sqrt2 \Xi_{\overline{10},3/2}^-, \qquad
T^{122} = \sqrt2 \Xi_{\overline{10},3/2}^0, \nonumber \\
&&
T^{222} = -\sqrt6 \Xi_{\overline{10},3/2}^+,
\qquad T^{113} = \sqrt2 \Sigma_{\overline{10}}^-, \qquad
T^{123} = -\Sigma_{\overline{10}}^0, \nonumber \\
&&
T^{223} = -\sqrt2 \Sigma_{\overline{10}}^+,
\qquad T^{133} = \sqrt2 N_{\overline{10}}^0, \qquad
T^{233} = -\sqrt2 N_{\overline{10}}^+, \nonumber \\
&&
T^{333} = \sqrt6 \Theta^+.
\end{eqnarray}

The observed $\Theta(1540)$ is identified as a member of antidecuplet.
Although it has to be confirmed by other experiments, $\Xi(1862)$ is
interpreted as $\Xi_{\overline{10},3/2}$.
The nucleon analog $N_{\overline{10}}$, however, is not identified yet.
Several ideas which suggest $N(1710)$ or $N(1440)$ as a pure
antidecuplet \cite{DPP97} or a mixture of octet and antidecuplet \cite{JW03}
have been advocated.
However, $N(1710)$ cannot be a pure antidecuplet member because of its
large coupling to $\Delta\pi$ \cite{OKL03b}, and OZI rule in the ideally mixed
pentaquark octet and antidecuplet also prohibits the coupling of the
nucleon analog with $\Delta\pi$ \cite{CD04,LKO04}, which excludes $N(1440)$
as a pentaquark state.
If this is true, the nucleon analog $N_8$ or $N_{\overline{10}}$ has not
been found yet.%
\footnote{It is interesting to note that SPHINX Collaboration claimed the
existence of $X(2000)$ which has hidden strangeness \cite{Lands99}.}
In Ref.~\cite{ZC04}, a possible way to search for $\Sigma_{\overline{10}}$
is discussed.

\subsection{$\bm{27}$-plet}

\begin{figure}[t]
\centering
\epsfig{file=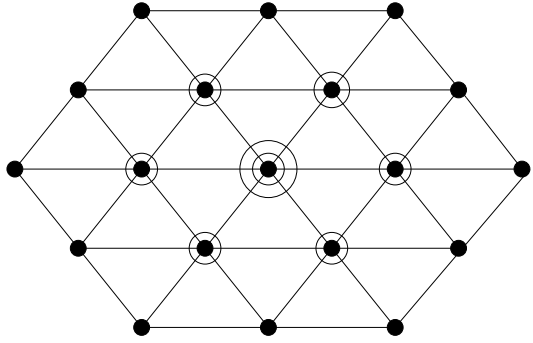, width=7cm}
\caption{Weight diagram for pentaquark $27$-plet.}
\label{fig:27}
\end{figure}

As can be read from its weight diagram (Fig.~\ref{fig:27}), the 27-plet
contains two $Y=1$ baryons with isospin $1/2$ and $3/2$, three $Y=0$
baryons with isospin $2$, $1$, and $0$, and two $Y=-1$ baryons with
isospin $1/2$ and $3/2$.
Such mixed states can be classified by the method described in the
previous subsection with the inner products of pentaquark states
obtained in Sect.~II.
As an example, let us consider the $Y=0$ and $I_3=0$ states, which are 
$T^{11}_{11}$, $T^{12}_{12}$, $T^{13}_{13}$, $T^{22}_{22}$,
$T^{23}_{23}$, $T^{33}_{33}$.
These can be written in terms of the physical states, $\Sigma_{27,2}^0$,
$\Sigma_{27}^0$, and $\Lambda_{27}^0$, as
\begin{eqnarray}
T^{11}_{11} &=& a_1 \Sigma_{27,2}^0 + b_1 \Sigma_{27}^0 + c_1
\Lambda_{27},
\nonumber \\
T^{12}_{12} &=& a_2 \Sigma_{27,2}^0 + b_2 \Sigma_{27}^0 + c_2
\Lambda_{27},
\nonumber \\
T^{13}_{13} &=& a_3 \Sigma_{27,2}^0 + b_3 \Sigma_{27}^0 + c_3
\Lambda_{27},
\nonumber \\
T^{22}_{22} &=& a_4 \Sigma_{27,2}^0 + b_4 \Sigma_{27}^0 + c_4
\Lambda_{27},
\nonumber \\
T^{23}_{23} &=& a_5 \Sigma_{27,2}^0 + b_5 \Sigma_{27}^0 + c_5
\Lambda_{27},
\nonumber \\
T^{33}_{33} &=& a_6 \Sigma_{27,2}^0 + b_6 \Sigma_{27}^0 + c_6
\Lambda_{27}.
\end{eqnarray}
First, we note that $T^{33}_{33}$ has isospin 0.
So we have $a_6 = b_6 = 0$.
The inner product relation (\ref{in-27}) gives
\begin{equation}
(T^{33}_{33}, T^{33}_{33}) = \frac{6}{5},
\end{equation}
which leads to $c_6 = \sqrt{\frac65}$.
We also have the inner products,
\begin{eqnarray}
&& (T^{11}_{11}, T^{33}_{33}) = \frac25, \qquad
(T^{12}_{12}, T^{33}_{33}) = \frac{1}{5}, \qquad
(T^{13}_{13}, T^{33}_{33}) = -\frac{3}{5}, \nonumber \\
&& (T^{22}_{22}, T^{33}_{33}) = \frac25, \qquad
(T^{23}_{23}, T^{33}_{33}) = -\frac{3}{5},
\end{eqnarray}
which allow us to fix the values of $c_i$.
The other inner products are
\begin{eqnarray}
&& (T^{11}_{11}, T^{11}_{11}) = \frac65, \qquad
(T^{12}_{12}, T^{12}_{12}) = \frac{7}{10}, \qquad
(T^{13}_{13}, T^{13}_{13}) = \frac{7}{10}, \qquad
(T^{22}_{22}, T^{22}_{22}) = \frac65, \nonumber \\
&& (T^{23}_{23}, T^{23}_{23}) = \frac{7}{10}, \qquad
(T^{11}_{11}, T^{12}_{12}) = -\frac{3}{5}, \qquad
(T^{11}_{11}, T^{13}_{13}) = -\frac35, \qquad
(T^{11}_{11}, T^{22}_{22}) = \frac25, \nonumber \\
&& (T^{11}_{11}, T^{23}_{23}) = \frac15, \qquad
(T^{12}_{12}, T^{13}_{13}) = -\frac{1}{10}, \qquad
(T^{12}_{12}, T^{22}_{22}) = -\frac{3}{5}, \qquad
(T^{12}_{12}, T^{23}_{23}) = -\frac{1}{10}, \nonumber \\
&& (T^{13}_{13}, T^{22}_{22}) = \frac{1}{5}, \qquad
(T^{13}_{13}, T^{23}_{23}) = -\frac{1}{10}, \qquad
(T^{22}_{22}, T^{23}_{23}) = -\frac{3}{5}.
\end{eqnarray}
Together with the traceless condition $T^{ij}_{ik} = 0$, which gives
\begin{equation}
a_1 + a_2 + a_3 = a_2 + a_4 + a_5 =
b_1 + b_2 + b_3 = b_2 + b_4 + b_5 = 0,
\end{equation}
and the fact that $T^{23}_{13}$ is the $\Sigma_{27}^+$ state, which
implies that $T^{12}_{12}$ does not contain $\Sigma_{27}^0$ state as it
cannot be obtained from $T^{23}_{13}$ by isospin lowering operator,
the above relations fix the constants $a_i$, $b_i$, and $c_i$ as given below.

After taking the conventional phase choice \cite{deS63}, we obtain as follows.

\medskip
\textbullet\ $Y = 2, I=1$
\begin{eqnarray}
T^{33}_{11} = -2 \Theta_1^{++}, \qquad
T^{33}_{12} = \sqrt2 \Theta_1^{+}, \qquad
T^{33}_{22} = 2 \Theta_1^{0},
\end{eqnarray}

\textbullet\ $Y = 1, I=3/2,1/2$
\begin{eqnarray}
&&
T^{23}_{11} = -\sqrt2 \Delta_{27}^{++}, \qquad
T^{13}_{11} = \sqrt{\frac23} \Delta_{27}^{+} + \sqrt{\frac{8}{15}}
N_{27}^+, \qquad
T^{23}_{12} = -\sqrt{\frac23} \Delta_{27}^{+} + \sqrt{\frac{2}{15}}
N_{27}^+,
\nonumber \\ &&
T^{33}_{13} = - \sqrt{\frac{6}{5}} N_{27}^+, \qquad
T^{13}_{12} = \sqrt{\frac23} \Delta_{27}^{0} + \sqrt{\frac{2}{15}}
N_{27}^0, \qquad
T^{23}_{22} = -\sqrt{\frac23} \Delta_{27}^{0} + \sqrt{\frac{8}{15}}
N_{27}^0,
\nonumber \\ &&
T^{33}_{23} = - \sqrt{\frac{6}{5}} N_{27}^0, \qquad
T^{13}_{22} = \sqrt2 \Delta_{27}^{-},
\end{eqnarray}

\textbullet\ $Y = 0, I=2,1,0$
\begin{eqnarray}
&&
T^{22}_{11} = 2\, \Sigma_{27,2}^{++}, \qquad
T^{12}_{11} = - \Sigma_{27,2}^+ + \frac{1}{\sqrt5} \Sigma_{27}^+, \qquad
T^{22}_{12} = \Sigma_{27,2}^+ + \frac{1}{\sqrt5} \Sigma_{27}^+, 
\nonumber \\ &&
T^{23}_{13} = - \frac{2}{\sqrt5} \Sigma_{27}^+, \qquad
T^{11}_{11} = \sqrt{\frac23} \Sigma_{27,2}^0 + \sqrt{\frac25}
\Sigma_{27}^0 + \sqrt{\frac{2}{15}} \Lambda_{27}^0,
\nonumber \\ &&
T^{12}_{12} = - \sqrt{\frac23} \Sigma_{27,2}^0 + \frac{1}{\sqrt{30}}
\Lambda_{27}^0, \qquad
T^{13}_{13} = - \sqrt{\frac25} \Sigma_{27}^0 - \sqrt{\frac{3}{10}}
\Lambda_{27}^0, \qquad
\nonumber \\ &&
T^{22}_{22} = \sqrt{\frac23} \Sigma_{27,2}^0 - \sqrt{\frac25}
\Sigma_{27}^0 + \sqrt{\frac{2}{15}} \Lambda_{27}^0,
\qquad
T^{23}_{23} = \sqrt{\frac25} \Sigma_{27}^0 - \sqrt{\frac{3}{10}}
\Lambda_{27}^0, \qquad
\nonumber \\ &&
T^{33}_{33} = \sqrt{\frac65} \Lambda_{27}^0, \qquad
T^{11}_{12} = \Sigma_{27,2}^- + \sqrt{\frac{1}{5}} \Sigma_{27}^-, \qquad
T^{12}_{22} = -\Sigma_{27,2}^- + \sqrt{\frac{1}{5}} \Sigma_{27}^-, \qquad
\nonumber \\ &&
T^{13}_{23} = -\frac{2}{\sqrt{5}} \Sigma_{27}^-, \qquad
T^{11}_{22} = 2 \Sigma_{27,2}^{--},
\end{eqnarray}

\textbullet\ $Y = -1, I=3/2,1/2$
\begin{eqnarray}
&&
T^{22}_{13} = -\sqrt2 \Xi_{27,3/2}^+, \qquad
T^{12}_{13} = \sqrt{\frac23} \Xi_{27,3/2}^0 + \sqrt{\frac{2}{15}}
\Xi_{27}^0, \qquad
\nonumber \\ &&
T^{22}_{23} = -\sqrt{\frac23} \Xi_{27,3/2}^0 + \sqrt{\frac{8}{15}}
\Xi_{27}^0, \qquad
T^{23}_{33} = -\sqrt{\frac65} \Xi_{27}^0, \qquad
\nonumber \\ &&
T^{11}_{13} = - \sqrt{\frac23} \Xi_{27,3/2}^- - \sqrt{\frac{8}{15}}
\Xi_{27}^-, \qquad
T^{12}_{23} = \sqrt{\frac23} \Xi_{27,3/2}^- - \sqrt{\frac{2}{15}}
\Xi_{27}^-, \qquad
\nonumber \\ &&
T^{13}_{33} = \sqrt{\frac65} \Xi_{27}^-, \qquad
T^{11}_{23} = -\sqrt{2} \Xi_{27,3/2}^{--}, \qquad
\end{eqnarray}

\textbullet\ $Y = -2, I=1$
\begin{eqnarray}
&&
T^{22}_{33} = -2 \Omega_{27,1}^0, \qquad
T^{12}_{33} = -\sqrt2 \Omega_{27,1}^-, \qquad
T^{11}_{33} = 2 \Omega_{27,1}^{--}.
\end{eqnarray}

\subsection{$\bm{35}$-plet}

\begin{figure}[t]
\centering
\epsfig{file=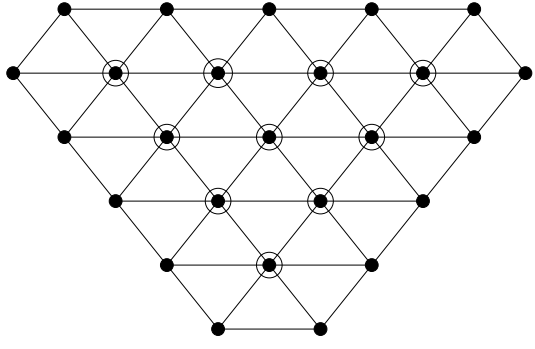, width=7cm}
\caption{Weight diagram for pentaquark $35$-plet.}
\label{fig:35}
\end{figure}

The weight diagram for $\bm{35}$-plet is given in Fig.~\ref{fig:35} and
the members are identified as follows.

\textbullet\ $Y = 2, I=2$
\begin{eqnarray}
&&
T^3_{1111} = 2\sqrt6 \Theta_2^{+++}, \qquad
T^3_{1112} = \sqrt6 \Theta_2^{++}, \qquad
T^3_{1122} = 2 \Theta_2^{+}, \qquad
\nonumber \\ &&
T^3_{1222} = \sqrt6 \Theta_2^{0}, \qquad
T^3_{2222} = 2\sqrt6 \Theta_2^{-}, \qquad
\end{eqnarray}

\textbullet\ $Y = 1, I=5/2,3/2$
\begin{eqnarray}
&&
T^2_{1111} = -2\sqrt6 \Delta_{5/2}^{+++}, \qquad
T^1_{1111} = 2\sqrt{\frac{6}{5}} \Delta_{5/2}^{++} + \frac{4}{\sqrt5}
\Delta_{35}^{++}, \qquad
\nonumber \\ &&
T^2_{1112} = -2\sqrt{\frac{6}{5}} \Delta_{5/2}^{++} + \frac{1}{\sqrt5}
\Delta_{35}^{++}, \qquad
T^{3}_{1113} = -\sqrt5 \Delta_{35}^{++}, \qquad
\nonumber \\ &&
T^1_{1112} = 2\sqrt{\frac{3}{5}} \Delta_{5/2}^{+} + \sqrt{\frac35}
\Delta_{35}^{+}, \qquad
T^2_{1122} = -2\sqrt{\frac{3}{5}} \Delta_{5/2}^{+} + \frac{2}{\sqrt{15}}
\Delta_{35}^{+}, \qquad
\nonumber \\ &&
T^{3}_{1123} = -\sqrt{\frac53} \Delta_{35}^{+}, \qquad
T^1_{1122} = 2\sqrt{\frac35} \Delta_{5/2}^0 + \frac{2}{\sqrt{15}}
\Delta_{35}^0, \qquad
\nonumber \\ &&
T^2_{1222} = -2\sqrt{\frac35} \Delta_{5/2}^0 + \sqrt{\frac35}
\Delta_{35}^0, \qquad
T^3_{1223} = - \sqrt{\frac53} \Delta_{35}^0,
\nonumber \\ &&
T^1_{1222} = 2 \sqrt{\frac65} \Delta_{5/2}^- + \frac{1}{\sqrt5}
\Delta_{35}^-, \qquad
T^2_{2222} = -2 \sqrt{\frac65} \Delta_{5/2}^- + \frac{4}{\sqrt5}
\Delta_{35}^-, \qquad
\nonumber \\ &&
T^3_{2223} = -\sqrt5 \Delta_{35}^-, \qquad
T^1_{2222} = 2\sqrt6 \Delta_{5/2}^{--}
\end{eqnarray}

\textbullet\ $Y = 0, I=2,1$
\begin{eqnarray}
&&
T^2_{1113} = -\sqrt6 \Sigma_{35,2}^{++}, \qquad
T^1_{1113} = \sqrt{\frac32} \Sigma_{35,2}^+ + \sqrt{\frac32}
\Sigma_{35}^+, \qquad
\nonumber \\ &&
T^2_{1123} = -\sqrt{\frac32} \Sigma_{35,2}^+ + \sqrt{\frac16}
\Sigma_{35}^+, \qquad
T^3_{1133} = - \sqrt{\frac83} \Sigma_{35}^+,
\nonumber \\ &&
T^1_{1123} = \Sigma_{35,2}^0 + \sqrt{\frac13} \Sigma_{35}^0, \qquad
T^2_{1223} = -\Sigma_{35,2}^0 + \sqrt{\frac13} \Sigma_{35}^0, \qquad
\nonumber \\ &&
T^3_{1233} = - \sqrt{\frac43} \Sigma_{35}^0, \qquad
T^1_{1223} = \sqrt{\frac32} \Sigma_{35,2}^- + \sqrt{\frac16}
\Sigma_{35}^-, \qquad
\nonumber \\ &&
T^2_{2223} = -\sqrt{\frac32} \Sigma_{35,2}^- + \sqrt{\frac32}
\Sigma_{35}^-, \qquad
T^3_{2233} = - \sqrt{\frac83} \Sigma_{35}^-, \qquad
\nonumber \\ &&
T^1_{2223} = \sqrt6 \Sigma_{35,2}^{--}.
\end{eqnarray}

\textbullet\ $Y = -1, I=3/2,1/2$
\begin{eqnarray}
&&
T^2_{1133} = -2 \Xi_{35,3/2}^{+}, \qquad
T^1_{1133} = \sqrt{\frac43} \Xi_{35,3/2}^0 + \sqrt{\frac43} \Xi_{35}^0,
\qquad
\nonumber \\ &&
T^2_{1233} = -\sqrt{\frac43} \Xi_{35,3/2}^0 + \sqrt{\frac13} \Xi_{35}^0,
\qquad
T^3_{1333} = -\sqrt3 \Xi_{35}^0,
\nonumber \\ &&
T^1_{1233} = \sqrt{\frac43} \Xi_{35,3/2}^- + \sqrt{\frac13} \Xi_{35}^-,
\qquad
T^2_{2233} = -\sqrt{\frac43} \Xi_{35,3/2}^- + \sqrt{\frac43} \Xi_{35}^-,
\qquad
\nonumber \\ &&
T^3_{2333} = -\sqrt3 \Xi_{35}^-, \qquad
T^1_{2233} = 2 \Xi_{35,3/2}^{--}
\end{eqnarray}

\textbullet\ $Y = -2, I=1,0$
\begin{eqnarray}
&&
T^2_{1333} = -\sqrt6 \Omega_{35,1}^0, \qquad
T^1_{1333} = -\sqrt3 \Omega_{35,1}^- - \sqrt2 \Omega_{35}^-, \qquad
\nonumber \\ &&
T^2_{2333} = \sqrt3 \Omega_{35,1}^- - \sqrt2 \Omega_{35}^-, \qquad
T^3_{3333} = 2\sqrt2 \Omega_{35}^-, \qquad
T^1_{2333} = -\sqrt6 \Omega_{35,1}^{--}.
\end{eqnarray}

\textbullet\ $Y = -3, I=1/2$
\begin{eqnarray}
T^2_{3333} = -\sqrt{24} X^-, \qquad
T^1_{3333} = \sqrt{24} X^{--}.
\end{eqnarray}

As we have identified the pentaquark irreducible tensors with the
physical baryon states, it is straightforward to obtain their flavor
wave functions.
For example, the flavor wave function of the singlet state,
$\Lambda_1^0$, is obtained as
\begin{equation}
\Lambda_1^0 = - \frac{c_1^{}}{\sqrt6} T_m \overline{q}^m
 - \frac{c_2^{}}{\sqrt6} Q_m \overline{q}^m
 - \frac{c_3^{}}{\sqrt6} \widetilde{Q}_m \overline{q}^m,
\end{equation}
where $T_m$, $Q_m$, and $\widetilde{Q}_m$ are defined in
Eqs.~(\ref{eq:3x3f}), (\ref{eq:6x3f}), and (\ref{eq:6x3f2}).
By noting that $q_1 = u$, $q_2 = d$, and $q_3 = s$, one can explicitly
write down the wave function in terms of quark flavors.
The coefficients $c_i^{}$ are constrained as $c_1^2 + c_2^2 + c_3^2
= 1$.
The octet, however, has eight coefficients as
\begin{eqnarray}
P^i_j &=& \frac{c_1^{}}{\sqrt2} \left( T_j \overline{q}^i - \frac13
\delta^i_j T_m \overline{q}^m \right)
+ \frac{c_2^{}}{\sqrt2} \left( Q_j \overline{q}^i - \frac13
\delta^i_j Q_m \overline{q}^m \right)
+ \frac{c_3^{}}{\sqrt2} \left( \widetilde{Q}_j \overline{q}^i - \frac13
\delta^i_j \widetilde{Q}_m \overline{q}^m \right)
\nonumber \\ && \mbox{}
+ \frac{c_4^{}}{\sqrt3} \epsilon_{jab} S^{ia} \overline{q}^b
+ \frac{c_5^{}}{\sqrt3} \epsilon_{jab} T^{ia} \overline{q}^b
+ \frac{c_6^{}}{\sqrt{15}} T^i_{jk} \overline{q}^k
+ \frac{c_7^{}}{\sqrt{15}} \widetilde{T}^i_{jk} \overline{q}^k
+ \frac{c_8^{}}{\sqrt{15}} S^i_{jk} \overline{q}^k.
\end{eqnarray}
The flavor wave functions of other multiplets are obtained as
\begin{eqnarray}
D_{ijk} &=& \frac{c_1^{}}{\sqrt6} T_{ijkl} \overline{q}^l
+ \frac{c_2^{}}{\sqrt{24}} \left ( \epsilon_{iab} T^a_{jk} \overline{q}^b
+\epsilon_{jab} T^a_{ki} \overline{q}^b+\epsilon_{kab} T^a_{ij} \overline{q}^b
\right )
\nonumber \\
&&
+ \frac{c_3^{}}{\sqrt{24}} \left ( \epsilon_{iab} \widetilde{T}^a_{jk} 
\overline{q}^b+\epsilon_{jab} \widetilde{T}^a_{ki} 
\overline{q}^b+\epsilon_{kab} \widetilde{T}^a_{ij} 
\overline{q}^b \right )
+ \frac{c_4^{}}{\sqrt{24}} \left ( 
\epsilon_{iab} S^a_{jk} \overline{q}^b
+\epsilon_{jab} S^a_{ki} \overline{q}^b
+\epsilon_{kab} S^a_{ij} \overline{q}^b
\right ) ,
\\
T^{ijk} &=& \frac{c_1^{}}{\sqrt3} \left( S^{ij} \overline{q}^k
+ S^{jk} \overline{q}^i + S^{ki} \overline{q}^j \right)
+ \frac{c_2^{}}{\sqrt3} \left( T^{ij} \overline{q}^k
+ T^{jk} \overline{q}^i + T^{ki} \overline{q}^j \right),
\\
T^{ij}_{kl} &=& c_1^{} \left\{
\frac{1}{2\sqrt{2}} \left( T^i_{kl} \bar{q}^j + T^j_{kl}
\bar{q}^i \right) - \frac{1}{10\sqrt{30}} \left( \delta^i_l T^j_{km}
\overline{q}^m + \delta^j_l T^i_{km} \overline{q}^m +
\delta^i_k T^j_{lm} \overline{q}^m
+ \delta^j_k T^i_{lm} \overline{q}^m \right) \right\}
\nonumber \\ && \mbox{}
+ c_2^{} \left\{
\frac{1}{2\sqrt{2}} \left( \widetilde{T}^i_{kl} \bar{q}^j + \widetilde{T}^j_{kl}
\bar{q}^i \right) - \frac{1}{10\sqrt{30}} \left( \delta^i_l \widetilde{T}^j_{km}
\overline{q}^m + \delta^j_l \widetilde{T}^i_{km} \overline{q}^m
+ \delta^i_k \widetilde{T}^j_{lm} \overline{q}^m
+ \delta^j_k \widetilde{T}^i_{lm} \overline{q}^m \right) \right\}
\nonumber \\ && \mbox{}
+ c_3^{} \left\{
\frac{1}{2\sqrt{2}} \left( S^i_{kl} \bar{q}^j + S^j_{kl}
\bar{q}^i \right) - \frac{1}{10\sqrt{30}} \left( \delta^i_l S^j_{km}
\overline{q}^m + \delta^j_l S^i_{km} \overline{q}^m +
\delta^i_k S^j_{lm} \overline{q}^m
+ \delta^j_k S^i_{lm} \overline{q}^m \right) \right\},
\\
T^a_{ijkl} &=& T_{ijkl} \bar{q}^a - \frac{1}{6} \left( \delta^a_i
T_{jklm} \bar{q}^m + \delta^a_j T_{iklm} \bar{q}^m +
\delta^a_k T_{ijlm} \bar{q}^m + \delta^a_l T_{ijkm} \bar{q}^m
\right),
\end{eqnarray}
where $S^{ij}$, $T_i$, $Q_i$, $T^i_{jk}$, $\widetilde{Q}_i$,
$\widetilde{T}^i_{jk}$, $T^{ij}$, $S^i_{jk}$, and $T_{ijkl}$ are defined
in Eqs.~(\ref{eq:3x3f}), (\ref{eq:6x3f}), (\ref{eq:6x3f2}), and
(\ref{eq:6x6f}).
This is consistent with the fact that
\begin{equation}
\bm{3} \otimes \bm{3} \otimes \bm{3} \otimes \bm{3} \otimes
\overline{\bm{3}} = \bm{35} \oplus (3)\bm{27} \oplus
(2)\overline{\bm{10}} \oplus
(4)\bm{10} \oplus (8)\bm{8} \oplus (3)\bm{1},
\end{equation}
where the numbers in parentheses are the number of multiplicity.
Here, we do not give the full list of the pentaquark wave function
explicitly in terms of $u$, $d$, and $s$ quarks, since they can be read
directly from the results presented in this Section.
The coefficients $c_i^{}$ are to be determined by quark dynamics.
For example, in the case of antidecuplet, Jaffe-Wilczek model favors
$\overline{\bm{3}}$ structure of a diquark, hence $c_2^{} = 0$
\cite{JW03,CD04}, while $c_2^{} \neq 0$ in SU(3) quark model of Carlson
{\it et al.\/} \cite{CCKN03c}.
Furthermore, if the diquark inside pentaquark baryons cannot form $\bm{6}$,
the pentaquark $\bm{10}$, $\bm{27}$, and $\bm{35}$ are hard to be
observed in the diquark-diquark-antiquark model.
Therefore, observation of other pentaquark states in higher multiplets
will help us to understand the structure of pentaquark baryons.

\section{Interactions of pentaquark baryons}

With the wave functions of pentaquark baryons constructed in the previous
Section, we now consider the SU(3) symmetric interaction Lagrangian of
pentaquark baryons. 
In this paper, we first consider pentaquark interactions with three-quark
baryons and meson octet.
Then we establish the pentaquark interactions with (other) pentaquarks
and meson octet.
Some of the results presented in this Section can be read from the
tables of Ref. \cite{deS63}, which, however, does not give all relations
for pentaquark-pentaquark interactions.
Since we are interested in the flavor SU(3) symmetric structure of the
interactions and the spin-parity of pentaquark baryons is yet to be
determined, we drop the Lorenz structure in this study.

\subsection{Interactions with 3-quark octet and meson octet}

We first consider the pentaquark interactions with normal baryon octet
($\bm{8}_3$) and meson octet ($\bm{8}_M$).
In this case, we note that
\begin{equation}
\bm{8} \otimes \bm{8} = \bm{27} \oplus \bm{10} \oplus \overline{\bm{10}}
\oplus \bm{8}_1 \oplus \bm{8}_2 \oplus \bm{1}.
\end{equation}
The SU(3)-invariant interactions can be obtained by constructing SU(3)
singlet.
In tensor notation, it is achieved by fully contracting the upper and
lower indices of the three tensors representing two baryon multiplets and
meson octet.
When the number of upper indices does not match that of lower indices,  
the Levi-Civita tensors $\epsilon_{ijk}$ are introduced to
make the interactions fully contracted. 
In this construction, $\bm{35}$-$\bm{8}$-$\bm{8}$ cannot be SU(3)-invariant
and the $\bm{35}$-plet cannot couple to 3-quark octet and meson
octet.
This property of the $\bm{35}$-plet would be useful to identify the
$\bm{35}$-plet members in experiments.

The 3-quark baryon octet and meson octet read
\begin{eqnarray}
&&
B^3_1 = p, \qquad
B^3_2 = n, \qquad
B^2_1 = \Sigma^+, \nonumber \\
&&
B^1_2 = \Sigma^-, \qquad
B^1_1 = 1/\sqrt2 \Sigma^0 + 1/\sqrt6 \Lambda^0, \qquad
B^2_2 = -1/\sqrt2 \Sigma^0 + 1/\sqrt6 \Lambda^0, \nonumber \\
&&
B^3_3 = -\sqrt{2/3} \Lambda^0, \qquad
B^2_3 = \Xi^0, \qquad
B^1_3 = -\Xi^-,
\label{boct}
\end{eqnarray}
and
\begin{eqnarray}
&&
M^3_1 = K^+, \qquad
M^3_2 = K^0, \qquad
M^2_1 = \pi^+, \nonumber \\
&&
M^1_2 = \pi^-, \qquad
M^1_1 = 1/\sqrt2 \pi^0 + 1/\sqrt6 \eta, \qquad
M^2_2 = -1/\sqrt2 \pi^0 + 1/\sqrt6 \eta, \nonumber \\
&&
M^3_3 = -\sqrt{2/3} \eta, \qquad
M^2_3 = \bar{K}^0, \qquad
M^1_3 = K^-.
\label{moct}
\end{eqnarray}
It is also useful to define isospin multiplets as
\begin{eqnarray}
&&
N = \left( \begin{array}{c} p \\ n \end{array} \right), \qquad
\Xi = \left( \begin{array}{c} \Xi^0 \\ \Xi^- \end{array} \right), \qquad
K = \left( \begin{array}{c} K^+ \\ K^0 \end{array} \right), \qquad
K_c = \left( \begin{array}{c} \bar{K}^0 \\ -K^- \end{array} \right),
\nonumber \\ &&
\overline{K} = ( K^-,\ \bar{K}^0), \qquad
\overline{K}_c = ( K^0,\ -K^+), \qquad
\Sigma = (\Sigma^+, \Sigma^0, \Sigma^-)^T,
\nonumber \\ &&
\bm{\tau} \cdot \bm{\Sigma} = \left( \begin{array}{cc} \Sigma^0 & \sqrt2
\Sigma^+ \\ \sqrt2 \Sigma^- & -\Sigma^0 \end{array} \right), \qquad
\bm{\tau} \cdot \bm{\pi} = \left( \begin{array}{cc} \pi^0 & \sqrt2
\pi^+ \\ \sqrt2 \pi^- & -\pi^0 \end{array} \right).
\label{iso}
\end{eqnarray}

We now construct the SU(3)-invariant Lagrangian for pentaquark baryons
up to the universal coupling constants.
The pentaquark baryon fields, $S$, $P_i^j$, $D_{ijk}$, $T^{ijk}$,
$T_{ij}^{kl}$, and $T^a_{ijkl}$ are defined in Section~III.

\subsubsection{$\bm{1}$-$\bm{8}_3$}

We start by considering the interaction of pentaquark singlet with
(three-quark) baryon octet and meson octet.  There is only one possible way
to contract the upper and lower indices in this case and 
the SU(3) invariant Lagrangian reads
\begin{equation}
\mathcal{L}_{\bm{1}\mbox{-}\bm{8}_3} = 
g_{\bm{1}\mbox{-}\bm{8}_3}^{}
\overline{S} \, B^i_k M^k_i
+ \mbox{(H.c.)}.
\end{equation}
Using Eqs.(\ref{sing}), (\ref{boct}), and (\ref{moct}), we obtain
\begin{equation}
\mathcal{L}_{\bm{1}\mbox{-}\bm{8}_3} /
g_{\bm{1}\mbox{-}\bm{8}_3}^{} =
- \overline{\Lambda}_1^0 \left( \overline K \, N + \overline{K}_c \, \Xi
+ \bm{\Sigma} \cdot \bm{\pi} + \Lambda\, \eta
\right)
+ \mbox{(H.c.)} ,
\label{Lag:1-8}
\end{equation}
where
\begin{equation}
\bm{\Sigma} \cdot \bm{\pi} = \Sigma^+ \pi^- + \Sigma^0 \pi^0 + \Sigma^-
\pi^+,
\end{equation}
and the isospin multiplets are defined in Eq.~(\ref{iso}).

\subsubsection{$\bm{8}$-$\bm{8}_3$}

For the interaction of pentaquark octet and normal baryon octet and
meson octet, there are two possible ways to contract the upper and lower
indices.
This is equivalent to the well-known $f$- and $d$-type interactions and the 
interaction Lagrangian reads
\begin{equation}
\mathcal{L}_{\bm{8}\mbox{-}\bm{8}_3} /
g_{\bm{8}\mbox{-}\bm{8}_3}^{} =
(d+f) \bar{P}_i^l B_k^i M_l^k +
(d-f) \bar{P}_i^l B_l^k M_k^i
+ \mbox{(H.c.)}.  
\end{equation}
Using identification of the pentaquark octet given in Sect.~III and 
Eqs.~(\ref{boct}) and (\ref{moct}), we obtain
\begin{eqnarray}
\mathcal{L}_{\bm{8}\mbox{-}\bm{8}_3} /
g_{\bm{8}\mbox{-}\bm{8}_3}^{}
&=&
\frac{1}{\sqrt2} (d+f)\,
\overline{N}_8^{} \bm{\tau} \cdot \bm{\pi} N
- \frac{1}{\sqrt6} (d-3f)\,
\overline{N}_8^{} N\, \eta
- \frac{1}{\sqrt6} (d+3f)\,
\overline{N}_8^{} K \Lambda^0
\nonumber \\ && \mbox{}
+ \frac{1}{\sqrt2} (d-f)\,
\overline{N}_8^{} \bm{\tau}\cdot \bm{\Sigma} \, K
- i \sqrt2 f \left( \overline{\bm \Sigma}_8^{}
\times \bm{\Sigma} \right) \cdot \bm{\pi}
+ \sqrt{\frac23} d \,
\overline{\Sigma}_8 \, \Sigma\, \eta
\nonumber \\ && \mbox{}
+ \sqrt{\frac23} d\,
\overline{\bm{\Sigma}}_8 \cdot \bm{\pi}\, \Lambda^0
- \frac{1}{\sqrt2} (d+f)\,
\overline{K}_c \overline{\bm{\Sigma}}_8 \cdot \bm{\tau} \,\Xi
+ \frac{1}{\sqrt2} (d-f)\,
\overline{K} \, \overline{\bm{\Sigma}}_8 \cdot \bm{\tau} N
\nonumber \\ && \mbox{}
+ \sqrt{\frac23} d\,
\overline{\Lambda}_8^0 \bm{\Sigma} \cdot \bm{\pi}
- \sqrt{\frac23} d\,
\overline{\Lambda}_8^0\, \Lambda^0\, \eta
- \frac{1}{\sqrt6} (d-3f)\,
\overline{\Lambda}_8^0 \overline{K}_c \Xi
\nonumber \\ && \mbox{}
- \frac{1}{\sqrt6} (d+3f)\,
\overline{\Lambda}_8^0 \overline{K} N
- \frac{1}{\sqrt2} (d-f) \,
\overline{\Xi}_8 \bm{\tau} \cdot \bm{\pi}\, \Xi
\nonumber \\ && \mbox{}
-\frac{1}{\sqrt6} (d+3f)\,
 \overline{\Xi}_8 \Xi\, \eta
-\frac{1}{\sqrt2} (d+f)\,
\overline{\Xi}_8 \bm{\tau} \cdot \bm{\Sigma} K_c
- \frac{1}{\sqrt6} (d-3f)\,
\overline{\Xi}_8 K_c \Lambda^0
\nonumber \\ && \mbox{}
+ \mbox{(H.c.)},
\label{Lag:8-8}
\end{eqnarray}
where the isospin multiplets of $\bm{8}_3$ are defined as in
Eq.~(\ref{iso}) and
\begin{eqnarray}
-i \left( \overline{\bm \Sigma}_8^{}
\times \bm{\Sigma} \right) \cdot \bm{\pi} &=& 
\overline{\Sigma}_8^+ \left( \Sigma^+ \pi^0 - \Sigma^0 \pi^+ \right)
+\overline{\Sigma}_8^0 \left( \Sigma^- \pi^+ - \Sigma^+ \pi^- \right)
+\overline{\Sigma}_8^- \left( \Sigma^0 \pi^- - \Sigma^- \pi^0 \right),
\nonumber \\
\overline{\bm{\Sigma}}_8 \cdot \bm{\Sigma} &=&
\overline{\Sigma}_8^+ \Sigma^+ + \overline{\Sigma}_8^0 \Sigma^0 +
\overline{\Sigma}_8^- \Sigma^- .
\end{eqnarray}
In Refs.~\cite{CD04,LKO04}, it has been claimed that the OZI rule leads
to $f/d = 1/3$ within the ideal mixing of pentaquark octet and
antidecuplet.

\subsubsection{$\bm{10}$-$\bm{8}_3$}

The SU(3) interaction Lagrangian of pentaquark decuplet with the normal
baryon octet and meson octet can be constructed as
\begin{equation}
\mathcal{L}_{\bm{10}\mbox{-}\bm{8}_3} =
g_{\bm{10}\mbox{-}\bm{8}_3}^{} \,
\epsilon_{ijk} \overline{D}^{jlm} B^i_l M^k_m + \mbox{(H.c.)}.
\label{10-83}
\end{equation}
Since the number of the upper indices is not the same with that of the lower
indices, $\epsilon_{ijk}$ has been introduced to form the fully contracted
interaction.
Other possible contractions are equivalent to Eq.~(\ref{10-83}) up to an
overall sign.

Noting that the indices of the pentaquark decuplet are symmetric, we work
out all the contractions and find 
\begin{eqnarray}
\mathcal{L}_{\bm{10}\mbox{-}\bm{8}_3} / g_{\bm{10}\mbox{-}\bm{8}_3}^{}
&=&
-\sqrt3 \overline{\Delta}_{10} \bm{T}(\textstyle\frac32,\frac12)
\cdot \bm{\pi} N + \sqrt3
\overline{\Delta}_{10} \bm{T}(\textstyle\frac32,\frac12) \cdot \bm{\Sigma}\, K
+ \overline{K}\, \overline{\bm{\Sigma}}_{10} \cdot \bm{\tau} N
\nonumber \\ && \mbox{}
-i \left( \overline{\bm{\Sigma}}_{10} \times \bm{\Sigma} \right) \cdot
\bm{\pi} 
+ \sqrt{3} \overline{\Sigma}_{10} \, {\Sigma}\, \eta
- \sqrt3 \overline{\bm{\Sigma}}_{10} \cdot \bm{\pi}\, \Lambda^0
\nonumber \\ && \mbox{}
+ \overline{K}_c \overline{\bm{\Sigma}}_{10} \cdot \bm{\tau}\, \Xi
- \overline{\Xi}_{10} \bm{\tau}\cdot \bm{\Sigma}\, K_c
+ \overline{\Xi}_{10} \bm{\tau} \cdot \bm{\pi}\, \Xi
\nonumber \\ && \mbox{}
+ \sqrt3 \overline{\Xi}_{10} \Xi\, \eta
- \sqrt3 \overline{\Xi}_{10} K_c\, \Lambda^0
- \sqrt6 \overline{\Omega}_{10}^- \bar{K}\, \Xi + \mbox{(H.c.)},
\label{Lag:10-8}
\end{eqnarray}
where
\begin{equation}
\Delta_{10} = (\Delta^{++}_{10},\ \Delta^+_{10}, \ \Delta_{10}^0, \
\Delta_{10}^- )^T.
\end{equation}
The other isospin multiplets are defined as in Eq.~(\ref{iso}) and
\begin{equation}
\bm{T}(\textstyle\frac32,\frac12) \cdot \bm{\pi}
 = - T^{(+1)}_{3/2,1/2} \pi^+ +
T^{(-1)}_{3/2,1/2} \pi^- +
T^{(0)}_{3/2,1/2} \pi^0,
\label{t31}
\end{equation}
with
\begin{eqnarray}
T^{(+1)}_{\textstyle\frac32,\frac12} = \frac{1}{\sqrt3}
\left( \begin{array}{cc} \sqrt6 & 0 \\ 0 & \sqrt2 \\ 0 & 0 \\ 0 & 0
\end{array} \right), \quad
T^{(0)}_{\textstyle\frac32,\frac12} =
\frac{1}{\sqrt3} \left( \begin{array}{cc} 0 & 0 \\ 2 & 0 \\ 0 & 2
\\ 0 & 0 \end{array} \right), \quad
T^{(-1)}_{\textstyle\frac32,\frac12} =
\frac{1}{\sqrt3} \left( \begin{array}{cc} 0 & 0 \\ 0 & 0 \\ \sqrt2
& 0 \\ 0 & \sqrt6 \end{array} \right).
\end{eqnarray}
Therefore, for example,
\begin{eqnarray}
\overline{\Delta}_{10}\,\bm{T}(\textstyle\frac32,\frac12) \cdot
\bm{\pi}\, N &=&
-\sqrt2 \, \overline{\Delta}_{10}^{++} N^+ \pi^+
- \sqrt{\frac23} \overline{\Delta}_{10}^+ \left( N^0 \pi^+ - \sqrt2 N^+
  \pi^0 \right)
\nonumber \\ && \mbox{}
+ \sqrt{\frac23} \overline{\Delta}_{10}^0 \left( N^+ \pi^- + \sqrt2 N^0
  \pi^0 \right)
+ \sqrt2 \overline{\Delta}_{10}^- N^0 \pi^-.
\end{eqnarray}
In a similar way, one can calculate $\overline{\Delta}_{10}
\bm{T}(\textstyle\frac32,\frac12) \cdot \bm{\Sigma}\, K$.
The generalized isospin matrices are defined as \cite{CN69}
\begin{equation}
T^{(\rho)}_{\nu\mu} (I',I) = \sqrt{\frac{(2I'+1)(2I+1)}{3}}
(-1)^{I-\mu}\, \langle I' \nu I -\mu | 1 \rho \rangle
\end{equation}
in the spherical basis.

\subsubsection{$\overline{\bm{10}}$-$\bm{8}_3$}

The SU(3) symmetric Lagrangian for this interaction was reported in
Ref.~\cite{OKL03b}, which reads
\begin{eqnarray}
\mathcal{L}_{\overline{\bm{10}}\mbox{-}\bm{8}_3} =
g^{}_{\overline{\bm{10}}\mbox{-}\bm{8}_3} \,
\epsilon^{ilm}
\overline{T}_{ijk} B_l^j M_m^k + \mbox{(H.c.)}. \label{int-1}
\end{eqnarray}
Then we have \cite{LKO04}
\begin{eqnarray}
\mathcal{L}_{\overline{\bm{10}}\mbox{-}\bm{8}_3} /
g^{}_{\overline{\bm{10}}\mbox{-}\bm{8}_3}
&=&
-\sqrt6 \, \overline{\Theta}^+ \, \overline{K}_c N
- \overline{N}_{\overline{10}}^{} \bm{\tau}\cdot \bm{\pi} N
+ \sqrt3 \, \overline{N}_{\overline{10}}^{}\, N\, \eta
\nonumber \\ && \mbox{}
- \sqrt3 \, \overline{N}_{\overline{10}}^{} K \Lambda^0
+ \overline{N}_{\overline{10}}^{} \bm{\tau} \cdot \bm{\Sigma} K
+ i \left( \overline{\bm{\Sigma}}_{\overline{10}} \times \bm{\Sigma} \right)
\cdot \bm{\pi}
\nonumber \\ && \mbox{}
- \sqrt3 \, \overline{\bm{\Sigma}}_{\overline{10}} \cdot \bm{\pi}
  \Lambda^0
+ \sqrt3 \, \overline{\Sigma}_{\overline{10}} \, \Sigma \, \eta
- \overline{K}_c \overline{\bm{\Sigma}}_{\overline{10}} \cdot \bm{\tau} \, \Xi
\nonumber \\ && \mbox{}
- \overline{K}\, \overline{\bm{\Sigma}}_{\overline{10}} \cdot \bm{\tau} N
- \sqrt3 \, \overline{\Xi}_{\overline{10}}
  \bm{T}(\textstyle\frac32,\frac12)
 \cdot \bm{\pi}\, \Xi
+ \sqrt3 \, \overline{\Xi}_{\overline{10}}
\bm{T}(\textstyle\frac32,\frac12)
 \cdot \bm{\Sigma} K_c
\nonumber \\ && \mbox{}
+ \mbox{(H.c.)},
\label{Lag:10bar-8}
\end{eqnarray}
where
\begin{equation}
\Xi_{\overline{10}} = (\Xi_{\overline{10}}^{+}, \
\Xi_{\overline{10}}^{0}, \ \Xi_{\overline{10}}^{-}, \
\Xi_{\overline{10}}^{--} )^T.
\end{equation}
Note that 
\begin{equation}
\overline{\Theta}^+ \, \overline{K}_c\, N = \overline{\Theta}^+ \left( p
K^0 - n K^+ \right),
\end{equation}
so that it has a relative negative sign for the $\Theta^+$ decays into
proton and neutron.

The universal coupling constant can be estimated from the decay width of
$\Theta^+(1540)$ as
\begin{equation}
g^{2}_{\overline{\bm{10}}\mbox{-}\bm{8}_3}= \frac{\pi M_\Theta
\Gamma_\Theta}{3 |\bm{p}_K| ( \sqrt{ M_N^2 + \bm{p}_K^2} \mp M_N)}
\end{equation}
where $M_\Theta$ and $\Gamma_\Theta$ are the mass and decay width of
$\Theta^+(1540)$, respectively, and $p_K$ is the momentum of kaon in the
$\Theta^+$ rest frame.
The upper and lower sign correspond to the case when the parity of
$\Theta^+$ is even and odd.
Numerically, we have $g^{2}_{\overline{\bm{10}}\mbox{-}\bm{8}_3}=
\Gamma_\Theta / (6.19 \mbox{ MeV})$ for the case that $\Theta(1540)$ has
even-parity, which gives
$g^{}_{\overline{\bm{10}}\mbox{-}\bm{8}_3}= 0.4$ if $\Gamma_\Theta = 1$
MeV is used, and $0.9$ if $\Gamma_\Theta = 5$ MeV. 
If $\Theta(1540)$ has odd-parity, then we have
$g^{2}_{\overline{\bm{10}}\mbox{-}\bm{8}_3}= \Gamma_\Theta / (318.2
\mbox{ MeV})$, which gives 
$g^{}_{\overline{\bm{10}}\mbox{-}\bm{8}_3}= 0.056$ and $0.125$ for
$\Gamma_\Theta = 1$ MeV and $5$ MeV, respectively.
We refer to Refs.~\cite{HK03,Nuss03-CN03,ASW03} for detailed discussions
on the decay width of $\Theta^+(1540)$.

\subsubsection{$\bm{27}$-$\bm{8}_3$}

In this case, the interaction can be written as
\begin{equation}
\mathcal{L}_{\bm{27}\mbox{-}\bm{8}_3} =
g_{\bm{27}\mbox{-}\bm{8}_3}^{} \,
\overline{T}^{kl}_{ij} B^i_k M^j_l + \mbox{(H.c.)}.
\end{equation}
Again, because of symmetric upper and lower indices of 
$\overline{T}^{kl}_{ij}$, other possible contractions are not independent
from the above expression.
In terms of the physical baryon states, we have 
\begin{eqnarray}
\mathcal{L}_{\bm{27}\mbox{-}\bm{8}_3} / g_{\bm{27}\mbox{-}\bm{8}_3}^{} &=&
\sqrt{2}\, \overline{K}_c \,\overline{\bm{\Theta}}_1 \cdot \bm{\tau}\, N
+ \overline{\Delta}_{27}\, \bm{T}(\textstyle\frac32,\frac12)
 \cdot \bm{\pi}\, N
+ \overline{\Delta}_{27}\, \bm{T}(\textstyle\frac32,\frac12)
 \cdot \bm{\Sigma}\, K
\nonumber \\ && \mbox{}
+ \frac{1}{\sqrt{15}} \overline{N}_{27} \bm{\tau} \cdot \bm{\pi} N
+ \frac{1}{\sqrt{15}} \overline{N}_{27} \bm{\tau} \cdot \bm{\Sigma} K
+ \frac{3}{\sqrt5} \overline{N}_{27} K \Lambda^0 
\nonumber \\ && \mbox{}
+ \frac{3}{\sqrt5} \overline{N}_{27}\, N \,\eta
+ \frac{2}{\sqrt3} \overline{\Sigma}_{27,2} \bm{T}(2,1) \cdot \bm{\pi}
\Sigma_c
- \sqrt{\frac25} \overline{K} \,\overline{\bm{\Sigma}}_{27} \cdot \bm{\tau} N
\nonumber \\ && \mbox{}
+ \sqrt{\frac25} \overline{K}_c \,\overline{\bm{\Sigma}}_{27} \cdot
\bm{\tau} \Xi
+ \sqrt{\frac65} \overline{\Sigma}_{27} \, {\Sigma} \,\eta
+ \sqrt{\frac65} \overline{\bm{\Sigma}}_{27} \cdot \bm{\pi} \Lambda^0
\nonumber \\ && \mbox{}
- \sqrt{\frac{3}{10}} \overline{\Lambda}_{27}^0 \overline{K} N
- \sqrt{\frac{3}{10}} \overline{\Lambda}_{27}^0 \overline{K}_c \Xi
+ \frac{1}{\sqrt{30}} \overline{\Lambda}_{27}^0 \bm{\Sigma} \cdot \bm{\pi}
\nonumber \\ && \mbox{}
+ 3\sqrt{\frac{3}{10}} \overline{\Lambda}_{27}^0 \Lambda^0 \eta
+ \overline{\Xi}_{27,3/2} \bm{T}(\textstyle\frac32,\frac12)
\cdot \bm{\Sigma} K_c
+ \overline{\Xi}_{27,3/2} \bm{T}(\textstyle\frac32,\frac12) \cdot \bm{\pi} \Xi
\nonumber \\ && \mbox{}
- \frac{1}{\sqrt{15}} \overline{\Xi}_{27} \bm{\tau} \cdot \bm{\Sigma}
  K_c 
- \frac{1}{\sqrt{15}} \overline{\Xi}_{27} \bm{\tau} \cdot \bm{\pi} \Xi
+ \frac{3}{\sqrt5} \overline{\Xi}_{27} K_c \Lambda^0
\nonumber \\ && \mbox{}
+ \frac{3}{\sqrt5} \overline{\Xi}_{27} \Xi \eta
- \sqrt2 \overline{K} \,\overline{\bm{\Omega}}_{27,1} \cdot \bm{\tau} \Xi
+ \mbox{(H.c.)},
\label{Lag:27-8}
\end{eqnarray}
where
\begin{eqnarray}
\Sigma_c = (-\Sigma^+, \Sigma^0, \Sigma^-)^T, \qquad
\Xi_{27,3/2} = (\Xi_{27,3/2}^+, \Xi_{27,3/2}^0, \Xi_{27,3/2}^-,
\Xi_{27,3/2}^{--})^T,
\nonumber \\
\bm{\tau} \cdot \bm{\Theta}_1 = \left( \begin{array}{cc} \Theta_1^+ &
\sqrt2 \Theta_1^{++} \\ \sqrt2 \Theta_1^0 & - \Theta_1^+ \end{array}
\right), \qquad
\bm{\tau} \cdot \bm{\Omega}_{27,1} = \left( \begin{array}{cc}
\Omega_{27,1}^- & \sqrt2 \Omega_{27,1}^{0} \\
\sqrt2 \Omega_{27,1}^{--} & - \Omega_{27,1}^- \end{array}
\right),
\end{eqnarray}
and
\begin{equation}
\bm{T}(2,1) \cdot \bm{\pi} = - T^{(+1)}_{2,1} \pi^+ + T^{(-1)}_{2,1}
\pi^- + T^{(0)}_{2,1} \pi^0,
\end{equation}
with
\begin{eqnarray}
&&
T^{(+1)}_{2,1} = \frac{1}{\sqrt2}
\left( \begin{array}{ccc} \sqrt6 & 0 & 0 \\
0 & \sqrt{3} & 0 \\
0 & 0 & 1 \\ 0 & 0 & 0 \\ 0 & 0 & 0 \end{array} \right),
\quad
T^{(0)}_{2,1} = \frac{1}{\sqrt2}
\left( \begin{array}{ccc} 0 & 0 & 0 \\
\sqrt{3} & 0 & 0 \\
0 & 2 & 0 \\ 0 & 0 & \sqrt{3} \\ 0 & 0 & 0 \end{array} \right),
\nonumber \\ &&
T^{(-1)}_{2,1} = \frac{1}{\sqrt2}
\left( \begin{array}{ccc} 0 & 0 & 0 \\
0 & 0 & 0 \\
1 & 0 & 0 \\ 0 & \sqrt3 & 0 \\ 0 & 0 & \sqrt6 \end{array} \right).
\end{eqnarray}
Thus,
\begin{eqnarray}
 \overline{\Sigma}_{27,2} \bm{T}(2,1) \cdot \bm{\pi}\, \Sigma_c
&=& \sqrt3\, \overline{\Sigma}_{27,2}^{++} \Sigma^+ \pi^+ 
- \sqrt{\frac32}\, \overline{\Sigma}_{27,2}^{+} \left( \Sigma^0 \pi^+ +
\sqrt{\frac32} \Sigma^+ \pi^0 \right)
- \frac{1}{\sqrt2}\,\overline{\Sigma}_{27,2}^{0} \left( \Sigma^- \pi^+ -
2 \Sigma^0 \pi^0 + \Sigma^+ \pi^- \right)
\nonumber \\ && \mbox{}
+ \sqrt{\frac32}\, \overline{\Sigma}_{27,2}^{-} \left( \Sigma^- \pi^0 +
 \Sigma^0 \pi^- \right)
+ \sqrt3 \, \overline{\Sigma}_{27,2}^{--} \Sigma^- \pi^- ,
\end{eqnarray}
and
\begin{equation}
\overline{K}_c \,\overline{\bm{\Theta}}_1 \cdot \bm{\tau}\, N
= -\sqrt2 \,\overline{\Theta}_1^{++} p K^+ + \overline{\Theta}_1^+ \left(
K^0 p + K^+ n \right) + \sqrt2 \,\overline{\Theta}_1^0 n K^0.
\end{equation}

Note that $\Theta_1^+$ decays into $KN$ quite similarly as the
anti-decuplet $\Theta^+$ does except the relative sign difference between
the proton and neutron channel. 
This would make it difficult to disentangle the two if their mass difference
is not big enough.
Thus, the decay of $\Theta_1^{++}$ and $\Theta_1^0$ would be useful to
identify the $\Theta_1$ isomultiplet in $KN$ channel.

\subsection{Interactions with 3-quark decuplet and meson octet}

We now move to interactions for pentaquarks with 3-quark decuplet and
meson octet.  
In this case, we have
\begin{equation}
\bm{10} \otimes \bm{8} = \bm{35} \oplus \bm{27} \oplus \bm{10} \oplus
\bm{8},
\end{equation}
so that pentaquark antidecuplet cannot couple to this channel.
Therefore this can give a useful constraint in finding antidecuplet
baryons \cite{OKL03b}.
For example, $N(1710)$ cannot be a pure antidecuplet state as we
discussed before.
As in the case of pentaquark decuplet, the three-quark decuplet is
represented by $D^{(3)}_{ijk}$ as
\begin{eqnarray}
&&
D^{(3)}_{111} = \sqrt6 \Delta^{++}, \qquad
D^{(3)}_{112} = \sqrt2 \Delta^+, \qquad
D^{(3)}_{122} = \sqrt2 \Delta^0, \nonumber \\
&&
D^{(3)}_{222} = \sqrt6 \Delta^-,
\qquad D^{(3)}_{113} = \sqrt2 \Sigma^{*+}, \qquad
D^{(3)}_{123} = -\Sigma^{*0}, \nonumber \\
&&
D^{(3)}_{223} = -\sqrt2 \Sigma^{*-}, \qquad
D^{(3)}_{133} = \sqrt2 \Xi^{*0}, \qquad
D^{(3)}_{233} = \sqrt2 \Xi^{*-}, \nonumber \\
&&
D^{(3)}_{333} = -\sqrt6 \Omega^-.
\end{eqnarray}

\subsubsection{$\bm{8}$-$\bm{10}_3$}

The interaction Lagrangian for pentaquark octet and normal baryon
decuplet reads
\begin{equation}
\mathcal{L}_{\bm{8}\mbox{-}\bm{10}_3} =
g_{\bm{8}\mbox{-}\bm{10}_3}^{} \,
\epsilon^{ijk} \overline{P}^l_i D^{(3)}_{jlm} M^m_k
+ \mbox{(H.c.)},
\end{equation}
which gives
\begin{eqnarray}
\mathcal{L}_{\bm{8}\mbox{-}\bm{10}_3} / g_{\bm{8}\mbox{-}\bm{10}_3}^{} &=&
\sqrt{3}\, \overline{N}_8\, \bm{T}(\textstyle \frac12, \frac32)
\cdot \bm{\pi}\, \Delta
+ \overline{N}_8\, \bm{\tau} \cdot \bm{\Sigma}^*\, K
- i \left( \overline{\bm{\Sigma}}_8 \times \bm{\Sigma}^* \right) \cdot
  \bm{\pi}
\nonumber \\ && \mbox{}
+ \sqrt3 \,\overline{\Sigma}_8 \, \Sigma^*\, \eta -
\overline{K}_c \,\overline{\bm{\Sigma}}_8 \cdot \bm{\tau}\, \Xi^*
- \sqrt3\, \overline{K}\, \overline{\bm{\Sigma}}_8\,
\cdot \bm{T}(\textstyle\frac12,\frac32)\, \Delta
\nonumber \\ && \mbox{}
-\sqrt3\, \overline{\Lambda}_8^0 \,\bm{\Sigma}^* \cdot \bm{\pi}
- \sqrt3\, \overline{\Lambda}_8^0\, \overline{K}_c \,\Xi^* +
  \overline{\Xi}_8\, \bm{\tau} \cdot \bm{\pi} \, \Xi^*
\nonumber \\ && \mbox{}
+ \sqrt3 \, \overline{\Xi}_8 \, {\Xi}^* \, \eta - \sqrt6 \,
\overline{\Xi}_8 \, K \Omega^- + \overline{\Xi}_8 \, \bm{\tau} \cdot
\bm{\Sigma}^* \, K_c
+ \mbox{(H.c.)},
\label{Lag:8-10}
\end{eqnarray}
where
\begin{equation}
\bm{T}(\textstyle\frac12,\frac32) \cdot \bm{\pi}
 = - T^{(+1)}_{1/2,3/2} \pi^- +
T^{(-1)}_{1/2,3/2} \pi^+ + T^{(0)}_{1/2,3/2} \pi^0,
\label{t13}
\end{equation}
and
\begin{eqnarray}
&&
T^{(+1)}_{1/2,3/2} = -\frac{1}{\sqrt3} \left( \begin{array}{cccc}
\sqrt6 & 0 & 0 & 0 \\ 0 & \sqrt2 & 0 & 0 \end{array} \right),
\quad
T^{(0)}_{1/2,3/2} = -\frac{1}{\sqrt3} \left( \begin{array}{cccc}
0 & 2 & 0 & 0 \\ 0 & 0 & 2 & 0 \end{array} \right),
\nonumber \\ &&
T^{(-1)}_{1/2,3/2} = -\frac{1}{\sqrt3} \left( \begin{array}{cccc}
0 & 0 & \sqrt2 & 0 \\ 0 & 0 & 0 & \sqrt6 \end{array} \right),
\end{eqnarray}
so that
\begin{eqnarray}
\overline{N}_8 \bm{T}({\textstyle\frac12,\frac32}) \cdot \bm{\pi}\,
\Delta &=&
\overline{N}_8^+ \left( \sqrt2 \Delta^{++} \pi^- - \sqrt{\frac23}
\Delta^0 \pi^+ - \frac{2}{\sqrt3} \Delta^+ \pi^0 \right)
\nonumber \\ && \mbox{}
+ \overline{N}_8^0 \left( \sqrt{\frac23} \Delta^{+} \pi^- - \sqrt{2}
\Delta^- \pi^+ - \frac{2}{\sqrt3} \Delta^0 \pi^0 \right).
\end{eqnarray}

Note that \cite{CN69}
\begin{equation}
\bm{T}({\textstyle\frac12,\frac32}) = -
\bm{T}({\textstyle\frac32,\frac12})^\dagger,
\end{equation}
and the difference between Eqs.~(\ref{t31}) and (\ref{t13}).

\subsubsection{$\bm{10}$-$\bm{10}_3$}

The interaction Lagrangian reads
\begin{equation}
\mathcal{L}_{\bm{10}\mbox{-}\bm{10}_3} =
 g_{\bm{10}\mbox{-}\bm{10}_3}^{} \,
\overline{D}^{jkl} D^{(3)}_{mkl} M^m_j
+ \mbox{(H.c.)},
\end{equation}
which leads to
\begin{eqnarray}
\mathcal{L}_{\bm{10}\mbox{-}\bm{10}_3} / g_{\bm{10}\mbox{-}\bm{10}_3}^{} &=&
\sqrt{\frac{15}{2}}\, \overline{\Delta}_{10} \bm{T}({\textstyle\frac32,
\frac32}) \cdot \bm{\pi} \, \Delta
+ \sqrt6\, \overline{\Delta}_{10} \Delta \, \eta
- \sqrt6
\, \overline{\Delta}_{10}\, \bm{T}({\textstyle\frac32,\frac12}) \cdot
  \bm{\Sigma}^*\, K
\nonumber \\ && \mbox{}
-i 2\sqrt{2}\, \left( \overline{\bm{\Sigma}}_{10} \times \bm{\Sigma}^*
\right) \cdot \bm{\pi}
- 2\sqrt2 \,\overline{K}_c \,\overline{\bm{\Sigma}}_{10} \cdot \bm{\tau}
  \,\Xi^*
+ \sqrt6 \,\overline{K} \, \overline{\bm{\Sigma}}_{10}
\cdot \bm{T}(\textstyle\frac12,\frac32) \, \Delta
\nonumber \\ && \mbox{}
+\sqrt2 \,\overline{\Xi}_{10}\, \bm{\tau} \cdot \bm{\pi}\, \Xi^* - \sqrt6
\,\overline{\Xi}_{10}\, \Xi^* \,\eta
- 2\sqrt3\, \overline{\Xi}_{10}\, K\, \Omega^-
\nonumber \\ && \mbox{}
-2\sqrt2\, \overline{\Xi}_{10}\, \bm{\tau} \cdot \bm{\Sigma}^*\, K_c - 2\sqrt6
\,\overline{\Omega}_{10}^- \, \Omega^-\, \eta
- 2\sqrt3 \,\overline{\Omega}_{10}^- \,\overline{K}\, \Xi^*
+ \mbox{(H.c.)},
\label{Lag:10-10}
\end{eqnarray}
where
\begin{equation}
\bm{T}(\textstyle\frac32,\frac32) \cdot \bm{\pi}
 = - T^{(+1)}_{3/2,3/2} \pi^+ +
T^{(-1)}_{3/2,3/2} \pi^- + T^{(0)}_{3/2,3/2} \pi^0,
\end{equation}
and
\begin{eqnarray}
&&
T^{(+1)}_{3/2,3/2} = - \frac{4}{\sqrt{30}} \left( \begin{array}{cccc} 0
& \sqrt3 & 0 & 0 \\ 0 & 0 & 2 & 0 \\ 0 & 0 & 0 & \sqrt3 \\ 0 & 0 & 0 & 0
\end{array} \right), \quad
T^{(0)}_{3/2,3/2} = \frac{2}{\sqrt{15}} \left( \begin{array}{cccc} 3
& 0 & 0 & 0 \\ 0 & 1 & 0 & 0 \\ 0 & 0 & -1 & 0 \\ 0 & 0 & 0 & -3
\end{array} \right),
\nonumber \\ &&
T^{(-1)}_{3/2,3/2} = \frac{4}{\sqrt{30}} \left( \begin{array}{cccc} 0
& 0 & 0 & 0 \\ \sqrt3 & 0 & 0 & 0 \\ 0 & 2 & 0 & 0 \\ 0 & 0 & \sqrt3 & 0
\end{array} \right).
\end{eqnarray}
Thus we have
\begin{eqnarray}
\overline{\Delta}_{10}\, \bm{T}(\textstyle\frac32,\frac32) \cdot
\bm{\pi}\, \Delta &=&
\sqrt{\frac65}\, \overline{\Delta}_{10}^{++} \left( \sqrt2
\Delta_{10}^{++} \pi^0 + \frac{2}{\sqrt3} \Delta^+ \pi^+ \right)
+ \sqrt{\frac65}\, \overline{\Delta}_{10}^{+} \left( \frac{2}{\sqrt3}
\Delta^{++} \pi^- + \frac{\sqrt2}{3} \Delta^+ \pi^0 + \frac43 \Delta^0
\pi^+ \right)
\nonumber \\ && \mbox{}
+ \sqrt{\frac65}\, \overline{\Delta}_{10}^{0} \left( \frac43 \Delta^+
\pi^- - \frac{\sqrt2}{3} \Delta^0 \pi^0 + \frac{2}{\sqrt3} \Delta^-
\pi^+ \right)
+ \sqrt{\frac65}\, \overline{\Delta}_{10}^{-} \left( \frac{2}{\sqrt3}
\Delta^0 \pi^- - \sqrt2 \Delta^- \pi^0 \right).
\nonumber \\ && \mbox{}
\end{eqnarray}

\subsubsection{$\bm{27}$-$\bm{10}_3$}

The Lagrangian of this interaction reads
\begin{equation}
\mathcal{L}_{\bm{27}\mbox{-}\bm{10}_3} =
g_{\bm{27}\mbox{-}\bm{10}_3}^{} \,
\epsilon^{imn} \overline{T}_{ij}^{kl} D^{(3)}_{mkl} M^j_n
+ \mbox{(H.c.)},
\end{equation}
which gives
\begin{eqnarray}
\mathcal{L}_{\bm{27}\mbox{-}\bm{10}_3} / g_{\bm{27}\mbox{-}\bm{10}_3}^{}
&=&
-2\sqrt3\, \overline{K}_c\, \overline{\bm{\Theta}}_1 \cdot
\bm{T}({\textstyle\frac12,\frac32})\, \Delta
+ \sqrt{\frac52}\, \overline{\Delta}_{27}\,
\bm{T}({\textstyle\frac32,\frac32}) \cdot \bm{\pi} \Delta\,
-3\sqrt2\, \overline{\Delta}_{27} \, {\Delta}\, \eta
\nonumber \\ && \mbox{}
+ \sqrt2\, \overline{\bm{\Delta}}_{27}\,
\bm{T}({\textstyle\frac32,\frac12}) \cdot \bm{\Sigma}^* \, K
+ \frac{2\sqrt2}{\sqrt5}\, \overline{N}_{27}\, \bm{T}({\textstyle\frac12,
\frac32}) \cdot \bm{\pi} \, \Delta
\nonumber \\ && \mbox{}
- \frac{8\sqrt2}{\sqrt{15}}\, \overline{N}_{27}\, \bm{\tau} \cdot
  \bm{\Sigma}^*\, K
- 3\, \overline{\Sigma}_{27,2}\, \bm{I}({\textstyle2,\frac32}) \cdot
  \bm{K}_c \, \Delta
\nonumber \\ && \mbox{}
+ \frac{2\sqrt2}{\sqrt3} \, \overline{\Sigma}_{27,2}\, \bm{T}(2,1) \cdot
\bm{\pi}\, \Sigma_c^*
- i \frac{6}{\sqrt5} \left( \overline{\bm{\Sigma}}_{27} \times
  \bm{\Sigma}^* \right) \cdot \bm{\pi}
\nonumber \\ && \mbox{}
+ \frac{4}{\sqrt5}\, \overline{K}_c \,\overline{\bm{\Sigma}}_{27} \cdot
\bm{\tau}\, \Xi^*
- \frac{4\sqrt3}{\sqrt5}\, \overline{\Sigma}_{27} \, {\Sigma}^*\, \eta
- \sqrt{\frac35}\, \overline{K}\, \overline{\bm{\Sigma}}_{27} \cdot
  \bm{T}({\textstyle\frac12,\frac32}) \, \Delta
\nonumber \\ && \mbox{}
- \frac{8}{\sqrt{15}}\, \overline{\Lambda}_{27}^0\, \bm{\Sigma}^* \cdot
  \bm{\pi}
+ \frac{4\sqrt{15}}{5}\, \overline{\Lambda}_{27}^0\, \overline{K}_c \,\Xi^*
+ 2\sqrt2\, \overline{\Xi}_{27,3/2}\, \bm{T}({\textstyle\frac32,\frac12})
\cdot \bm{\pi}\, \Xi^*
\nonumber \\ && \mbox{}
- 2\sqrt2\, \overline{\Xi}_{27,3/2}\, \bm{T}({\textstyle\frac32,\frac12})
\cdot \bm{\Sigma}^*\, K_c
+ \frac{7\sqrt2}{\sqrt{15}}\, \overline{\Xi}_{27}\, \bm{\tau} \cdot \bm{\pi}
\,\Xi^*
- \frac{3\sqrt2}{\sqrt5}\, \overline{\Xi}_{27} \, {\Xi}^* \, \eta
\nonumber \\ && \mbox{}
+ \frac{6}{\sqrt5}\, \overline{\Xi}_{27}\, K\, \Omega^-
+ \frac{2\sqrt2}{\sqrt{15}}\, \overline{\Xi}_{27}\, \bm{\tau} \cdot
\bm{\Sigma}^* \, K_c
+ 2\sqrt6\, \overline{\bm{\Omega}}_{27,1} \cdot \bm{\pi}\, \Omega^-
\nonumber \\ && \mbox{}
+ 2\, \overline{K}\, \overline{\bm{\Omega}}_{27,1} \cdot \bm{\tau}\, \Xi^*
+ \mbox{(H.c.)},
\label{Lag:27-10}
\end{eqnarray}
where
\begin{equation}
\bm{I}({\textstyle 2,\frac32}) \cdot \bm{K}_c = I^{1/2}_{2,3/2}\,
\overline{K}^0 - I^{-1/2}_{2,3/2}\, K^-,
\end{equation}
with
\begin{eqnarray}
I^{1/2}_{2,3/2} = \sqrt{\frac23} \left( \begin{array}{cccc}
2 & 0 & 0 & 0 \\ 0 & \sqrt3 & 0 & 0 \\ 0 & 0 & \sqrt2 & 0 \\
0 & 0 & 0 & 1 \\ 0 & 0 & 0 & 0 \end{array} \right),
\qquad
I^{-1/2}_{2,3/2} = \sqrt{\frac23} \left( \begin{array}{cccc}
0 & 0 & 0 & 0 \\ 1 & 0 & 0 & 0 \\ 0 & \sqrt2 & 0 & 0 \\
0 & 0 & \sqrt3 & 0 \\ 0 & 0 & 0 & 2 \end{array} \right).
\end{eqnarray}
So we have
\begin{eqnarray}
\overline{K}_c\, \overline{\bm{\Theta}}_1 \cdot
\bm{T}({\textstyle\frac12,\frac32})\, \Delta &=&
-\sqrt{\frac23}\, \overline{\Theta}_{1}^{++} \left( \Delta^+ K^+ - \sqrt3
\Delta^{++} K^0 \right)
+\frac{2}{\sqrt3}\, \overline{\Theta}_{1}^{+} \left( \Delta^0 K^+ -
\Delta^{+} K^0 \right)
\nonumber \\ && \mbox{}
+ \sqrt{\frac23}\, \overline{\Theta}_{1}^{0} \left( \sqrt{3} \Delta^- K^+ -
\Delta^{0} K^0 \right),
\nonumber \\
\overline{\Sigma}_{27,2}\, \bm{I}({\textstyle2,\frac32}) \cdot
  \bm{K}_c \, \Delta &=&
\frac{2\sqrt2}{\sqrt3} \overline{\Sigma}_{27,2}^{++} \Delta^{++}
\overline{K}^0
- \sqrt{\frac23} \left( \Delta^{++} K^- - \sqrt3 \Delta^+ \overline{K}^0
  \right)
- \frac{2}{\sqrt3} \left( \Delta^{+} K^- - \Delta^0 \overline{K}^0
  \right)
\nonumber \\ && \mbox{}
- \sqrt{\frac23} \left( \sqrt3 \Delta^{0} K^- - \Delta^- \overline{K}^0
  \right)
- \frac{2\sqrt2}{\sqrt3} \overline{\Sigma}_{27,2}^{--} \Delta^{-} K^-.
\end{eqnarray}
As $\Theta_1$ can decay into $K\Delta$, this decay channel can
distinguish $\Theta_1^+$ from $\Theta^+$.

\subsubsection{$\bm{35}$-$\bm{10}_3$}

The interaction Lagrangian reads
\begin{equation}
\mathcal{L}_{\bm{35}\mbox{-}\bm{10}_3} =
g_{\bm{35}\mbox{-}\bm{10}_3}^{} \,
\overline{T}^{ijkl}_a D_{ijk}^{(3)} M^a_l
+ \mbox{(H.c.)},
\end{equation}
so that we have
\begin{eqnarray}
\mathcal{L}_{\bm{35}\mbox{-}\bm{10}_3} /
g_{\bm{35}\mbox{-}\bm{10}_3}^{}  &=&
\frac{6\sqrt3}{\sqrt2}\, \overline{\Theta}_2 \bm{I}
({\textstyle 2,\frac32}) \cdot \bm{K} \Delta
+ 6 \, \overline{\Delta}_{5/2} \bm{T} ({\textstyle\frac52,\frac32})
\cdot \bm{\pi} \Delta
+ \frac32 \, \overline{\Delta}_{35} \bm{T} ({\textstyle\frac32,\frac32})
\cdot \bm{\pi} \Delta
\nonumber \\ && \mbox{}
+ 3\sqrt5\, \overline{\Delta}_{35} \, {\Delta} \, \eta
+ 3 \sqrt5\, \overline{\Delta}_{35} \bm{T}({\textstyle\frac32,\frac12})
\cdot \bm{\Sigma}^* K
+ 6 \, \overline{\Sigma}_{35,2} \bm{T} (2,1) \cdot \bm{\pi} \Sigma^*
\nonumber \\ && \mbox{}
+ \frac{3\sqrt3}{\sqrt2}\, \overline{\Sigma}_{35,2}
\bm{I}(2,{\textstyle\frac32}) \cdot \bm{K}_c \Delta
-i \sqrt6 \left( \overline{\bm{\Sigma}}_{35} \times \bm{\Sigma}^*
\right) \cdot \bm{\pi}
+ 6\sqrt2\, \overline{\Sigma}_{35} \, {\Sigma}^*\, \eta
\nonumber \\ && \mbox{}
+ 2\sqrt6\, \overline{K}_c \, \overline{\bm{\Sigma}}_{35} \cdot \bm{\tau}
\Xi^*
+ \frac{3}{\sqrt2} \, \overline{K} \,\overline{\bm{\Sigma}}_{35} \cdot
\bm{T} ({\textstyle\frac12,\frac32}) \Delta
+ 6\, \overline{\Xi}_{35,3/2} \bm{T}({\textstyle\frac32,\frac12}) \cdot
\bm{\pi} \,\Xi^*
\nonumber \\ && \mbox{}
+ 6\, \overline{\Xi}_{35,3/2} \bm{T}({\textstyle\frac32,\frac12}) \cdot
\bm{\Sigma}^* K_c
+ \sqrt3\, \overline{\Xi}_{35} \bm{\tau} \cdot \bm{\pi} \Xi^*
+ 9\, \overline{\Xi}_{35} \, {\Xi}\, \eta
\nonumber \\ && \mbox{}
+ 3\sqrt2 \, \overline{\Xi}_{35} \, K\, \Omega^-
- 2\sqrt3\, \overline{\Xi}_{35} \bm{\tau} \cdot \bm{\Sigma}^* K_c
+ 6\, \overline{\bm{\Omega}}_{35,1} \cdot \bm{\pi} \, \Omega^-
\nonumber \\ && \mbox{}
- 3\sqrt6\, \overline{K} \, \overline{\bm{\Omega}}_{35,1} \cdot
\bm{\tau}\, \Xi^*
+ 6 \sqrt2\, \overline{\Omega}_{35} \, \Omega^-\, \eta
- 6\, \overline{\Omega}_{35}^- \, \overline{K} \Xi^*
\nonumber \\ && \mbox{}
+ 12 \, \overline{X} \, K_c\, \Omega^-
+ \mbox{(H.c.)},
\label{Lag:35-10}
\end{eqnarray}
where
\begin{eqnarray}
\Delta_{5/2} &=& (\Delta_{5/2}^{+++},\ \Delta_{5/2}^{++},\ 
\Delta_{5/2}^{+},\ \Delta_{5/2}^{0},\ \Delta_{5/2}^{-},\
\Delta_{5/2}^{--})^T, 
\nonumber \\
\bm{T}(\textstyle\frac52,\frac32) \cdot \bm{\pi}
 &=& - T^{(+1)}_{5/2,3/2} \pi^+ +
T^{(-1)}_{5/2,3/2} \pi^- + T^{(0)}_{5/2,3/2} \pi^0,
\end{eqnarray}
and
\begin{eqnarray}
&&
T^{(+1)}_{5/2,3/2} = \frac{1}{\sqrt5} \left( \begin{array}{cccc}
\sqrt{20} & 0 & 0 & 0 \\
0 & \sqrt{12} & 0 & 0 \\
0 & 0 & \sqrt6 & 0 \\
0 & 0 & 0 & \sqrt2 \\
0 & 0 & 0 & 0 \\
0 & 0 & 0 & 0 \end{array} \right), \qquad
T^{(0)}_{5/2,3/2} = \frac{2}{\sqrt5} \left( \begin{array}{cccc}
0 & 0 & 0 & 0 \\
\sqrt{2} & 0 & 0 & 0 \\
0 & \sqrt{3} & 0 & 0 \\
0 & 0 & \sqrt3 & 0 \\
0 & 0 & 0 & \sqrt2 \\
0 & 0 & 0 & 0 \end{array} \right),
\nonumber \\ &&
T^{(-1)}_{5/2,3/2} = \frac{1}{\sqrt5} \left( \begin{array}{cccc}
0 & 0 & 0 & 0 \\
0 & 0 & 0 & 0 \\
\sqrt{2} & 0 & 0 & 0 \\
0 & \sqrt{6} & 0 & 0 \\
0 & 0 & \sqrt{12} & 0 \\
0 & 0 & 0 & \sqrt{20} \\
\end{array} \right),
\end{eqnarray}
which gives
\begin{eqnarray}
\overline{\Delta}_{5/2}\, \bm{T} ({\textstyle\frac52,\frac32})
\cdot \bm{\pi}\, \Delta &=&
-2\, \overline{\Delta}_{5/2}^{+++} \Delta^{++} \pi^+
-\frac{2}{\sqrt5}\, \overline{\Delta}_{5/2}^{++} \left( \sqrt3
\Delta^{+} \pi^+ - \sqrt2 \Delta^{++} \pi^0 \right)
\nonumber \\ && \mbox{}
-\sqrt{\frac25}\, \overline{\Delta}_{5/2}^{+} \left( \sqrt3 \Delta^{0} \pi^+
- \sqrt6 \Delta^+ \pi^0 - \Delta^{++} \pi^- \right)
-\sqrt{\frac25}\, \overline{\Delta}_{5/2}^{0} \left( \Delta^{-} \pi^+
- \sqrt6 \Delta^0 \pi^0 - \sqrt3 \Delta^- \pi^- \right)
\nonumber \\ && \mbox{}
+ \frac{2}{\sqrt5} \, \overline{\Delta}_{5/2}^{-} \left( \sqrt2 \Delta^{-} \pi^+
+\sqrt3 \Delta^0 \pi^- \right)
+2\, \overline{\Delta}_{5/2}^{--} \Delta^{-} \pi^-.
\end{eqnarray}

Since $\bm{35}$-plet is the highest multiplet among pentaquark multiplets,
some resonances located in the boundary of the weight diagram have unique
decay channels that do not suffer from mixing with other multiplets due to
their quantum numbers.
In particular, $X^-$ ($X^{--}$) decays into $\overline{K}^0 \Omega^-$ 
($K^- \Omega^-$) and this is the unique mode that can be searched
in experiments.
Similarly, the isotensor ($I=2$) $\Theta^{++}$, $\Theta^-$
as well as $\Delta^{+++}_{5/2}$, $\Delta^{--}_{5/2}$ with $I=5/2$
can be measured without suffering from the mixing.
In addition, $\Theta_2$ can decay into $K\Delta$.
This isospin selection rule may be useful to identify the $Y=2$ baryons.
Namely, $\Theta^+$ (and its higher spin resonances) can decay into $KN$ only,
while $\Theta_2$ can decay into $K\Delta$ only.
But $\Theta_1$ can decay into both $KN$ and $K\Delta$, if energetically
allowed.

\subsection{Interactions with pentaquark baryons}

Now we consider the interactions among pentaquarks with meson octet.
These interactions can be constructed by noting that
\begin{eqnarray}
\bm{8} \otimes \bm{8} &=& \bm{27} \oplus \bm{10} \oplus \overline{\bm{10}}
\oplus \bm{8}_1 \oplus \bm{8}_2 \oplus \bm{1}, \nonumber \\
\bm{10} \otimes \bm{8} &=& \bm{35} \oplus \bm{27} \oplus \bm{10} \oplus
\bm{8}, \nonumber \\
\overline{\bm{10}} \otimes \bm{8} &=& \overline{\bm{35}} \oplus \bm{27}
\oplus \overline{\bm{10}} \oplus \bm{8},
\nonumber \\
\bm{27} \otimes \bm{8} &=& \bm{64} \oplus \bm{35} \oplus
\overline{\bm{35}} \oplus \bm{27}_1 \oplus \bm{27}_2 \oplus \bm{10} \oplus
\overline{\bm{10}} \oplus \bm{8},
\nonumber \\
\bm{35} \otimes \bm{8} &=& \bm{81} \oplus \bm{64} \oplus \bm{35}_1 \oplus
\bm{35}_2 \oplus \bm{28} \oplus \bm{27} \oplus \bm{10}.
\end{eqnarray}
Thus we can see the followings.
First, the pentaquark singlet can couple to pentaquark octet only.
Second, the $\bm{27}$-$\bm{27}$ and $\bm{35}$-$\bm{35}$ interactions
have two types like $\bm{8}$-$\bm{8}$ interaction.
Third, the interactions including $\bm{10}$-$\overline{\bm{10}}$,
$\bm{35}$-$\bm{8}$, and $\bm{35}$-$\overline{\bm{10}}$ are not allowed
as they cannot form SU(3)-invariant interactions.
Thus, $\bm{35}$-plet couplings are limited to the interactions with
$\bm{27}$-plet and decuplet.

\subsubsection{$\bm{8}$-$\bm{8}$}

This interaction Lagrangian contains the well-known $f$- and $d$-type
interactions as in Eq.~(\ref{Lag:8-8}):
\begin{equation}
\mathcal{L}_{\bm{8}\mbox{-}\bm{8}} /
g_{\bm{8}\mbox{-}\bm{8}}^{} =
(d+f) \bar{P}_i^l M_l^k P_k^i +
(d-f) \bar{P}_i^l P_l^k M_k^i ,
\end{equation}
so that
\begin{eqnarray}
\mathcal{L}_{\bm{8}\mbox{-}\bm{8}} /
g_{\bm{8}\mbox{-}\bm{8}}^{} =
&=&
\frac{1}{\sqrt2} (d+f)\,
\overline{N}_8^{} \bm{\tau} \cdot \bm{\pi} N_8^{}
- \frac{1}{\sqrt6} (d-3f)\,
\overline{N}_8^{} N_8^{} \, \eta
- \frac{1}{\sqrt6} (d+3f)\,
\overline{N}_8^{} K \Lambda_8^{0}
\nonumber \\ && \mbox{}
+ \frac{1}{\sqrt2} (d-f)\,
\overline{N}_8^{} \bm{\tau}\cdot \bm{\Sigma}_8^{} \, K
- i \sqrt2 f \left( \overline{\bm \Sigma}_8^{}
\times \bm{\Sigma}_8^{} \right) \cdot \bm{\pi}
+ \sqrt{\frac23} d \,
\overline{\Sigma}_8 \, {\Sigma}_8^{}\, \eta
\nonumber \\ && \mbox{}
+ \sqrt{\frac23} d\,
\overline{\bm{\Sigma}}_8 \cdot \bm{\pi}\, \Lambda_8^{0}
- \frac{1}{\sqrt2} (d+f)\,
\overline{K}_c \overline{\bm{\Sigma}}_8 \cdot \bm{\tau} \,\Xi_8^{}
+ \frac{1}{\sqrt2} (d-f)\,
\overline{K} \, \overline{\bm{\Sigma}}_8 \cdot \bm{\tau} N_8^{}
\nonumber \\ && \mbox{}
+ \sqrt{\frac23} d\,
\overline{\Lambda}_8^0 \bm{\Sigma}_8^{} \cdot \bm{\pi}
- \sqrt{\frac23} d\,
\overline{\Lambda}_8^0\, \Lambda_8^{0}\, \eta
- \frac{1}{\sqrt6} (d-3f)\,
\overline{\Lambda}_8^0 \overline{K}_c \Xi_8^{}
\nonumber \\ && \mbox{}
- \frac{1}{\sqrt6} (d+3f)\,
\overline{\Lambda}_8^0 \overline{K} N_8^{}
- \frac{1}{\sqrt2} (d-f) \,
\overline{\Xi}_8 \bm{\tau} \cdot \bm{\pi}\, \Xi_8^{}
-\frac{1}{\sqrt6} (d+3f)\,
 \overline{\Xi}_8 \Xi_8^{} \, \eta
\nonumber \\ && \mbox{}
-\frac{1}{\sqrt2} (d+f)\,
\overline{\Xi}_8 \bm{\tau} \cdot \bm{\Sigma}_8^{}  K_c
- \frac{1}{\sqrt6} (d-3f)\,
\overline{\Xi}_8 K_c \Lambda_8^{0}.
\end{eqnarray}
Note that the values of $d$ and $f$ are different from those in
$\mathcal{L}_{\bm{8}\mbox{-}\bm{8}_3}$ of Eq.~(\ref{Lag:8-8}).

\subsubsection{$\bm{10}$-$\bm{10}$}

The interaction Lagrangian reads
\begin{equation}
\mathcal{L}_{\bm{10}\mbox{-}\bm{10}}
= g_{\bm{10}\mbox{-}\bm{10}}^{} 
\overline{D}^{jkl} D_{mkl} M^m_j,
\end{equation}
which gives
\begin{eqnarray}
\mathcal{L}_{\bm{10}\mbox{-}\bm{10}} / g_{\bm{10}\mbox{-}\bm{10}}^{} &=&
\sqrt{\frac{15}{2}}\, \overline{\Delta}_{10} \bm{T}({\textstyle\frac32,
\frac32}) \cdot \bm{\pi} \, \Delta_{10}
+ \sqrt6\, \overline{\Delta}_{10} \Delta_{10} \, \eta
- \sqrt6
\, \overline{\Delta}_{10}\, \bm{T}({\textstyle\frac32,\frac12}) \cdot
  \bm{\Sigma}_{10}\, K
\nonumber \\ && \mbox{}
-i 2\sqrt{2}\, \left( \overline{\bm{\Sigma}}_{10} \times
\bm{\Sigma}_{10}
\right) \cdot \bm{\pi}
- 2\sqrt2 \,\overline{K}_c \,\overline{\bm{\Sigma}}_{10} \cdot \bm{\tau}
  \,\Xi_{10}
+ \sqrt6 \,\overline{K} \,\bm{T}(\textstyle\frac12,\frac32) \cdot
  \overline{\bm{\Sigma}}_{10}\, \Delta_{10}
\nonumber \\ && \mbox{}
+\sqrt2 \,\overline{\Xi}_{10}\, \bm{\tau} \cdot \bm{\pi}\, \Xi_{10}
- \sqrt6 \,\overline{\Xi}_{10}\, \Xi_{10} \,\eta
- 2\sqrt3\, \overline{\Xi}_{10}\, K\, \Omega_{10}^-
\nonumber \\ && \mbox{}
-2\sqrt2\, \overline{\Xi}_{10}\, \bm{\tau} \cdot \bm{\Sigma}_{10}\, K_c
- 2\sqrt6 \,\overline{\Omega}_{10}^- \, \Omega_{10}^-\, \eta
- 2\sqrt3 \,\overline{\Omega}_{10}^- \,\overline{K}\, \Xi_{10}.
\end{eqnarray}

\subsubsection{$\overline{\bm{10}}$-$\overline{\bm{10}}$}

The interaction Lagrangian reads
\begin{equation}
\mathcal{L}_{\overline{\bm{10}}\mbox{-}\overline{\bm{10}}}
= g_{\overline{\bm{10}}\mbox{-}\overline{\bm{10}}}^{} 
\overline{T}_{jkl} T^{mkl} M_m^j,
\end{equation}
which gives
\begin{eqnarray}
\mathcal{L}_{\overline{\bm{10}}\mbox{-}\overline{\bm{10}}}
/ g_{\overline{\bm{10}}\mbox{-}\overline{\bm{10}}}^{} &=&
-2\sqrt6 \, \overline{\Theta}\, \Theta \eta
- 2\sqrt3 \, \overline{\Theta}\, \overline{K}_c \, N_{\overline{10}}
- 2\sqrt3 \, \overline{N}_{\overline{10}} \, K_c \, \Theta
-\sqrt2\, \overline{N}_{\overline{10}}\, \bm{\tau} \cdot \bm{\pi}
N_{\overline{10}}
\nonumber \\ && \mbox{}
- \sqrt6\, \overline{N}_{\overline{10}}\, N_{\overline{10}}\, \eta
+ 2\sqrt2\, \overline{N}_{\overline{10}}\, \bm{\tau} \cdot
\bm{\Sigma}_{\overline{10}} \, K
+ 2\sqrt2\, \overline{K}\, \overline{\bm{\Sigma}}_{\overline{10}} \cdot
\bm{\tau} N_{\overline{10}}
\nonumber \\ && \mbox{}
+ i 2\sqrt2 \left( \overline{\bm{\Sigma}}_{\overline{10}} \times
\bm{\Sigma}_{\overline{10}} \right) \cdot \bm{\pi}
+ \sqrt6\, \overline{K}_c \overline{\bm{\Sigma}}_{\overline{10}} \cdot
\bm{T}({\textstyle\frac12,\frac32}) \, \Xi_{\overline{10},3/2}
\nonumber \\ && \mbox{}
- \sqrt6\, \overline{\Xi_{\overline{10},3/2}} 
\bm{T}({\textstyle\frac32,\frac12}) \cdot \bm{\Sigma}_{\overline{10}}\,
K_c
+ \sqrt6\, \overline{\Xi}_{\overline{10},3/2} \,
{\Xi}_{\overline{10},3/2}\, \eta
\nonumber \\ && \mbox{}
- \sqrt{\frac{15}{2}}\, \overline{\Xi}_{\overline{10},3/2}
  \bm{T}({\textstyle\frac32,\frac32}) \cdot \bm{\pi}
\Xi_{\overline{10},3/2}.
\end{eqnarray}

\subsubsection{$\bm{27}$-$\bm{27}$}

This interaction contains two types as there are two ways to fully
contract the upper and lower indices.
As in the case of $\mathcal{L}_{\bm{8}\mbox{-}\bm{8}}$ we define $f$- and
$d$-type interactions as
\begin{equation}
\mathcal{L}_{\bm{27}\mbox{-}\bm{27}}
/ g_{\bm{27}\mbox{-}\bm{27}}^{} =
(d+f) \overline{T}^{kl}_{ij} T^{ij}_{km} M^m_l
+ (d-f) \overline{T}^{kl}_{ij} T^{im}_{kl} M^j_m.
\end{equation}
Then we have
\begin{equation}
\mathcal{L}_{\bm{27}\mbox{-}\bm{27}}
/ g_{\bm{27}\mbox{-}\bm{27}}^{} =
\mathcal{L}_{|\Delta S| = 0}^{\bm{27}\mbox{-}\bm{27}} +
\mathcal{L}_{|\Delta S|=1}^{\bm{27}\mbox{-}\bm{27}},
\end{equation}
where
\begin{eqnarray}
\mathcal{L}_{|\Delta S| = 0}^{\bm{27}\mbox{-}\bm{27}} &=&
- \frac{2\sqrt6}{3} (d-3f)\, \overline{\Theta}_1 \, {\Theta}_1
  \, \eta
+ i 2\sqrt2 (d+f)\, \left( \overline{\bm{\Theta}}_1 \times \bm{\Theta}_1
\right) \cdot \bm{\pi}
+ \frac{\sqrt6}{3}(d+3f)\, \overline{\Delta}_{27} \Delta_{27} \, \eta
\nonumber \\ && \mbox{}
+ \sqrt{\frac56}(d+3f)\, \overline{\Delta}_{27}
\bm{T}({\textstyle\frac32,\frac32}) \cdot \bm{\pi}\, \Delta_{27}
+ \frac{8}{\sqrt{30}} d\, \overline{\Delta}_{27}
\bm{T}({\textstyle\frac32,\frac12}) \cdot \bm{\pi}\, N_{27}
- \frac{8}{\sqrt{30}}d\, \overline{N}_{27}
  \bm{T}({\textstyle\frac12,\frac32}) \cdot \bm{\pi}\, \Delta_{27}
\nonumber \\ && \mbox{}
- \frac{\sqrt6}{15} (13d - 15f)\, \overline{N}_{27} N_{27}\, \eta
+ \frac{\sqrt2}{15} (19d + 15f) \overline{N}_{27} \bm{\tau} \cdot \bm{\pi}
N_{27}
+ \frac{4\sqrt6}{3} d\, \overline{\Sigma}_{27,2} \Sigma_{27,2} \, \eta
\nonumber \\ && \mbox{}
+ \frac{4\sqrt3}{\sqrt5} f\, \overline{\Sigma}_{27,2} \bm{T}(2,2) \cdot
\bm{\pi} \, \Sigma_{27,2}
+ \frac{4}{\sqrt{15}} d\, \overline{\Sigma}_{27,2} \bm{T}(2,1) \cdot
\bm{\pi} \, \Sigma_{27}^c
- \frac{4}{\sqrt{15}}d\, \overline{\Sigma}_{27}^c \bm{T}(1,2) \cdot
  \bm{\pi} \, \Sigma_{27,2}
\nonumber \\ && \mbox{}
- \frac{4\sqrt6}{15} d\, \overline{\Sigma}_{27} \, {\Sigma}_{27}\, \eta
-i 2\sqrt2 f\, \left( \overline{\bm{\Sigma}}_{27} \times \bm{\Sigma}_{27}
\right) \cdot \bm{\pi}
+ \frac{16\sqrt6}{15} d \, \overline{\bm{\Sigma}}_{27} \cdot \bm{\pi} \,
\Lambda_{27}^0
+ \frac{16\sqrt6}{15}d\, \overline{\Lambda}_{27}^0 \bm{\pi} \cdot
\bm{\Sigma}_{27}
\nonumber \\ && \mbox{}
- \frac{16\sqrt6}{15}d\, \overline{\Lambda}_{27}^0 \Lambda_{27}^0 \, \eta
+ \sqrt{\frac23} (d-3f)\, \overline{\Xi}_{27,3/2}\, \Xi_{27,3/2}\, \eta
- \sqrt{\frac56} (d-3f)\, \overline{\Xi}_{27,3/2}\,
  \bm{T}({\textstyle\frac32,\frac32}) \cdot \bm{\pi}\, \Xi_{27,3/2}
\nonumber \\ && \mbox{}
+ \frac{8}{\sqrt{30}} d\, \overline{\Xi}_{27,3/2}
\bm{T}({\textstyle\frac32,\frac12}) \cdot \bm{\pi}\, \Xi_{27}
-\frac{8}{\sqrt{30}}d\, \overline{\Xi}_{27}
\bm{T}({\textstyle\frac12,\frac32}) \cdot \bm{\pi}\, \Xi_{27,3/2}
- \frac{\sqrt2}{15} (19d-15f) \,\overline{\Xi}_{27}\, \bm{\tau} \cdot \bm{\pi}
  \, \Xi_{27}
\nonumber \\ && \mbox{}
- \frac{\sqrt6}{15} (13d+15f)\, \overline{\Xi}_{27}\, \Xi_{27}\, \eta
-\frac{2\sqrt2}{\sqrt3} (d+3f)\, \overline{\Omega}_{27,1} \,
{\Omega}_{27,1} \, \eta
+ i 2\sqrt2 (d-f) \left( \overline{\bm{\Omega}}_{27,1} \times
\bm{\Omega}_{27,1} \right) \cdot \bm{\pi},
\nonumber \\ && \mbox{}
\end{eqnarray}
and
\begin{eqnarray}
\mathcal{L}_{|\Delta S| = 1}^{\bm{27}\mbox{-}\bm{27}} &=&
2(d-f)\, \overline{K}_c\, \overline{\bm{\Theta}}_1 \cdot
\bm{T}({\textstyle\frac12,\frac32})\, \Delta_{27}
-2(d-f)\, \overline{\Delta}_{27} \bm{T}({\textstyle\frac32,\frac12}) \cdot
\bm{\Theta}_1\, K_c
- \frac{2}{\sqrt{15}} (d+5f)\, \overline{K}_c \, \overline{\bm{\Theta}}_1
  \cdot \bm{\tau} \, N_{27}
\nonumber \\ && \mbox{}
- \frac{2}{\sqrt{15}} (d+5f) \overline{N}_{27} \bm{\tau} \cdot
  \bm{\Theta}_1\, K_c
-\sqrt3 (d-f)\, \overline{\Delta}_{27} \overline{\bm{K}}_c \cdot
\bm{I}({\textstyle\frac32,2}) \, \Sigma_{27,2}
-\sqrt3 (d-f) \overline{\Sigma}_{27,2} \bm{I}({\textstyle2,\frac32}) \cdot
\bm{K}_c\, \Delta_{27}
\nonumber \\ && \mbox{}
- \frac{1}{\sqrt5}(3d+5f) \overline{\Delta}_{27}
  \bm{T}({\textstyle\frac32,\frac12}) \cdot \bm{\Sigma}_{27}\, K
+ \frac{1}{\sqrt5}(3d+5f)\, \overline{K}\, \overline{\bm{\Sigma}}_{27} \cdot
\bm{T}({\textstyle\frac12,\frac32})\, \Delta_{27}
- \frac45 (d+5f) \overline{N}_{27} K \Lambda_{27}^0
\nonumber \\ && \mbox{}
- \frac45 (d+5f)\, \overline{\Lambda}_{27}^0 \overline{K}\, N
+ \frac{4\sqrt3}{15} (3d-5f) \overline{N}_{27} \bm{\tau} \cdot
\bm{\Sigma}_{27}\, K
+ \frac{4\sqrt3}{15} (3d-5f)\, \overline{K} \,\overline{\bm{\Sigma}}_{27}
\cdot \bm{\tau}\, N_{27}
\nonumber \\ && \mbox{}
-\sqrt3 (d+f)\, \overline{\Sigma}_{27,2} \bm{I}({\textstyle2,\frac32})
\cdot \bm{K} \, \Xi_{27,3/2}
- \sqrt3 (d+f)\, \overline{\Xi}_{27,3/2} \,\overline{\bm{K}} \cdot
  \bm{I}({\textstyle\frac32,2})\, \Sigma_{27,2}
\nonumber \\ && \mbox{}
+ \frac{1}{\sqrt5} (3d-5f)\, \overline{K}_c \overline{\bm{\Sigma}}_{27}
\cdot \bm{T} ({\textstyle\frac12,\frac32}) \, \Xi_{27,3/2}
- \frac{1}{\sqrt5} (3d-5f)\, \overline{\Xi}_{27,3/2}\,
  \bm{T}({\textstyle\frac32,\frac12}) \cdot \bm{\Sigma}_{27}\, K_c
\nonumber \\ && \mbox{}
- \frac{4\sqrt3}{15} (3d+5f)\, \overline{K}_c \overline{\bm{\Sigma}}_{27}
  \cdot \bm{\tau} \, \Xi_{27}
- \frac{4\sqrt3}{15} (3d+5f)\, \overline{\Xi}_{27} \,\bm{\tau} \cdot
  \bm{\Sigma}_{27}\, K_c
- \frac45 (d-5f)\, \overline{\Lambda}_{27}^0 \overline{K}_c \, \Xi_{27}
\nonumber \\ && \mbox{}
- \frac45 (d-5f) \overline{\Xi}_{27}\, K_c \,\Lambda_{27}^0
-2 (d+f)\, \overline{\Xi}_{27,3/2}\, \bm{T}({\textstyle\frac32,\frac12})
\cdot \bm{\Omega}_{27,1}\, K
+ 2(d+f)\, \overline{K}\, \overline{\bm{\Omega}}_{27,1} \cdot
\bm{T}({\textstyle\frac12,\frac32}) \, \Xi_{27,3/2}
\nonumber \\ && \mbox{}
+ \frac{2}{\sqrt{15}} (d-5f)\, \overline{\Xi}_{27}\, \bm{\tau} \cdot
\bm{\Omega}_{27,1}\, K
+ \frac{2}{\sqrt{15}} (d-5f)\, \overline{K}\,
\overline{\bm{\Omega}}_{27,1} \cdot \bm{\tau}\, \Xi_{27}.
\end{eqnarray}
Here,
\begin{eqnarray}
\bm{I}({\textstyle\frac32,2}) \cdot \overline{\bm{K}}_c &=&
 K^0 I^{1/2}_{3/2,2} - K^+ I^{-1/2}_{3/2,2} ,
\nonumber \\
\bm{T}(2,2) \cdot \bm{\pi}
 &=& - T^{(+1)}_{2,2} \pi^+ +
T^{(-1)}_{2,2} \pi^- + T^{(0)}_{2,2} \pi^0,
\end{eqnarray}
with
\begin{eqnarray}
I^{1/2}_{3/2,2} = \sqrt{\frac23} \left( \begin{array}{ccccc}
2 & 0 & 0 & 0 & 0 \\
0 & \sqrt3 & 0 & 0 & 0 \\
0 & 0 & \sqrt2 & 0 & 0 \\
0 & 0 & 0 & 1 & 0 \end{array} \right), \qquad
I^{-1/2}_{3/2,2} = \sqrt{\frac23} \left( \begin{array}{ccccc}
0 & 1 & 0 & 0 & 0 \\
0 & 0 & \sqrt2 & 0 & 0 \\
0 & 0 & 0 & \sqrt3 & 0 \\
0 & 0 & 0 & 0 & 2 \end{array} \right),
\end{eqnarray}
and
\begin{eqnarray}
&&
T^{+1}_{2,2} = - \sqrt{\frac56} \left( \begin{array}{ccccc}
0 & \sqrt2 & 0 & 0 & 0 \\
0 & 0 & \sqrt3 & 0 & 0 \\
0 & 0 & 0 & \sqrt3 & 0 \\
0 & 0 & 0 & 0 & \sqrt2 \\
0 & 0 & 0 & 0 & 0 \end{array} \right), \qquad
T^{0}_{2,2} = \sqrt{\frac56} \left( \begin{array}{ccccc}
2 & 0 & 0 & 0 & 0 \\
0 & 1 & 0 & 0 & 0 \\
0 & 0 & 0 & 0 & 0 \\
0 & 0 & 0 & -1 & 0 \\
0 & 0 & 0 & 0 & -2 \end{array} \right), \qquad
\nonumber \\ &&
T^{-1}_{2,2} = \sqrt{\frac56} \left( \begin{array}{ccccc}
0 & 0 & 0 & 0 & 0 \\
\sqrt2 & 0 & 0 & 0 & 0 \\
0 & \sqrt3 & 0 & 0 & 0 \\
0 & 0 & \sqrt3 & 0 & 0 \\
0 & 0 & 0 & \sqrt2 & 0 \end{array} \right).
\end{eqnarray}
Therefore, we have
\begin{eqnarray}
\overline{\Delta}_{27}\,
\bm{I}({\textstyle\frac32,2}) \cdot \overline{\bm{K}}_c\,
\Sigma_{27,2} &=&
\sqrt{\frac23} \, \overline{\Delta}^{++} \left( 2 \Sigma_{27,2}^{++} K^0
- \Sigma_{27,2}^+ K^+ \right)
+ \sqrt{\frac23} \, \overline{\Delta}^{+} \left( \sqrt3 \Sigma_{27,2}^{+} K^0
- \sqrt2 \Sigma_{27,2}^0 K^+ \right)
\nonumber \\ && \mbox{}
+ \sqrt{\frac23} \, \overline{\Delta}^{0} \left( \sqrt2 \Sigma_{27,2}^{0} K^0
- \sqrt3 \Sigma_{27,2}^- K^+ \right)
+ \sqrt{\frac23} \, \overline{\Delta}^{-} \left( \Sigma_{27,2}^{-} K^0
- 2 \Sigma_{27,2}^{--} K^+ \right),
\nonumber \\
\overline{\Sigma}_{27,2} \bm{T}(2,2) \cdot
\bm{\pi}\, \Sigma_{27,2} &=&
\sqrt{\frac56} \overline{\Sigma}_{27,2}^{++} \left( \sqrt2
\Sigma_{27,2}^{+} \pi^+ + 2 \Sigma_{27,2}^{++} \pi^0 \right)
+ \sqrt{\frac56} \overline{\Sigma}_{27,2}^{+} \left( \sqrt3
\Sigma_{27,2}^{0} \pi^+ + \Sigma_{27,2}^{+} \pi^0 + \sqrt2
\Sigma_{27,2}^{++} \pi^- \right)
\nonumber \\ && \mbox{}
+ \sqrt{\frac56} \overline{\Sigma}_{27,2}^{0} \left( \sqrt3
\Sigma_{27,2}^{-} \pi^+ + \sqrt3 \Sigma_{27,2}^{+} \pi^- \right)
+ \sqrt{\frac56} \overline{\Sigma}_{27,2}^{+} \left( \sqrt2
\Sigma_{27,2}^{--} \pi^+ - \Sigma_{27,2}^{+} \pi^0 + \sqrt3
\Sigma_{27,2}^{0} \pi^- \right)
\nonumber \\ && \mbox{}
-\sqrt{\frac56} \overline{\Sigma}_{27,2}^{--} \left( 2
\Sigma_{27,2}^{--} \pi^0 - \sqrt2 \Sigma_{27,2}^{-} \pi^- \right).
\end{eqnarray}

\subsubsection{$\bm{35}$-$\bm{35}$}

This interaction Lagrangian also contains $f$- and $d$-type and reads
\begin{eqnarray}
\mathcal{L}_{\bm{35}\mbox{-}\bm{35}}
/ g_{\bm{35}\mbox{-}\bm{35}}^{} &=&
(d+f) \overline{T}_a^{ijkl} T^a_{ijkm} M^m_l
+ (d-f) \overline{T}_a^{ijkl} T^m_{ijkl} M^a_m \nonumber \\ &=&
\mathcal{L}_{|\Delta S| = 0}^{\bm{35}\mbox{-}\bm{35}} +
\mathcal{L}_{|\Delta S|=1}^{\bm{35}\mbox{-}\bm{35}},
\end{eqnarray}
where
\begin{eqnarray}
\mathcal{L}_{|\Delta S| = 0}^{\bm{35}\mbox{-}\bm{35}} &=&
-4\sqrt6 (d-3f) \, \overline{\Theta}_2 \, \Theta_2\, \eta
+ \frac{6\sqrt{12}}{\sqrt5} (d+f)\, \overline{\Theta}_2\, \bm{T}(2,2)
\cdot \bm{\pi} \,\Theta_2
+ 8\sqrt6\, d\, \overline{\Delta}_{5/2} \, \Delta_{5/2} \, \eta
\nonumber \\ && \mbox{}
+ \frac{24\sqrt{7}}{\sqrt{15}}\, f\, \overline{\Delta}_{5/2}
\bm{T}({\textstyle\frac52,\frac52}) \cdot \bm{\pi}\, \Delta_{5/2}
+ \frac{5\sqrt{30}}{9} (35d + 27f)\, \overline{\Delta}_{35}
\bm{T}({\textstyle\frac32,\frac32}) \cdot \bm{\pi}\, \Delta_{35}
\nonumber \\ && \mbox{}
+ \sqrt{\frac65} (5d-3f)\, \overline{\Delta}_{5/2}
\bm{T}({\textstyle\frac52,\frac32}) \cdot \bm{\pi} \, \Delta_{35}
- \sqrt{\frac65} (5d-3f)\, \overline{\Delta}_{35}
  \bm{T}({\textstyle\frac32,\frac52}) \cdot \bm{\pi} \Delta_{5/2}
\nonumber \\ && \mbox{}
- \frac{3\sqrt6}{2} (3d-5f) \overline{\Delta}_{35}\, \Delta_{35}\, \eta
+ \sqrt6 (5d-3f)\, \overline{\Sigma}_{35,2}\, \Sigma_{35,2}\, \eta
\nonumber \\ && \mbox{}
- \frac{3\sqrt3}{\sqrt5} (d+7f)\, \overline{\Sigma}_{35,2}
  \bm{T}({\textstyle2,2}) \cdot \bm{\pi}\, \Sigma_{35,2}
+ \sqrt3 (5d-3f) \overline{\Sigma}_{35,2} \bm{T}(2,1) \cdot \bm{\pi}
\Sigma_{35}^c
\nonumber \\ && \mbox{}
- \sqrt3 (5d-3f)\, \overline{\Sigma}_{35}^c \bm{T}(2,1) \cdot \bm{\pi}\,
  \Sigma_{35,2}
- \frac{i}{\sqrt2} (17d+9f)\, \left( \overline{\bm{\Sigma}}_{35} \times 
\bm{\Sigma}_{35} \right) \cdot \bm{\pi}
\nonumber \\ && \mbox{}
-\sqrt6 (5d-3f)\, \overline{\Sigma}_{35}\, \Sigma_{35}\, \eta
+ 2\sqrt6 (d-3f)\, \overline{\Xi}_{35,3/2}\, \Xi_{35,3/2}\, \eta
\nonumber \\ && \mbox{}
- \sqrt{30} (d-3f)\, \overline{\Xi}_{35,3/2}
  \bm{T}({\textstyle\frac32,\frac32}) \cdot \bm{\pi} \,\Xi_{35,3/2}
+ \sqrt3 (5d-3f)\, \overline{\Xi}_{35,3/2}
\bm{T}({\textstyle\frac32,\frac12}) \cdot \bm{\pi}\, \Xi_{35}
\nonumber \\ && \mbox{}
-\sqrt3 (5d-3f) \overline{\Xi}_{35} \bm{T}({\textstyle\frac32,\frac12})
\cdot \bm{\pi}\, \Xi_{35,3/2}
+ \frac{1}{\sqrt2} (11d+3f)\, \overline{\Xi}_{35} \bm{\tau} \cdot
\bm{\pi} \Xi_{35}
\nonumber \\ && \mbox{}
- \frac{\sqrt6}{2} (11d+3f)\, \overline{\Xi}_{35}\, \Xi_{35}\, \eta
+ i 3\sqrt2 (3d-5f)\, \left( \overline{\bm{\Omega}}_{35,1} \times
\bm{\Omega}_{35,1} \right) \cdot \bm{\pi}
\nonumber \\ && \mbox{}
- \sqrt6 (d+9f)\, \Omega_{35,1}\, \Omega_{35,1} \, \eta
+ 2\sqrt3 (5d-3f) \,  \overline{\bm{\Omega}}_{35,1} \cdot \bm{\pi}\,
\Omega^-_{35}
+ 2\sqrt3 (5d-3f)\, \overline{\Omega}_{35}^- \bm{\pi} \cdot
\bm{\Omega}_{35,1}
\nonumber \\ && \mbox{}
- 6\sqrt6 (d+f)\, \overline{\Omega}_{35}^-\, \Omega_{35}^- \, \eta
- 4\sqrt6 (d+3f)\, \overline{X}\, X\, \eta
- 12\sqrt2 (d-f)\, \overline{X} \, \bm{\tau} \cdot \bm{\pi}\, X.
\end{eqnarray}
and
\begin{eqnarray}
\mathcal{L}_{|\Delta S| = 1}^{\bm{35}\mbox{-}\bm{35}} &=&
- \frac{24\sqrt3}{\sqrt{10}} (d-f)\, \overline{\Theta}_2\,
  \bm{I}({\textstyle2,\frac52}) \cdot \bm{K}\, \Delta_{5/2}
+ \frac{24\sqrt3}{\sqrt{10}} (d-f)\, \overline{\Delta}_{25}\,
\overline{\bm{K}} \cdot \bm{I}({\textstyle\frac52,2})\, \Theta_2
\nonumber \\ && \mbox{}
- \frac{3}{\sqrt5} (d+9f)\, \overline{\Theta}_2\,
  \bm{I}({\textstyle2,\frac32}) \cdot \bm{K} \, \Delta_{35}
- \frac{3}{\sqrt5}(d+9f)\, \overline{\Delta}_{35} \overline{\bm{K}} \cdot
  \bm{I}({\textstyle\frac32,2})\, \Theta_2
\nonumber \\ && \mbox{}
- \frac{12\sqrt3}{\sqrt{10}} (d+f)\, \overline{\Delta}_{5/2}
  \overline{\bm{K}}_c \cdot \bm{I}({\textstyle\frac52,2})\, \Sigma_{27,2}
+ \frac{12\sqrt3}{\sqrt{10}} (d+f)\, \overline{\Sigma}_{35,2}\,
\bm{I}({\textstyle2,\frac52}) \cdot \bm{K}_c\, \Delta_{5/2}
\nonumber \\ && \mbox{}
- \frac{3}{\sqrt{20}} (19d - 21f)\, \overline{\Delta}_{35}
  \overline{\bm{K}}_c \cdot \bm{I}({\textstyle\frac32,2}) \, \Sigma_{35,2}
- \frac{3}{\sqrt{20}} (19d - 21f) \overline{\Sigma}_{35,2}
  \bm{I}({\textstyle2,\frac32}) \cdot \bm{K}_c \,\Delta_{35}
\nonumber \\ && \mbox{}
- \frac{\sqrt{15}}{2} (d+9f)\, \overline{\Delta}_{35}
  \bm{T}({\textstyle\frac32,\frac12}) \cdot \bm{\Sigma}\, K
+ \frac{\sqrt{15}}{2} (d+9f)\, \overline{K} \,\overline{\bm{\Sigma}}_{35}
\cdot \bm{T}({\textstyle\frac12,\frac32})\, \Delta_{35}
\nonumber \\ && \mbox{}
-9 (d+f)\, \overline{\Sigma}_{35,2} \bm{I}({\textstyle2,\frac32}) \cdot
\bm{K}\, \Xi_{35,3/2}
-9 (d+f) \overline{\Xi}_{35,3/2} \overline{\bm{K}} \cdot
\bm{I}({\textstyle\frac32,2}) \, \Sigma_{35,2}
\nonumber \\ && \mbox{}
-2 (d+9f)\, \overline{K}_c \overline{\bm{\Sigma}}_{35} \cdot \bm{\tau} \,
\Xi_{35}
-2(d+9f)\, \overline{\Xi}_{35} \bm{\tau} \cdot \bm{\Sigma}_{35} \, K_c
\nonumber \\ && \mbox{}
+ \sqrt3 (7d-9f)\, \overline{K}_c \overline{\bm{\Sigma}}_{35} \cdot
\bm{T}({\textstyle\frac12,\frac32})\, \Xi_{35,3/2}
-\sqrt3 (7d-9f)\, \overline{\Xi}_{35,3/2}
\bm{T}({\textstyle\frac32,\frac12}) \cdot \bm{\Sigma}_{35}\, K_c
\nonumber \\ && \mbox{}
- 6\sqrt3 (d+f)\, \overline{\Xi}_{35,3/2}
  \bm{T}({\textstyle\frac32,\frac12}) \cdot \bm{\Omega}_{35,1}\, K
+ 6 \sqrt3 (d+f)\, \overline{K}\, \overline{\bm{\Omega}}_{35,1} \cdot
\bm{T}({\textstyle\frac12,\frac32})\, \Xi_{35,3/2}
\nonumber \\ && \mbox{}
+ 3(3d+5f)\, \overline{\Xi}_{35} \bm{\tau} \cdot \bm{\Omega}_{35,1}, K
+ 3 (3d+5f)\, \overline{K}\, \overline{\bm{\Omega}}_{35,1} \cdot
\bm{\tau}\, \Xi_{35}
\nonumber \\ && \mbox{}
- \sqrt6 (d+9f)\, \overline{\Xi}_{35} \, K \, \Omega^-_{35}
- \sqrt6 (d+9f)\, \overline{\Omega}_{35}^- \, \overline{K}\, \Xi_{35}
- 6\sqrt2 (d+f)\, \overline{K}_c \, \overline{\bm{\Omega}}_{35,1} \cdot
\bm{\tau}\, X
\nonumber \\ && \mbox{}
- 6\sqrt2 (d+f)\, \overline{X} \bm{\tau} \cdot \bm{\Omega}_{35,1}\, K_c
- 4\sqrt3 (d-3f) \overline{\Omega}_{35}^-\, \overline{K}_c\, X
- 4\sqrt3 (d-3f)\, \overline{X} \, K_c\, \Omega_{35}^-.
\end{eqnarray}
Here,
\begin{equation}
\bm{T}(\textstyle\frac52,\frac52) \cdot \bm{\pi}
 = - T^{(+1)}_{5/2,5/2} \pi^+ +
T^{(-1)}_{5/2,5/2} \pi^- + T^{(0)}_{5/2,5/2} \pi^0,
\end{equation}
with
\begin{eqnarray}
&&
T^{+1}_{5/2,5/2} = - \frac{2\sqrt3}{\sqrt{35}} \left(
\begin{array}{cccccc}
0 & \sqrt5 & 0 & 0 & 0 & 0 \\
0 & 0 & 2\sqrt2 & 0 & 0 & 0 \\
0 & 0 & 0 & 3 & 0 & 0 \\
0 & 0 & 0 & 0 & 2\sqrt2 & 0 \\
0 & 0 & 0 & 0 & 0 & \sqrt5 \\
0 & 0 & 0 & 0 & 0 & 0 \end{array} \right), 
\qquad
T^{0}_{5/2,5/2} = \sqrt{\frac{6}{35}} \left(
\begin{array}{cccccc}
5 & 0 & 0 & 0 & 0 & 0 \\
0 & 3 & 0 & 0 & 0 & 0 \\
0 & 0 & 1 & 0 & 0 & 0 \\
0 & 0 & 0 & -1 & 0 & 0 \\
0 & 0 & 0 & 0 & -3 & 0 \\
0 & 0 & 0 & 0 & 0 & -5 \end{array} \right),
\nonumber \\ &&
T^{-1}_{5/2,5/2} = \frac{2\sqrt3}{\sqrt{35}} \left(
\begin{array}{cccccc}
0 & 0 & 0 & 0 & 0 & 0 \\
\sqrt5 & 0 & 0 & 0 & 0 & 0 \\
0 & 2\sqrt2 & 0 & 0 & 0 & 0 \\
0 & 0 & 3 & 0 & 0 & 0 \\
0 & 0 & 0 & 2\sqrt2 & 0 & 0 \\
0 & 0 & 0 & 0 & \sqrt5 & 0 
\end{array} \right), \qquad
\end{eqnarray}
which leads to
\begin{eqnarray}
\overline{\Delta}_{5/2} \bm{T}(\textstyle\frac52,\frac52) \cdot
\bm{\pi}\, \Delta_{5/2}
&=& \sqrt{\frac{6}{35}} \Biggl\{
\overline{\Delta}_{5/2}^{+++} \left( 5 \Delta_{5/2}^{+++} \pi^0
+\sqrt{10} \Delta^{++}_{5/2} \pi^+ + \right)
+ \overline{\Delta}_{5/2}^{++} \left( \sqrt{10} \Delta^{+++}_{5/2} \pi^- +
3 \Delta_{5/2}^{++} \pi^0 + 4 \Delta^+ \pi^+ \right)
\nonumber \\ && \quad \mbox{}
+ \overline{\Delta}_{5/2}^{+} \left( 4 \Delta^{++}_{5/2} \pi^- +
 \Delta_{5/2}^{+} \pi^0 + 3\sqrt2 \Delta^0 \pi^+ \right)
+ \overline{\Delta}_{5/2}^{0} \left( 3\sqrt2 \Delta^{+}_{5/2} \pi^- -
 \Delta_{5/2}^{0} \pi^0 + 4 \Delta^- \pi^+ \right)
\nonumber \\ && \quad \mbox{}
+ \overline{\Delta}_{5/2}^{-} \left( 4 \Delta^{0}_{5/2} \pi^- -
3 \Delta_{5/2}^{-} \pi^0 + \sqrt10 \Delta^{--} \pi^+ \right)
+ \overline{\Delta}_{5/2}^{--} \left( \sqrt{10} \Delta^{-}_{5/2} \pi^- -
5 \Delta_{5/2}^{--} \pi^0 \right) \Biggr\}.
\end{eqnarray}

\subsubsection{$\bm{27}$-$\overline{\bm{10}}$}

The interaction Lagrangian of $\bm{27}$ and $\overline{\bm{10}}$ reads
\begin{equation}
\mathcal{L}_{\bm{27}\mbox{-}\overline{\bm{10}}}
/ g_{\bm{27}\mbox{-}\overline{\bm{10}}}^{} =
\epsilon_{imn} \overline{T}^{ij}_{kl} T^{mkl} M^n_j
+ \mbox{(H.c.)},
\end{equation}
which gives
\begin{eqnarray}
\mathcal{L}_{\bm{27}\mbox{-}\overline{\bm{10}}}
/ g_{\bm{27}\mbox{-}\overline{\bm{10}}}^{} &=&
2\sqrt6\, \overline{\bm{\Theta}}_1 \cdot \bm{\pi} \Theta
-2 \, \overline{K}_c \, \overline{\bm{\Theta}}_1 \cdot \bm{\tau}\,
N_{\overline{10}}
+ 2\sqrt2\, \overline{\Delta}_{27} \, \bm{T}({\textstyle\frac32,\frac12})
\cdot \bm{\pi} \, N_{\overline{10}}
\nonumber \\ && \mbox{}
- 2\sqrt2\, \overline{\Delta}_{27} \, \bm{T}({\textstyle\frac32,\frac12})
\cdot \bm{\Sigma}_{\overline{10}}\, K
- \frac{7\sqrt2}{\sqrt{15}} \, \overline{N}_{27} \, \bm{\tau} \cdot
\bm{\pi}\, N_{\overline{10}}
- \frac{3\sqrt{10}}{5} \overline{N}_{27}\, N_{\overline{10}}\, \eta
\nonumber \\ && \mbox{}
+ \frac{6\sqrt5}{5}\, \overline{N}_{27}\, K_c\, \Theta 
- \frac{2\sqrt2}{\sqrt{15}} \overline{N}_{27} \, \bm{\tau} \cdot
\bm{\Sigma}_{\overline{10}}\, K
+ \frac{2\sqrt2}{\sqrt3} \overline{\Sigma}_{27,2} \, \bm{T}(2,1) \cdot
\bm{\pi}\, \Sigma_{\overline{10}}^c
\nonumber \\ && \mbox{}
-3\, \overline{\Sigma}_{27,2}\, \bm{I}({\textstyle2,\frac32}) \cdot
\bm{K}\, \Xi_{\overline{10},3/2}
+ i \frac{6}{\sqrt5} \left( \overline{\bm{\Sigma}}_{27} \times
\bm{\Sigma}_{\overline{10}} \right) \cdot \bm{\pi}
- \frac{4\sqrt3}{\sqrt5} \, \overline{\Sigma}_{27}\,
\Sigma_{\overline{10}} \, \eta
\nonumber \\ && \mbox{}
- \frac{4}{\sqrt5}\, \overline{K}\, \overline{\bm{\Sigma}}_{27} \cdot
\bm{\tau}\, N_{\overline{10}}
- \sqrt{\frac35} \, \overline{K}_c \overline{\bm{\Sigma}}_{27} \cdot
\bm{T}({\textstyle\frac12,\frac32}) \, \Xi_{\overline{10},3/2}
- \frac{8}{\sqrt{15}} \, \overline{\Lambda}_{27}^0 \bm{\pi} \cdot
\bm{\Sigma}_{\overline{10}}
\nonumber \\ && \mbox{}
+ \frac{4\sqrt3}{\sqrt5}\, \overline{\Lambda}_{27}^0\, \overline{K}\,
N_{\overline{10}}
- \sqrt{\frac52}\, \overline{\Xi}_{27,3/2}\,
\bm{T}({\textstyle\frac32,\frac32}) \cdot \bm{\pi} \,
\Xi_{\overline{10},3/2}
- 3\sqrt2\, \overline{\Xi}_{27,3/2} \, \Xi_{\overline{10},3/2}\, \eta
\nonumber \\ && \mbox{}
+ \sqrt2 \, \overline{\Xi}_{27,3/2} \,
\bm{T}({\textstyle\frac32,\frac12}) \cdot \bm{\Sigma}_{\overline{10}}\,
K_c
+ \frac{2\sqrt6}{\sqrt{15}}\, \overline{\Xi}_{27}\,
\bm{T}({\textstyle\frac12,\frac32}) \cdot \bm{\pi}\,
\Xi_{\overline{10},3/2}
\nonumber \\ && \mbox{}
+ \frac{8\sqrt2}{\sqrt{15}} \, \overline{\Xi}_{27}\, \bm{\tau} \cdot
\bm{\Sigma}_{\overline{10}} \, K_c
- 2\sqrt3\, \overline{K}\, \overline{\bm{\Omega}}_{27,1} \cdot
\bm{T}({\textstyle\frac12,\frac32})\, \Xi_{\overline{10},3/2}
\nonumber \\ && \mbox{}
+ \mbox{(H.c.)}.
\end{eqnarray}

\subsubsection{$\bm{35}$-$\bm{27}$}

The interaction Lagrangian in this case is obtained as
\begin{equation}
\mathcal{L}_{\bm{35}\mbox{-}\bm{27}}
/ g_{\bm{35}\mbox{-}\bm{27}}^{} =
\epsilon_{imn} \overline{T}_{a}^{ijkl} T^{ma}_{jk} M^n_l
+ \mbox{(H.c.)},
\end{equation}
which gives
\begin{eqnarray}
\mathcal{L}_{\bm{35}\mbox{-}\bm{27}}
/ g_{\bm{35}\mbox{-}\bm{27}}^{} &=&
4\sqrt2\, \overline{\Theta}_2\, \bm{T}(2,1) \cdot \bm{\pi}\, \Theta_1^c
- 3\sqrt2 \, \overline{\Theta}_2\, \bm{I}({\textstyle2,\frac32}) \cdot
\bm{K}\, \Delta_{27}
\nonumber \\ && \mbox{}
- \frac{12}{\sqrt5}\, \overline{\Delta}_{5/2}\, \overline{\bm{K}}_c \cdot
\bm{I}({\textstyle\frac52,2})\, \Sigma_{27,2}
+ 2\sqrt3\, \overline{\Delta}_{5/2}\,
\bm{T}({\textstyle\frac52,\frac32}) \cdot \bm{\pi}\, \Delta_{27}
\nonumber \\ && \mbox{}
+ \sqrt{10}\, \overline{\Delta}_{35}\,
\bm{T}({\textstyle\frac32,\frac12}) \cdot \bm{\Theta}_1\, K_c
+ \sqrt{\frac{3}{10}} \, \overline{\Delta}_{35}\, \overline{\bm{K}}_c \cdot
\bm{I}({\textstyle\frac32,2})\, \Sigma_{27,2}
\nonumber \\ && \mbox{}
- \frac{5}{\sqrt2}\, \overline{\Delta}_{35}\,
\bm{T}({\textstyle\frac32,\frac12}) \cdot \bm{\Sigma}_{27}\, K
- \frac{7}{\sqrt{12}} \, \overline{\Delta}_{35} \,
\bm{T}({\textstyle\frac32,\frac32}) \cdot \bm{\pi}\, \Delta_{27}
\nonumber \\ && \mbox{}
+ \frac{10}{\sqrt3}\, \overline{\Delta}_{35}\,
\bm{T}({\textstyle\frac32,\frac12}) \cdot \bm{\pi}\, N_{27}
- 6 \, \overline{\Sigma}_{35,2} \, \Sigma_{27,2}\, \eta
- \frac{3\sqrt2}{\sqrt5}\, \overline{\Sigma}_{35,2}\, \bm{T}(2,2) \cdot
\bm{\pi}\, \Sigma_{27,2}
\nonumber \\ && \mbox{}
+ \sqrt{10}\, \overline{\Sigma}_{35,2}\, \bm{T}(2,1) \cdot \bm{\pi}\,
\Sigma_{27}^c
- 3\sqrt2\, \overline{\Sigma}_{35,2} \, \bm{I}({\textstyle2,\frac32})
\cdot \bm{K} \, \Xi_{27,3/2}
\nonumber \\ && \mbox{}
+ \frac{3}{\sqrt2} \overline{\Sigma}_{35,2} \, \bm{I}({\textstyle2,\frac32})
\cdot \bm{K}_c \, \Delta_{27}
- \frac{2\sqrt{10}}{3}\, \overline{K}_c \, \overline{\bm{\Sigma}}_{35}
\cdot \bm{\tau}\, \Xi_{27}
\nonumber \\ && \mbox{}
- 2\sqrt5\, \overline{\Sigma}_{35}\, \Sigma_{27}\, \eta
- \frac{5}{\sqrt6} \, \overline{K} \, \overline{\bm{\Sigma}}_{35}
\cdot \bm{T}({\textstyle\frac12,\frac32}) \, \Delta_{27}
- \sqrt{\frac23}\, \overline{K}_c\, \overline{\bm{\Sigma}}_{35} \cdot
\bm{T}({\textstyle\frac12,\frac32}) \, \Xi_{27,3/2}
\nonumber \\ && \mbox{}
+ i \sqrt{15}\, \left( \overline{\bm{\Sigma}}_{35} \times
\bm{\Sigma}_{27} \right) \cdot \bm{\pi}
+ \frac{8\sqrt5}{3} \overline{\bm{\Sigma}}_{35} \cdot \bm{\pi}\,
\Lambda_{27}^0
- \frac{4\sqrt{10}}{3} \overline{K}\, \overline{\bm{\Sigma}}_{35} \cdot
\bm{\tau}\, N_{27}
\nonumber \\ && \mbox{}
+ \frac{\sqrt2}{3}\, \overline{\Sigma}^c_{35} \, \bm{T}(1,2) \cdot
\bm{\pi}\, \Sigma_{27,2}
- 4\sqrt3\, \overline{\Xi}_{35,3/2}\, \Xi_{27,3/2}\, \eta
\nonumber \\ && \mbox{}
- \frac{2\sqrt5}{\sqrt3}\, \overline{\Xi}_{35,3/2}\,
\bm{T}({\textstyle\frac32,\frac32}) \cdot \bm{\pi}\, \Xi_{27,3/2}
+ \sqrt6 \, \overline{\Xi}_{35,3/2}\, \overline{\bm{K}} \cdot
\bm{I}({\textstyle\frac32,2}) \, \Sigma_{27,2}
\nonumber \\ && \mbox{}
+ \sqrt{10}\, \overline{\Xi}_{35,3/2}\,
\bm{T}({\textstyle\frac32,\frac12}) \cdot \bm{\Sigma}_{27}\, K_c
+ \frac{2\sqrt{15}}{3}\, \overline{\Xi}_{35,3/2}\,
\bm{T}({\textstyle\frac32,\frac12}) \cdot \bm{\pi}\, \Xi_{27}
\nonumber \\ && \mbox{}
- 2\sqrt2\, \overline{\Xi}_{35,3/2}\,
\bm{T}({\textstyle\frac32,\frac12}) \cdot \bm{\Omega}_{27,1}\, K
- \frac{5\sqrt5}{3}\, \overline{\Xi}_{35}\, \bm{\tau} \cdot \bm{\pi} \,
\Xi_{27}
\nonumber \\ && \mbox{}
+ \frac{2}{\sqrt3}\, \overline{\Xi}_{35}\,
\bm{T}({\textstyle\frac12,\frac32}) \cdot \bm{\pi}\, \Xi_{27,3/2}
+ 2\sqrt{10}\, \overline{\Xi}_{35} \, K_c \, \Lambda_{27}^0
+ \frac{2\sqrt{10}}{\sqrt3} \, \overline{\Xi}_{35}\, \bm{\tau} \cdot
\bm{\Sigma}_{27}\, K_c
\nonumber \\ && \mbox{}
- \sqrt{\frac23}\, \overline{\Xi}_{35} \bm{\tau} \cdot
\bm{\Omega}_{27,1} \, K
- 6 \, \overline{\Omega}_{35,1}\, \Omega_{27,1}\, \eta
+ i 2\sqrt3 \left( \overline{\bm{\Omega}}_{35,1} \times
\bm{\Omega}_{27,1} \right) \cdot \bm{\pi}
\nonumber \\ && \mbox{}
- 2\sqrt6 \, \overline{K}\, \overline{\bm{\Omega}}_{35,1} \cdot
\bm{T}({\textstyle\frac12,\frac32})\, \Xi_{27,3/2}
- \sqrt{10} \, \overline{K} \, \overline{\bm{\Omega}}_{35,1} \cdot
\bm{\tau}\, \Xi_{27}
- 2\sqrt2\, \overline{\Omega}_{35}^-\, \bm{\pi} \cdot \bm{\Omega}_{27,1}
\nonumber \\ && \mbox{}
+ 2\sqrt{15}\, \overline{\Omega}_{35}^-\, \overline{K}\, \Xi_{27}
+ 4\sqrt3\, \overline{X} \, \bm{\tau} \cdot \bm{\Omega}_{27,1}\, K_c
\nonumber \\ && \mbox{}
+ \mbox{(H.c.)}.
\end{eqnarray}

\subsubsection{Other interactions}

Other interactions of pentaquarks can be obtained directly from the
results in the previous subsections.
Namely, the interactions for $\bm{1}$-$\bm{8}$, $\bm{10}$-$\bm{8}$,
$\overline{\bm{10}}$-$\bm{8}$, and $\bm{27}$-$\bm{8}$ can be obtained from
Eqs.~(\ref{Lag:1-8}), (\ref{Lag:10-8}), (\ref{Lag:10bar-8}), and
(\ref{Lag:27-8}) by replacing the three-quark octet by pentaquark octet.
Also the interactions for $\bm{27}$-$\bm{10}$ and $\bm{35}$-$\bm{10}$ can be
read from Eqs.~(\ref{Lag:27-10}) and (\ref{Lag:35-10}), respectively, by
replacing the three-quark decuplet by pentaquark decuplet.
This completes the interactions of pentaquark baryons with pentaquark baryons
and meson octet.

\section{Mass relations}

In this Section, we derive several relations for the mass differences
among pentaquark baryons.
This can be obtained by using the Gell-Mann--Okubo mass formula,
\begin{equation}
M = M_0 + \alpha Y + \beta D^3_3,
\label{mass-form}
\end{equation}
where $M_0$ is a common mass of a given multiplet and $D^3_3 =
I(I+1) - Y^2/4 - C/6$ with $C = 2(p+q) + \frac23 (p^2 + pq +
q^2)$ for the $(p,q)$ representation.
In the tensor representation, one can obtain the same results as
follows.
Since all the particles belonging to an irreducible representation of
SU(3) are degenerate in the SU(3) symmetry limit, it is required to
include SU(3) symmetry breaking to obtain the mass splittings.
It is well-known that the Hamiltonian which breaks SU(3) symmetry but
still preserves the isospin symmetry and hypercharge is proportional to
the Gell-Mann matrix $\lambda_8$.
Thus we introduce the hypercharge tensor as
\begin{equation}
\mathcal{Y} = \left( \begin{array}{ccc} 1 & 0 & 0 \\ 0 & 1 & 0 \\ 0 & 0
& -2 \end{array} \right).
\end{equation}
Then the baryon masses can be obtained by constructing all possible
contractions among irreducible tensors and the hypercharge tensor.
As the mass formulas contain several parameters which take different
values depending on the multiplet, we can obtain only the relations
between the masses.
Also note that we do not consider mixing among baryon multiplets in
this work.

\subsection{Octet}

For the pentaquark octet, the mass term is given by
\begin{eqnarray}
H_{\bf 8} = a \overline{P}^i_j P^j_i + b\overline{P}^i_j \mathcal{Y}^l_i P^j_l
    + c \overline{P}^i_j \mathcal{Y}^j_l P^l_i.
\label{H8}
\end{eqnarray}
Using the particle identifications for the octet tensor,
we obtain the octet mass as
\begin{eqnarray}
&&
N_8 = a+ b -2c , \qquad 
\Lambda_8 = a- b-c,
\nonumber \\ &&
\Sigma_8 = a + b+c , \qquad 
\Xi_8 = a - 2b + c.
\end{eqnarray} 
The particle name in each equation indicates its mass in this Section.
One can then easily find that Eq.~(\ref{H8}) leads to the Gell-Mann--Okubo
mass relation for the pentaquark octet,
\begin{eqnarray}
2(N_8 + \Xi_8) = 3 \Lambda_8 + \Sigma_8.
\end{eqnarray}

\subsection{Decuplet}

In this case, the fully contracted Hamiltonian reads
\begin{eqnarray}
H_{\bf 10}= a \overline{D}^{ijk} D_{ijk} + b \overline{D}^{ijk}
\mathcal{Y}^l_k D_{ijl}.
\end{eqnarray}
Working out the contraction, we obtain
\begin{eqnarray}
&&
\Delta_{10} = 6(a+b), \qquad \Sigma_{10} = 6 a ,
\nonumber \\ &&
\Xi_{10} = 6(a-b) , \qquad \Omega_{10} = 6(a-2b).
\end{eqnarray}
Note that we have the well-known equal spacing for the decuplet spectrum,
\begin{equation}
\Omega_{10} - \Xi_{10} = \Xi_{10} - \Sigma_{10} = \Sigma_{10} -
\Delta_{10}.
\end{equation}

\subsection{Antidecuplet}

The mass spectrum of antidecuplet is very similar to that of decuplet,
namely,
\begin{eqnarray}
H_{\bf \overline{10}}= a \overline{T}_{ijk} T^{ijk} + 
b \overline{T}_{ijk} \mathcal{Y}_l^k T^{ijl},
\end{eqnarray}
which gives
\begin{eqnarray}
&&
\Theta = 6(a-2b), \qquad
N_{\overline{10}} = 6(a-b),
\nonumber \\ &&
\Sigma_{\overline{10}} = 6 a, \qquad
\Xi_{\overline{10},3/2} = 6(a+b).
\end{eqnarray}
Again, we have equal spacing rule in the spectrum \cite{DPP97},
\begin{equation}
\Xi_{\overline{10},3/2} - \Sigma_{\overline{10}} =
\Sigma_{\overline{10}} - N_{\overline{10}} =
N_{\overline{10}} - \Theta.
\end{equation}

\subsection{27-plet}

There are three possible ways to contract upper and lower indices as in
the case of octet.
We have
\begin{eqnarray}
H_{\bf 27}= a \overline{T}^{ij}_{kl} T^{kl}_{ij} + 
b \overline{T}^{ij}_{kl} \mathcal{Y}^l_m T^{km}_{ij} + 
c \overline{T}^{ij}_{kl} \mathcal{Y}^m_j T^{kl}_{im}.
\end{eqnarray}
The masses are then given by
\begin{eqnarray}
&& \Theta_1 = 4a - 8b + 4c ,
\nonumber \\
&& \Delta_{27} = 4a -2b + 4c , \qquad 
N_{27}= 4a - \frac{28}{5} b + \frac{2}{5}c,
\nonumber \\
&&\Sigma_{27,2} = 4a + 4b + 4c , \qquad 
\Sigma_{27} = 4a -\frac{4}{5} b -\frac{4}{5} c , \qquad
\Lambda_{27} = 4a -\frac{16}{5} b -\frac{16}{5} c,
\nonumber \\
&&\Xi_{27,3/2}=4a+4b-2c , \qquad
\Xi_{27} = 4a+\frac25 b - \frac{28}{5} c,
\nonumber \\
&&
\Omega_{27,1} = 4a+4b -8c.
\end{eqnarray}
{}From the above results, we can obtain several mass relations,
\begin{eqnarray}
&&
3(\Sigma_{27} +\Theta_1) = 2(\Delta_{27} + 2 N_{27}), \qquad
\nonumber \\
&&
3(\Xi_{27,3/2} +2 \Theta_1) = 4\Delta_{27}+ 5 N_{27}, \qquad
3(\Xi_{27} +2 \Theta_1) =  \Delta_{27} + 8 N_{27},
\nonumber \\
&&
3(\Omega_{27,1} +3 \Theta_1) = 2 (\Delta_{27} + 5 N_{27}).
\end{eqnarray}
Some interesting relations can be found. 
First, one can find the analog of Gell-Mann--Okubo relation,
\begin{eqnarray}
2(N_{27}+\Xi_{27}) = 3\Lambda_{27} +\Sigma_{27}.
\end{eqnarray}
Second, we note that some of the {\bf 27}-plet members, i.e., $\Theta_1$,
$\Delta_{27}$, $\Sigma_{27,2}$, $\Xi_{27,3/2}$, and $\Omega_{27,1}$,
satisfy two independent equal-spacing rules,
\begin{eqnarray}
\Omega_{27,1} - \Xi_{27,3/2} &=&
\Xi_{27,3/2} - \Sigma_{27,2},
\nonumber \\ 
\Sigma_{27,2} - \Delta_{27} &=&
\Delta_{27} - \Theta_1.
\end{eqnarray}
Note that they are the states with maximum isospin for a given
hypercharge and the equal-spacing rule holds independently for the upper
half of the $\bm{27}$-plet weight diagram and for the lower half of the
weight diagram (Fig.~\ref{fig:27}).

\subsection{35-plet}

In this case, we have 
\begin{eqnarray}
H_{\bf 35}= a \overline{T}^{jklm}_i T^{i}_{jklm} + 
b \overline{T}^{jklm}_i \mathcal{Y}^n_j T^{i}_{nklm} + 
c \overline{T}^{jklm}_i \mathcal{Y}^i_n T^{n}_{jklm}.
\end{eqnarray}
The obtained masses are
\begin{eqnarray}
&&\Theta_2 = 24a + 24 b - 48 c,
\nonumber \\
&&\Delta_{5/2} = 24 a + 24 b + 24 c, \qquad
\Delta_{35} = 24a +9b -36 c,
\nonumber \\
&&\Sigma_{35,2} = 24a + 6b + 24c , \qquad
\Sigma_{35} = 24 a -6b -24c,
\nonumber \\
&& \Xi_{35,3/2} = 24a - 12 b + 24c , \qquad
\Xi_{35} = 24a -21 b -12 c,
\nonumber \\
&&\Omega_{35,1} = 24a -30b+24c , \qquad
\Omega_{35} = 24a - 36b,
\nonumber \\
&& X = 24a -48b + 24c.
\end{eqnarray}
Here we observe that there are two sets of baryons which satisfy the
equal-spacing rule separately.
Namely, ($\Theta_2$, $\Delta_{35}$, $\Sigma_{35}$, $\Xi_{35}$,
$\Omega_{35}$) are equally spaced and another equal spacing rule holds for
($\Delta_{5/2}$, $\Sigma_{35,2}$, $\Xi_{35,3/2}$, $\Omega_{35,1}$, $X$),
\begin{eqnarray}
&&
\Omega_{35} - \Xi_{35} = \Xi_{35} - \Sigma_{35} =
\Sigma_{35} - \Delta_{35} = \Delta_{35} - \Theta_2, 
\nonumber \\ &&
X - \Omega_{35,1} = \Omega_{35,1} - \Xi_{35,3/2} = \Xi_{35,3/2} -
\Sigma_{35,2} = \Sigma_{35,2} - \Delta_{5/2}.
\end{eqnarray}
One can derive other mass relations such as
\begin{eqnarray}
&&
5(\Theta_2 +\Sigma_{35,2}) = 2 (2 \Delta_{5/2} + 3 \Delta_{35}),
\qquad
5(\Sigma_{35} - \Sigma_{35,2}) = -4 (\Delta_{5/2} - \Delta_{35}),
\nonumber \\ &&
5(\Xi_{35} - 2 \Sigma_{35,2}) = -8 \Delta_{5/2} +3 \Delta_{35},
\qquad
5(\Omega_{35} - 3 \Sigma_{35,2}) = -2 (6 \Delta_{5/2} -  \Delta_{35}).
\end{eqnarray}

\section{Summary}

In summary, we have obtained the flavor wave functions of all pentaquark
baryons in SU(3) quark model including $\bm{1}$, $\bm{8}$, $\bm{10}$,
$\overline{\bm{10}}$, $\bm{27}$, and $\bm{35}$.
Then the SU(3) symmetric Lagrangian for the interactions involving
pentaquark baryons is constructed and the mass relations among the
pentaquark baryons are derived.
Together with the mass sum rules, the obtained SU(3) selection rules and
SU(3) symmetry relations would be useful in searching for the pentaquark
baryons in future experiments and studying their production processes as
well as developing more sophisticated models for pentaquark structure.
The flavor-spin wave function of pentaquark baryons can be constructed by
extending our results to SU(6) \cite{BGS03} and the symmetry breaking
effects can be introduced to our SU(3) symmetric Lagrangian in a standard way.

\acknowledgments

We are grateful to Su Houng Lee and K. Nakayama for fruitful
discussions.
This work was supported by Forschungszentrum-J{\"u}lich, contract
No. 41445282 (COSY-058) and the Brain Korea 21 project of Korean
Ministry of Education.


\begin{thebibliography}{100}

\bibitem{LEPS03}
\mbox{LEPS Collaboration,} T.~Nakano {\em et~al.\/},
  Phys. Rev. Lett. {\bf 91}, 012002 (2003).
%%CITATION = HEP-EX 0301020;%%

\bibitem{DIANA03}
\mbox{DIANA Collaboration,} V.~V.~Barmin, {\em et~al.\/},
  Yad. Fiz. {\bf 66}, 1763 (2003),
  [Phys. At. Nucl. {\bf 66}, 1715 (2003)].
%%CITATION = HEP-EX 0304040;%%

\bibitem{CLAS03-b}
\mbox{CLAS Collaboration,} S.~Stepanyan {\em et~al.\/},
  Phys. Rev. Lett. {\bf 91}, 252001 (2003).
%%CITATION = HEP-EX 0307018;%%

\bibitem{SAPHIR03}
\mbox{SAPHIR Collaboration,} J.~Barth {\em et~al.\/},
  Phys. Lett. B {\bf 572}, 127 (2003).
%%CITATION = HEP-EX 0307083;%%

\bibitem{CLAS03-c}
\mbox{CLAS Collaboration,} V.~Kubarovsky {\em et~al.\/},
  Talk at CIPANP 2003, New York, 2003, hep-ex/0307088.
%%CITATION = HEP-EX 0307088;%%

\bibitem{ADK03}
A.~E. Asratyan, A.~G. Dolgolenko, and M.~A. Kubantsev,
  hep-ex/0309042.
%%CITATION = HEP-EX 0309042;%%

\bibitem{CLAS03-d}
\mbox{CLAS Collaboration,} V.~Kubarovsky {\em et~al.\/},
  Phys. Rev. Lett. {\bf 92}, 032001 (2004).
%%CITATION = HEP-EX 0311046;%%

\bibitem{AMW03}
S.~Armstrong, B.~Mellado, and S.~L. Wu,
  hep-ph/0312344.
%%CITATION = HEP-PH 0312344;%%

\bibitem{HERMES-04}
\mbox{HERMES Collaboration,} A.~Airapetian {\em et~al.\/},
  Phys. Lett. B {\bf 585}, 213 (2004).
%%CITATION = HEP-EX 0312044;%%

\bibitem{SVD04}
\mbox{SVD Collaboration,} A.~Aleev {\em et~al.\/},
  hep-ex/0401024.
%%CITATION = HEP-EX 0401024;%%

\bibitem{AER04}
P.~\mbox{Zh}. Aslanyan, V.~N. Emelyanenko, and G.~G. Rikhkvitzkaya,
  hep-ex/0403044.
%%CITATION = HEP-EX 0403044;%%

\bibitem{Salur04}
S.~Salur,
  nucl-ex/0403009.
%%CITATION = NUCL-EX 0403009;%%

\bibitem{PHENIX-04}
C.~Pinkenburg,
  nucl-ex/0404001.
%%CITATION = NUCL-EX 0404001;%%

\bibitem{NA49-03}
\mbox{NA49 Collaboration,} C.~Alt {\em et~al.\/},
  Phys. Rev. Lett. {\bf 92}, 042003 (2004).
%%CITATION = HEP-EX 0310014;%%

\bibitem{FW04}
H.~G. Fischer and S.~Wenig,
  hep-ex/0401014.
%%CITATION = HEP-EX 0401014;%%

\bibitem{WA89-04}
WA89 Collaboration, M.~I. Adamovich {\em et~al.\/},
  hep-ex/0405042.
%%CITATION = HEP-EX 0405042;%%

\bibitem{Pocho04}
J.~Pochodzalla,
  hep-ex/0406077.
%%CITATION = HEP-EX 0406077;%%

\bibitem{H1-04}
\mbox{H1 Collaboration,} A.~Aktas {\em et~al.\/},
  hep-ex/0403017.
%%CITATION = HEP-EX 0403017;%%

\bibitem{DKST03}
A.~R. Dzierba, D.~Krop, M.~Swat, S.~Teige, and A.~P. Szczepaniak,
  Phys. Rev. D {\bf 69}, 051901 (2004).
%%CITATION = HEP-PH 0311125;%%

\bibitem{Golo71}
E.~Golowich,
  Phys. Rev. D {\bf 4}, 262 (1971).
%%CITATION = PHRVA,D4,262;%%

\bibitem{PDG86}
\mbox{Particle Data Group,} M.~Aguilar-Benitez {\em et~al.\/},
  Phys. Lett. B {\bf 170}, 1 (1986).
%%CITATION = PHLTA,B170,1;%%

\bibitem{GM99}
H.~Gao and B.-Q. Ma,
  Mod. Phys. Lett. A {\bf 14}, 2313 (1999).
%%CITATION = HEP-PH 0305294;%%

\bibitem{Lip87}
H.~J. Lipkin,
  Phys. Lett. B {\bf 195}, 484 (1987).
%%CITATION = PHLTA,B195,484;%%

\bibitem{GSR87}
C.~Gignoux, B.~Silvestre-Brac, and J.~M. Richard,
  Phys. Lett. B {\bf 193}, 323 (1987).
%%CITATION = PHLTA,B193,323;%%

\bibitem{FGRS89-ZR94}
S.~Fleck, C.~Gignoux, J.~M. Richard, and B.~Silvestre-Brac,
  Phys. Lett. B {\bf 220}, 616 (1989);
S.~Zouzou and J.-M. Richard,
  Few-Body Syst. {\bf 16}, 1 (1994).
%%CITATION = PHLTA,B220,616;%%
%%CITATION = HEP-PH 9309303;%%

\bibitem{Stan98-GRSP97}
M.~Genovese, J.-M. Richard, \mbox{Fl}. Stancu, and S.~Pepin,
  Phys. Lett. B {\bf 425}, 171 (1998);
\mbox{Fl}. Stancu,
  Phys. Rev. D {\bf 58}, 111501 (1998).
%%CITATION = HEP-PH 9712452;%%
%%CITATION = HEP-PH 9803442;%%

\bibitem{RS93}
D.~O. Riska and N.~N. Scoccola,
  Phys. Lett. B {\bf 299}, 338 (1993).
%%CITATION = PHLTA,B299,338;%%

\bibitem{OPM94b-OPM94c-OP95}
Y.~Oh, B.-Y. Park, and D.-P. Min,
  Phys. Lett. B {\bf 331}, 362 (1994);
  Phys. Rev. D {\bf 50}, 3350 (1994);
Y.~Oh and B.-Y. Park,
  Phys. Rev. D {\bf 51}, 5016 (1995).
%%CITATION = PHLTA,B331,362;%%
%%CITATION = PHRVA,D50,3350;%%
%%CITATION = PHRVA,D51,5016;%%

\bibitem{E791-98-E791-99}
\mbox{E791 Collaboration,} E.~M.~Aitala, {\em et~al.\/},
  Phys. Rev. Lett. {\bf 81}, 44 (1998);
  Phys. Lett. B {\bf 448}, 303 (1999).
%%%%CITATION = PRLTA,81,44;%%
%%%%CITATION = PHLTA,B448,303;%%

\bibitem{Cheung03}
K.~Cheung,
  hep-ph/0308176.
%%CITATION = HEP-PH 0308176;%%

\bibitem{CCH04-HL04}
H.-Y. Cheng, C.-K. Chua, and C.-W. Hwang,
  hep-ph/0403232;
X.-G. He and X.-Q. Li,
  hep-ph/0403191.
%%CITATION = HEP-PH 0403232;%%
%%CITATION = HEP-PH 0403191;%%

\bibitem{BKM04}
T.~E. Browder, I.~R. Klebanov, and D.~R. Marlow, Phys. Lett. B {\bf 587}, 62
  (2004).
%%CITATION = HEP-PH 0401115;%%

\bibitem{SWW04}
I.~W. Stewart, M.~E. Wessling, and M.~B. Wise, hep-ph/0402076.
%%CITATION = HEP-PH 0402076;%%

\bibitem{WM04c}
B.~Wu and B.-Q. Ma, hep-ph/0402244.
%%CITATION = HEP-PH 0402244;%%

\bibitem{KLO04}
H.~Kim, S.~H. Lee, and Y.~Oh, hep-ph/0404170.
%%CITATION = HEP-PH 0404170;%%

\bibitem{HKKK79}
T.~Hirose, K.~Kanai, S.~Kitamura, and T.~Kobayashi,
  Nuovo Cim. {\bf 50A}, 120 (1979);
C.~Fukunaga, R.~Hamatsu, T.~Hirose, W.~Kitamura, and T.~Yamagata,
  Nuovo Cim. {\bf 58A}, 199 (1980).
%%CITATION = NUCIA,A50,120;%%}
%%CITATION = NUCIA,A58,199;%%}

\bibitem{KMCM91}
V.~M. Karnaukhov, V.~I. Moroz, C.~Coca, and A. Mihul,
  Phys. Lett. B {\bf 281}, 148 (1992).
%%CITATION = PHLTA,B281,148;%%

\bibitem{ABGG90}
A.~V. Aref'ev, {\em et~al.\/},
  Yad. Fiz. {\bf 51}, 414 (1990)
  [Sov. J. Nucl. Phys. {\bf 51}, 264 (1990)].
%%CITATION = SJNCA,51,414;%%

\bibitem{BCGMP79}
J.~Amirzadeh, {\em et~al.\/},
  Phys. Lett. {\bf 89B}, 125 (1979).
%%CITATION = PHLTA,B89,125;%%

\bibitem{AGDD91}
B.~M. Abramov, {\em et~al.\/},
  Yad. Fiz. {\bf 53}, 179 (1991)
  [Sov. J. Nucl. Phys. {\bf 53}, 114 (1991)].
%%CITATION = SJNCA,51,414;%%

\bibitem{ACDD85}
D.~Aston, {\em et~al.\/},
  Phys. Rev. D {\bf 32}, 2270 (1985).
%%CITATION = PHRVA,D32,2270;%%

\bibitem{Lands99}
L.~G. Landsberg,
  Phys. Rep. {\bf 320}, 223 (1999);
\mbox{SPHINX Collaboration,} \mbox{Yu}.~M. Antipov, {\em et~al.\/},
  Yad. Fiz. {\bf 65}, 2131 (2002)
  [Phys. Atom. Nucl. {\bf 65}, 2070 (2002)].
%%CITATION = PRPLC,320,223;%%
%%CITATION = PANUE,65,2070;%%

\bibitem{early}
R.~L. Jaffe,
  Phys. Rev. D {\bf 15}, 267 (1977);
D. Strottman,
  {\it ibid.\/} {\bf 20}, 748 (1979);
H. Hogaasen and P. Sorba,
  Nucl. Phys. {\bf B145}, 119 (1978);
M. De Crombrugghe, H. Hogaasen, and P. Sorba,
  {\it ibid.\/} {\bf B156}, 347 (1979);
K. Maltman and S. Godfrey,
  {\it ibid.\/} {\bf A452}, 669 (1986);
A.~Ferrer, V.~F. Perepelitsa, and A.~A. Grigoryan,
  Z. Phys. C {\bf 56}, 215 (1992).
%%CITATION = PHRVA,D15,267;%%
%%CITATION = PHRVA,D20,748;%%
%%CITATION = NUPHA,B145,119;%%
%%CITATION = NUPHA,B156,347;%%
%%CITATION = NUPHA,A452,669;%%
%%CITATION = ZEPYA,C56,215;%%

\bibitem{Chem85-Man84-Pras87}
M.~Chemtob,
  Nucl. Phys. {\bf B256}, 600 (1985);
%%CITATION = NUPHA,B256,600;%%
A.~V. Manohar,
  {\it ibid.\/} {\bf B248}, 19 (1984);
%%CITATION = NUPHA,B248,19;%%
M.~Prasza{\l}owicz,
  in {\em Skyrmions and Anomalies\/}, edited by M.~Jezabeck
  and M.~Prasza{\l}owicz, (World Scientific, Singapore, 1987), p. 112.

\bibitem{Weig98}
H.~Weigel,
  Eur. Phys. J. A {\bf 2}, 391 (1998).
%%CITATION = HEP-PH 9804260;%%

\bibitem{DPP97}
D.~Diakonov, V.~Petrov, and M.~Polyakov,
  Z. Phys. A {\bf 359}, 305 (1997).
%%CITATION = HEP-PH 9703373;%%

\bibitem{PSTCG00}
M.~V. Polyakov, A.~Sibirtsev, K.~Tsushima, W.~Cassing, and K.~Goeke,
  Eur. Phys.  J. A {\bf 9}, 115 (2000).
%%CITATION = EPHJA,A9,115;%%

\bibitem{WK03}
H.~Walliser and V.~B. Kopeliovich,
  Zh. Eksp. Teor. Fiz. {\bf 124}, 483 (2003),
  [J. Exp. Theor. Phys. {\bf 97}, 433 (2003)].
%%CITATION = HEP-PH 0304058;%%

\bibitem{Pras03}
M.~Prasza{\l}owicz,
  Phys. Lett. B {\bf 575}, 234 (2003).
%%CITATION = HEP-PH 0308114;%%

\bibitem{IKOR03}
N.~Itzhaki, I.~R. Klebanov, P.~Ouyang, and L.~Rastelli,
  Nucl. Phys. {\bf B684}, 264 (2004).
%%CITATION = HEP-LAT 0309305;%%

\bibitem{JM03}
B.~K. Jennings and K.~Maltman,
  hep-ph/0308286.
%%CITATION = HEP-PH 0308286;%%

\bibitem{BFK03}
D.~Borisyuk, M.~Faber, and A.~Kobushkin,
  hep-ph/0307370.
%%CITATION = HEP-PH 0307370;%%

\bibitem{Zhu03-MNNRL03-SDO03-Eidem04}
S.-L. Zhu,
  Phys. Rev. Lett. {\bf 91}, 232002 (2003);
%%CITATION = HEP-PH 0307345;%%
R.~D. Matheus, F.~S. Navarra, M.~Nielsen, R.~da~Silva, and S.~H. Lee,
  Phys.  Lett. B {\bf 578}, 323 (2003);
%%CITATION = HEP-PH 0309001;%%
J.~Sugiyama, T.~Doi, and M.~Oka,
  Phys. Lett. B {\bf 581}, 167 (2004);
%%CITATION = HEP-PH 0309271;%%
M.~Eidem{\"u}ller,
  hep-ph/0404126.
%%CITATION = HEP-PH 0404126;%%

\bibitem{Cohen03-CL04}
T.~D. Cohen,
  Phys. Lett. B {\bf 581}, 175 (2004);
%%CITATION = HEP-PH 0309111;%%
T.~D. Cohen and R.~F. Lebed,
  Phys. Lett. B {\bf 578}, 150 (2004).
%%CITATION = HEP-PH 0309150;%%

\bibitem{Man04}
E.~Jenkins and A.~V. Manohar,
  hep-ph/0401190; hep-ph/0402024;
A.~V. Manohar,
  hep-ph/0404122.
%%CITATION = HEP-PH 0401190;%%
%%CITATION = HEP-PH 0402024;%%
%%CITATION = HEP-PH 0404122;%%

\bibitem{CFKK03}
F.~Csikor, Z.~Fodor, S.~D. Katz, and T.~G. Kov{\'a}cs,
  JHEP {\bf 0311}, 070 (2003).
%%CITATION = HEP-LAT 0309090;%%

\bibitem{Sasaki03}
S.~Sasaki,
  hep-lat/0310014.
%%CITATION = HEP-LAT 0310014;%%

\bibitem{CH04a}
T.-W. Chiu and T.-H. Hsieh,
  hep-ph/0403020.
%%CITATION = HEP-PH 0404007;%%

\bibitem{KL03a}
M.~Karliner and H.~J. Lipkin,
  hep-ph/0307243.
%%CITATION = HEP-PH 0307243;%%

\bibitem{JW03}
R.~Jaffe and F.~Wilczek,
  Phys. Rev. Lett. {\bf 91}, 232003 (2003).
%%CITATION = HEP-PH 0307341;%%

\bibitem{OKL03b}
Y.~Oh, H.~Kim, and S.~H. Lee,
  Phys. Rev. D {\bf 69} 094009.
%%CITATION = HEP-PH 0310117;%%

\bibitem{CD04}
F.~E. Close and J.~J. Dudek,
  Phys. Lett. B {\bf 586}, 75 (2004).
%%CITATION = HEP-PH 0401192;%%

\bibitem{LKO04}
S.~H. Lee, H.~Kim, and Y.~Oh,
  hep-ph/0402135.
%%CITATION = HEP-PH 0402135;%%

\bibitem{DP03b}
D.~Diakonov and V.~Petrov,
  hep-ph/0310212.
%%CITATION = HEP-PH 0310212;%%

\bibitem{BGS03}
R.~Bijker, M.~M. Giannini, and E.~Santopinto,
  hep-ph/0310281.
%%CITATION = HEP-PH 0310281;%%

\bibitem{SR03}
\mbox{Fl}. Stancu and D.~O. Riska,
  Phys. Lett. B {\bf 575}, 242 (2003).
%%CITATION = HEP-PH 0307010;%%

\bibitem{CCKN03a}
C.~E. Carlson, C.~D. Carone, H.~J. Kwee, and V.~Nazaryan,
  Phys. Lett. B {\bf 573}, 101 (2003).
%%CITATION = HEP-PH 0307396;%%

\bibitem{CCKN04b}
C.~E. Carlson, C.~D. Carone, H.~J. Kwee, and V.~Nazaryan,
  Phys. Lett. B {\bf 579}, 52 (2004).
%%CITATION = HEP-PH 0310038;%%

\bibitem{GK03b}
S.~M. Gerasyuta and V.~I. Kochkin,
  hep-ph/0310227.
%%CITATION = HEP-PH 0310227;%%

\bibitem{KLV04}
N.~I. Kochelev, H.~J. Lee, and V.~Vento,
  hep-ph/0404065.
%%CITATION = HEP-PH 0404065;%%

\bibitem{Pras04}
M.~Prasza{\l}owicz,
  Acta Phys. Pol. {\bf B35}, 1625 (2004).
%%CITATION = HEP-PH 0402038;%%

\bibitem{WM04a-WM04b}
B.~Wu and B.-Q. Ma,
  Phys. Rev. D {\bf 69}, 077501 (2004);
  Phys. Lett. B {\bf 586}, 62 (2004).
%%CITATION = HEP-PH 0312041;%%
%%CITATION = HEP-PH 0312326;%%

\bibitem{EKP04}
J.~Ellis, M.~Karliner, and M.~Prasza{\l}owicz,
  hep-ph/0401127.
%%CITATION = HEP-PH 0401127;%%

\bibitem{DS04}
V.~Dmitra\v{s}inovi{\'c} and \mbox{Fl}. Stancu,
  hep-ph/0402190.
%%CITATION = HEP-PH 0402190;%%

\bibitem{LK03a-LK03b}
W.~Liu and C.~M. Ko,
  Phys. Rev. C {\bf 68}, 045203 (2003);
  nucl-th/0309023.
%%CITATION = NUCL-TH 0308034;%%
%%CITATION = NUCL-TH 0309023;%%

\bibitem{NHK03}
S.~I. Nam, A.~Hosaka, and H.-C. Kim,
  Phys. Lett. B {\bf 579}, 43 (2003).
%%CITATION = HEP-PH 0308313;%%

\bibitem{OKL03a}
Y.~Oh, H.~Kim, and S.~H. Lee,
  Phys. Rev. D {\bf 69}, 014009 (2004).
%%CITATION = HEP-PH 0310019;%%

\bibitem{ZA03}
Q.~Zhao and J.~S. Al-Khalili,
  Phys. Lett. B {\bf 585}, 91 (2004).
%%CITATION = HEP-PH 0312348;%%

\bibitem{NL04}
K.~Nakayama and W.~G. Love,
  hep-ph/0404011.
%%CITATION = HEP-PH 0404011;%%

\bibitem{YCJ03}
B.-G. Yu, T.-K. Choi, and C.-R. Ji,
  nucl-th/0312075.
%%CITATION = NUCL-TH 0312075;%%

\bibitem{OKL03d}
Y.~Oh, H.~Kim, and S.~H. Lee,
  hep-ph/0312229.
%%CITATION = HEP-PH 0312229;%%

\bibitem{LKK03}
W.~Liu, C.~M. Ko, and V.~Kubarovsky,
  Phys. Rev. C {\bf 69}, 025202 (2004).
%%CITATION = NUCL-TH 0310087;%%

\bibitem{NT03}
K.~Nakayama and K.~Tsushima,
  Phys. Lett. B {\bf 583}, 269 (2004).
%%CITATION = HEP-PH 0311112;%%

\bibitem{OKL03c}
Y.~Oh, H.~Kim, and S.~H. Lee,
  Phys. Rev. D {\bf 69}, 074016 (2004).
%%CITATION = HEP-PH 0311054;%%

\bibitem{THH03}
A.~W. Thomas, K.~Hicks, and A.~Hosaka,
  Prog. Theor. Phys. {\bf 111}, 291 (2004).
%%CITATION = HEP-PH 0312083;%%

\bibitem{HBEK03}
C.~Hanhart {\it et al.\/},
  hep-ph/0312236.
%%CITATION = HEP-PH 0312236;%%

\bibitem{HHO03}
T.~Hyodo, A.~Hosaka, and E.~Oset,
  Phys. Lett. B {\bf 579}, 290 (2004).
%%CITATION = NUCL-TH 0307105;%%

\bibitem{CCKN03c}
C.~E. Carlson, C.~D. Carone, H.~J. Kwee, and V.~Nazaryan,
  hep-ph/0312325.
%%CITATION = HEP-PH 0312325;%%

\bibitem{KLLP03}
P.~Ko, J.~Lee, T.~Lee, and J.-H. Park,
  hep-ph/0312147.
%%CITATION HEP-PH 0312147;%%

\bibitem{LZHD04}
Y.-R. Liu, A.~Zhang, P.-Z. Huang, W.-Z. Deng, X.-L. Chen, and S.-L. Zhu,
  hep-ph/0404123.
%%CITATION = HEP-PH 0404123;%%

\bibitem{PS04}
S.~Pakvasa and M.~Suzuki,
  hep-ph/0402079.
%%CITATION = HEP-PH 0402079;%%

\bibitem{HL64}
H.~Harari and H.~J. Lipkin,
  Phys. Rev. Lett. {\bf 13}, 345 (1964).
%%CITATION = PRLTA,13,345;%%

\bibitem{Low}
F.~E. Low,
  {\em Symmetries and Elementary Particles\/}, (Gordon and Breach, New
  York, 1967).

\bibitem{deS63}
J.~J. de~Swart,
  Rev. Mod. Phys. {\bf 35}, 916 (1963), {\bf 37}, 326(E) (1965).
%%CITATION = RMPHA,35,916;%%

\bibitem{Close}
F. E. Close,
  {\em An Introduction to Quarks and Partons\/}, (Academic Press,
  London, 1979).

\bibitem{FR}
Fayyazuddin and Riazuddin,
  {\em A Modern Introduction to Particle Physics\/},
  (World Scientific, Singapore, 1992).

\bibitem{ZC04}
Q.~Zhao and F.~E. Close,
  hep-ph/0404075.
%%CITATION = HEP-PH 0404075;%%

\bibitem{CN69}
P.~Carruthers and M.~M. Nieto,
  Ann. Phys. (N.Y.) {\bf 51}, 359 (1969).
%%CITATION = APNYA,51,359;%%

\bibitem{HK03}
J.~Haidenbauer and G.~Krein,
  Phys. Rev. C {\bf 68}, 052201 (2003).
%%CITATION = HEP-PH 0309243;%%

\bibitem{Nuss03-CN03}
S.~Nussinov,
  hep-ph/0307357;
%%CITATION = HEP-PH 0307357;%%
A.~Casher and S.~Nussinov,
  Phys. Lett. B {\bf 578}, 124 (2004).
%%CITATION = HEP-PH 0309208;%%

\bibitem{ASW03}
R.~A. Arndt, I.~I. Strakovsky, and R.~L. Workman,
  Phys. Rev. C {\bf 68}, 042201 (2003), {\bf 69}, 019901(E) (2004);
%%CITATION = NUCL-TH 0308012;%%
  nucl-th/0311030.
%%CITATION = NUCL-TH 0311030;%%

\end{thebibliography}
\end{document}